\documentclass[aps,pra,amsmath,amssymb,showpacs,twocolumn,preprintnumbers,
floatfix,letterpaper]{revtex4-1}

\usepackage{bm}
\usepackage{epsfig}
\usepackage{graphicx}
\usepackage{graphics}
\usepackage{amsmath}
\usepackage{amssymb}
\usepackage{amsfonts}
\usepackage{color}

\newcommand{\be}{\begin{equation}}
\newcommand{\ee}{\end{equation}}
\newcommand{\ba}{\begin{eqnarray}}
\newcommand{\ea}{\end{eqnarray}}
\newcommand{\br}{\begin{array}}
\newcommand{\er}{\end{array}}
\newcommand{\Schro}{Schr\"o\-din\-ger }

\definecolor{darkgreen}{rgb}{0,0.55,0}

\begin{document}
\preprint{submitted to Phys. Rev. A; 7 January 2011}
\author{Xiaoxu~Guan$^1$}
\author{Klaus~Bartschat$^1$}
\author{Barry I. Schneider$^2$}
\affiliation{$^1$Department of Physics and Astronomy, Drake University, Des Moines, Iowa 50311, USA}
\affiliation{$^2$Physics Division, National Science Foundation, Arlington, Virgina 22230, USA}

\date{\today}

\title{\bf Breakup of the aligned H$_2$ molecule by xuv laser pulses: 
         A time-dependent treatment in prolate spheroidal coordinates}

\begin{abstract}
We have carried out calculations of the triple-differential cross section for one-photon
double ionization of molecular hydrogen for a central photon energy of $75$~eV,
using a fully {\it ab initio}, non\-perturbative approach to solve the
time-dependent \Schro equation in prolate spheroidal coordinates. The spatial
coordinates $\xi$ and $\eta$ are discretized in a finite-element 
discrete-variable representation.  The wave packet of the laser-driven 
two-electron system is propagated in time through 
an effective short iterative Lanczos method to simulate the double ionization of the
hydrogen molecule.  For both symmetric and asymmetric energy sharing, the present results 
agree to a satisfactory level with most earlier predictions for the absolute magnitude and the shape
of the angular distributions.  A notable exception, however, concerns the predictions of the recent time-independent calculations
based on the exterior complex scaling method in prolate spheroidal coordinates
[Phys.~Rev.~A~{\bf 82}, 023423 (2010)]. Extensive tests of the numerical implementation were performed, 
including the effect of truncating the Neumann 
expansion for the dielectronic interaction on the description of the initial bound state and the predicted
cross sections. We observe that the dominant escape mode
of the two photo\-electrons dramatically depends upon the energy sharing. 
In the parallel geometry, when the ejected electrons are collected along the direction
of the laser polarization axis, back-to-back escape is the dominant channel for strongly
asymmetric energy sharing, while it is completely forbidden if the two electrons share the excess energy equally. 
\end{abstract}

\maketitle

\section{Introduction}
A measurement of the complete breakup of the atomic helium target by xuv 
radiation
was achieved over $10$ years ago~{\cite{Brauning1998}}. Since
then, rapid
developments in strong xuv light sources and momentum imaging techniques have
made it 
possible to record all the
reaction fragments, nuclei and electrons,
in double photo\-ionization of the simplest two-electron hydrogen/deuterium molecule by  
one-photon absorption~\cite{Weber-1,Weber-2,Weber-3,Gisselbrecht2006}, and most recently also for
two-photon absorption~\cite{Jiang2010}. 
For the double
ionization of H$_2$ by single-photon absorption, only randomly oriented
molecules were investigated in earlier experiments (e.g.~\cite{Wightman1998}).
Using ``fixed-in-space" molecules, more  
recent experimental
efforts include the measurements of energy- and angle-resolved
differential cross sections
by Weber {\it et al.}\ 
for either equal energy sharing~\cite{Weber-1,Weber-2} or asymmetric energy
sharing~\cite{Weber-3}, and by Gisselbrecht {\it et al.}
\cite{Gisselbrecht2006}
for equal energy sharing at a photon energy of $76$~eV. 
These experimental studies were at least partially 
stimulated by the goal of understanding the similarities and differences between the hydrogen molecule
and its atomic counterpart helium. However, all recorded fully differential cross
sections to date suffer from some experimental uncertainties regarding the alignment angle of the molecule with
respect to the polarization vector of the laser and the emission angles of the photo\-electrons. 
 
From a theoretical point of view, the hydrogen molecule exhibits a significant complexity compared to helium and,
therefore, provides an enormous challenge to a fully {\it ab
initio} description inherent in a multi-center, multi-electron system.  The single-center
convergent close-coupling method was used to model the double
ionization of H$_2$ by Kheifets and Bray \cite{Kheifets2005-1,Kheifets2005-2}.
Later McCurdy, Rescigno, Mart\'in and their
collaborators~\cite{Van2004,Van2006-1,Van2006-2} implemented a formulation
based on time-independent exterior complex scaling (ECS)
in spherical coordinates, with the origin of the coordinate system placed at the center of the molecule, to
treat the double photo\-ionization at a photon energy of $75$~eV. The
radial coordinates of the two electrons are measured from the center,
and the radial parts of the 
wave function were either expanded in $B$-splines or
using a finite-element discrete-variable representation (FE-DVR).  

The time-dependent
close-coupling (TDCC) method~\cite{Colgan2007}, again in spherical coordinates, was also
extended to calculate the triple-differential cross section (TDCS) for
double photo\-ionization of the H$_2$ molecule. While the agreement between the published TDCSs  
from the ECS~\cite{Van2006-1,Van2006-2} and TDCC~\cite{Colgan2007} calculations is basically acceptable, 
noticeable discrepancies remained for a few particular geometries. In
the parallel geometry, for instance, where the molecular axis $\bm{\zeta}$ is chosen along the laser polarization
vector~$\bm{\epsilon}$, the coplanar TDCS predictions
from the ECS and TDCC calculations differ by up to $40$~percent when 
one of the electrons (we will refer to it as the ``fixed electron'' below) is observed along the direction 
perpendicular to the $\bm{\epsilon}$-axis.
In some other cases, there exists a noticeable ``wing" structure in the published TDCC predictions for equal
energy sharing. Additional TDCC calculations~\cite{Colgan2010} suggest that the agreement can be 
systematically improved, albeit the above-mentioned discrepancy still exists at a somewhat 
reduced level. 

Another independent approach~\cite{Tao2010} to this problem is the very
recent time-independent ECS treatment, formulated -- as in the current work -- in prolate spheroidal coordinates.
Quite surprisingly, the results of that calculation differed from both the earlier ECS and also the TDCC predictions, both obtained in spherical coordinates.  Specifically, the ECS prolate spheroidal calculations showed differences from the earlier spherical coordinate calculation for the TDCSs, at a level of about $20\%$ depending on the details of the energy sharing.  As will be demonstrated below, we have gone to considerable lengths in an attempt to resolve these discrepancies. However, significant differences between the present results and those of the ECS~\cite{Tao2010} still remain. 

Both the ECS~\cite{Van2006-1,Van2006-2} and the TDCC~\cite{Colgan2007} calculations made some attempt to deal with the experimental uncertainties in the scattering angles.  Given the experimental uncertainties and the differences in the previous calculations, however, it appeared worthwhile to investigate the computational  effort required to obtain accurate \hbox{TDCSs} before averaging over any experimental acceptance angles.  Consequently, the present calculation represents an independent implementation of the time-dependent FE-DVR approach in prolate spheroidal coordinates. As in the other approaches mentioned above, the inter\-nuclear separation ($R$) was held fixed at its equilibrium distance of $1.4$~bohr.  The two-center prolate system, with the foci located on the nuclei, provides  a suitable description for the two-center characteristics of the H$_2$ molecule.  The formulation of the \Schro equation in prolate spheroidal coordinates for diatomic molecules is not new. The pioneering work of Bates, \"Opik, and Poots~\cite{Bates1953} 
for the H$_2^+$ ion, which is exactly separable in prolate spheroidal coordinates,  already revealed the appealing features of the prolate system.  In particular, the electron-nuclear interaction is rendered benign in this coordinate system.  A partial list of recent applications of prolate spheroidal coordinates to diatomic molecules can be found in~\cite{Bachau2003,Barmaki2004,Vanne2004,Bandrauk2005,Tao2009}. 

As has been demonstrated in a number of recent publications, a grid-based approach provides a very appropriate description of laser-driven atomic and molecular physics when combined with an efficient time-propagation method such as the short iterative Lanczos (SIL) method~\cite{Park1986,Tannor2007}.   In the present  work, we employ the FE-DVR/SIL approach in prolate spheroidal coordinates to study the correlated response of a two-electron molecule in the double ionization process.

The remainder of this manuscript is organized as follows. 
After presenting the Hamiltonian of the hydrogen molecule in Section~\ref{sec.2} and
providing some details about the 
discretization of the system in an FE-DVR basis in Sec.~\ref{sec.3},
the solution of the two-center Poisson equation is presented in Sec.~\ref{sec.4}.  This is
followed by a description of the procedure for extracting the cross sections of interest in Sec.~\ref{sec.5}.
The results are presented and discussed in Sec.~\ref{sec.6}, before we finish with a summary in 
Sec.~\ref{sec.7}. 

\section{The Schr\"odinger equation in prolate spheroidal
coordinates}
\label{sec.2}
The prolate spheroidal coordinates with the two foci separated by a distance~$R$ are defined by
\begin{equation}
\xi=\frac{r_1+r_2}{R}, \hspace{3mm}
\eta=\frac{r_1-r_2}{R},
\end{equation}
and the azimuthal angle~$\varphi$. Here $r_1$ and $r_2$ are the distances
measured from the two nuclei, respectively. These coordinates are specified in
the ranges $\xi\in[1,+\infty)$, $\eta\in[-1,+1]$, and $\varphi\in[0,2\pi]$.
The volume element is $dV=(R/2)^3(\xi^2-\eta^2)d\xi d\eta d\varphi$.
According to the asymptotic behaviors as
$r_1,r_2\rightarrow + \infty$, $\xi$ and $\eta$ approach $2r/R$
and $\cos\theta$, respectively, where $r$ and $\theta$ are the standard spherical coordinates.
Consequently, $\xi$ is the ``quasi-radial"
coordinate, while $\eta$ is ``quasi-angular". The
Hamiltonian of a single electron, ${\cal H}_q$, ($q=1,2$ for the two electrons in H$_2$ below), is written as

\begin{align}
{\cal H}_q = &
-\frac{2}{R^2(\xi_q^2-\eta_q^2)}\Bigg[\frac{\partial}{\partial
\xi_q}(\xi_q^2-1)\frac{\partial}{\partial \xi_q} +
\frac{\partial}{\partial\eta_q}(1-\eta_q^2)\frac{\partial}{\partial \eta_q}\\  \notag
+ &\frac{1}{(\xi_q^2-1)}\frac{\partial^2}{\partial \varphi_q^2} +
\frac{1}{(1-\eta_q^2)}\frac{\partial^2}{\partial\varphi_q^2}
\Bigg] 
-\frac{4\xi_q}{R(\xi_q^2-\eta_q^2)}.
\end{align}

We solve the time-dependent \Schro equation (TDSE) of
the laser-driven 
H$_2$ molecule (with two electrons) in the dipole length gauge:
\begin{equation}
i\frac{\partial}{\partial t}\Psi(1,2,t)=\Big[{\cal H}_1+{\cal H}_2
+\frac{1}{r_{12}} +\bm{E}(t) \cdot(\bm{r}_1+\bm{r}_2) \Big]\Psi(1,2,t).
\end{equation}
Here ${\bm r}_q$ is the coordinate of the $q$-th electron measured relative to the center of the molecule and $r_{12}=|\bm{r}_1-\bm{r}_2|$ is the interelectronic distance.
Without loss of generality, we choose the molecular axis along the $z$~axis, and 
the plane formed by the molecular axis and the polarization vector as
the $xz$~plane. Generally, we can decompose the polarization vector into its two components, 
 $\boldsymbol{\epsilon}=\cos\theta_N
\boldsymbol{e}_z + \sin\theta_N\boldsymbol{e}_x$, where $\theta_N$ is the angle
between the $\boldsymbol{\varsigma}$ and $\boldsymbol{\epsilon}$ axes, and
$\boldsymbol{e}_z$ and $\boldsymbol{e}_x$ are the unit vectors along the $z$ and $x$ axes, respectively.
The dipole interaction is therefore given as
\begin{align}
\bm{E}(t) \cdot(\bm{r}_1+\bm{r}_2)=&E(t)\Big[(z_1+z_2)\cos\theta_N \notag \\
+&(x_1+x_2)
\sin\theta_N\Big],
\end{align}
where the rectangular coordinates $x$ and $z$ and the prolate spheroidal 
coordinates are related through
\begin{equation}
x= \frac{R}{2}\sqrt{(\xi^2-1)(1-\eta^2)}\cos\varphi,  \hspace{3mm}
z = \frac{R}{2}\xi\eta.
\end{equation}
 We expand the wave function for the H$_2$ molecule in the body-frame as
\begin{align}
\Psi(1,2,t)
=&\sum_{m_1m_2}\Pi_{m_1m_2}(\xi_1,\eta_1,\xi_2,\eta_2,t)  \notag \\
\times &\Phi_{m_1m_2}(\varphi_1,\varphi_2). 
\end{align}
Here 
$ \Phi_{m_1m_2}(\varphi_1,\varphi_2) =e^{ i ( m_1\varphi_1 + m_2
\varphi_2 )}/(2\pi)$ is the angular function, where
$m_1$ and $m_2$ denote the magnetic quantum numbers of the two electrons
along the molecular axis.

Next, $\Pi_{m_1m_2}(\xi_1,\eta_1,\xi_2,\eta_2,t)$ is 
expanded in a product of  normalized ``radial" $\{f_{i}(\xi)\}$ and ``angular"
$\{g_{k}(\eta)\}$ DVR bases:
\begin{equation}
\begin{split}
\Pi_{m_1m_2}(\xi_1,\eta_1,\xi_2,\eta_2,t)  = \\ 
& \hspace{-3cm}\sum_{ijk\ell} f_i(\xi_1)
f_j(\xi_2)   g_k(\eta_1)g_\ell(\eta_2) C_{ijk\ell}^{m_1m_2}(t).
\end{split}
\end{equation}
Note that the basis is not symmetrized with respect to the coordinates of the two electrons.  Since we always begin the calculation with a properly symmetrized initial state, however, that symmetry will be preserved in the calculation.

To discretize this partial differential equation, we employ the 
FE-DVR approach for both the $\xi$ and $\eta$ variables~\cite{Tao2009,Guan2010}.  
If we normalize the DVR bases according to 
\begin{equation}
\int d\xi f_i(\xi) f_{i'}(\xi) = \delta_{ii'} \hspace{2mm} \mbox{and}
\hspace{1mm} 
\int d\eta g_k(\eta) g_{k'}(\eta) = \delta_{kk'},
\end{equation}
respectively, the overall $(\xi,\eta)$ DVR basis is not normalized with respect to the volume element in the prolate coordinate system.  This is corrected by defining the two-electron basis
\begin{align}
\label{eq:overall-basis}
b_{ijk\ell}^{m_1m_2}(1,2) = &
\Big(\frac{2}{R}\Big)^3\frac{1}{\sqrt{(\xi_i^2-\eta_k^2)(\xi_j^2-\eta_\ell^2)}}
\\
& \hspace{-4mm} \times f_i(\xi_1)f_j(\xi_2) g_k(\eta_1)g_\ell(\eta_2)
\Phi_{m_1m_2}(\varphi_1,\varphi_2) \nonumber,
\end{align} 
which satisfies the desired normalization
\begin{align}
\iint dV_1 dV_2  b_{ijk\ell}^{m_1m_2*}(1,2) b_{i'j'k'\ell'}^{m'_1m'_2}(1,2)
=&  \notag \\
&\hspace{-4cm}\delta_{ii'}\delta_{jj'}\delta_{kk'}\delta_{\ell\ell'}
\delta_{m_1m'_1}\delta_{m_2m'_2},
\end{align}
to expand $\Pi_{m_1m_2}(\xi_1,\eta_1,\xi_2,\eta_2,t)$. Specifically, we have
\begin{equation}
\Psi(1,2,t) = \sum_{m_1m_2}\sum_{ijk\ell}b_{ijk\ell}^{m_1m_2}(1,2)
X_{ijk\ell}^{m_1m_2}(t). 
\end{equation}
Introducing the normalized $(\xi,\eta)$ DVR basis eliminates the
complexities of matrix operations related to the overlap matrix at each time,
and hence makes the standard SIL algorithm directly applicable to
study the temporal response of the molecule to laser pulses.

\section{The FE-DVR basis}
\label{sec.3}
In our current implementation of the FE-DVR approach for the time-dependent wave
function in prolate spheroidal coordinates, we have chosen to work directly in the FE-DVR basis.  This differs from what is usually done for atoms in spherical coordinates, where spherical harmonics are used for the angular variables and an FE-DVR for the radial coordinates.   Consequently, the boundary
conditions in prolate spheroidal coordinates require some more discussion. 
Analyzing the asymptotics reveals that in the region near the boundaries of $\xi=1$ and $\eta=\pm
1$, which correspond to the molecular axis, the single-electron wave function
behaves like $(\xi^2-1)^{|m|/2}(1-\eta^2)^{|m|/2}$.  This indicates that the
physical wave function is finite for $|m|=0$, whereas it goes to zero in the region close to the
molecular axis for $|m|\neq 0$.  More importantly, the behavior of the wave function for odd $|m|$ contains a square-root factor, giving a decidedly non\-polynomial behavior to the wave function that is impossible to capture in a straightforward fashion using a DVR basis.  

The former problem is readily treated by using a Gauss-Radau quadrature in the first DVR element for~$\xi$, where only the right-most point is constrained to lie on the boundary between the first and second finite element.  The volume element ensures that the integrand is well behaved near the end points and makes it unnecessary to invoke a separate quadrature for different $m$ values.  For all the other elements, a Gauss-Lobatto quadrature is employed.  This allows us to make the FE-DVR
basis continuous everywhere and to satisfy the $|m|$-dependent boundary condition. 

To overcome the non\-analytic behavior of the basis for odd $m$,  Bachau and collaborators~\cite{Bachau2003} explicitly factored out the $(\xi^2-1)^{|m|/2}(1-\eta^2)^{|m|/2}$ part before the wave function was expanded in terms of $B$-splines in the discretization approach. 
We have adopted a similar idea in our FE-DVR treatment of the two-center problem to achieve much faster
convergence, as was also done in Ref.~\cite{Tao2009}. For the case of even $|m|$, no changes
need to be made to define the DVR basis, i.e., the normalized basis is written as
\begin{equation}
f_i(\xi) =\frac{1}{\sqrt{\omega_\xi^i}} \prod_{k\neq
i}\frac{\xi-\xi_k}{\xi_i-\xi_k} \hspace{3mm}\text{and} \hspace{3mm}
g_i(\eta) =\frac{1}{\sqrt{\omega_\eta^i}} \prod_{k\neq
i}\frac{\eta-\eta_k}{\eta_i-\eta_k}.
\end{equation}
For odd $|m|$, however, we define the DVR basis as
\begin{equation}
f_i(\xi) =\frac{1}{\sqrt{\omega_\xi^i}}\frac{(\xi^2-1)^{1/2}}{(\xi^2_i-1)^{1/2}}
\prod_{k\neq
i}\frac{\xi-\xi_k}{\xi_i-\xi_k}
\end{equation}
and
\begin{equation}
g_i(\eta)
=\frac{1}{\sqrt{\omega_\eta^i}}\frac{(1-\eta^2)^{1/2}}{(1-\eta^2_i)^{1/2}}
\prod_{k\neq
i}\frac{\eta-\eta_k}{\eta_i-\eta_k}.
\end{equation}
Here $\omega_\xi^i$ and $\omega_\eta^i$ are the weight factors related to the DVR bases $f_i(\xi)$ 
and $g_i(\eta)$, respectively.
The goal of using a unique set of mesh points, which are $|m|$-independent, to discretize the $(\xi,\eta)$ coordinates has now been achieved in this scheme. The same technique was employed in recent calculations of one- 
and two-photon double ionization of H$_2$~\cite{Tao2010,Guan2010}.
In principle, it is also possible to introduce the factors \hbox{$(\xi^2-1)^{|m|/2}(1-\eta^2)^{|m|/2}$} into the
DVR bases to circumvent the difficulties related to the non\-analytic behavior near the boundary. However, this results in an $|m|$-dependence of the DVR bases and quadrature points.  This, in turn, leads to a number of unnecessary complications in the practical implementation of the computational methodology.   One might argue that an  \hbox{$|m|$-dependent} discretization procedure could be useful for a system in which the magnetic quantum number $m$ is conserved.  An example is the H$_2^+$ ion in external magnetic fields along the molecular axis~\cite{Guan2003}.  However, that is not the situation in the current calculation.

\section{The Electron-Electron Coulomb Interaction in prolate spheroidal coordinates}
\label{sec.4}
Similar to the expansion of the electron-electron interaction in terms of
spherical coordinates, a counterpart exists in prolate spheroidal coordinates
\cite{Morse1953} through the Neumann expansion
\begin{align}
\label{ele}
\frac{1}{r_{12}}=&
\frac{1}{a}\sum_{l=0}^{\infty}\sum_{m=-l}^{l}(-1)^{|m|}(2l+1)
\bigg(\frac{(l-|m|)!}{(l+|m|)!} \bigg)^2 \\ \notag
\times& 
P_{l}^{|m|}(\xi_<)Q_{l}^{|m|}(\xi_>)P_{l}^{|m|}(\eta_1)P_{l}^{|m|}
(\eta_2)
e^{im(\varphi_1-\varphi_2)},
\end{align}
where $a\equiv R/2$.
The two nuclei are located at $\pm R/2$ along the $z$~axis and
$\xi_{>(<)}=\max(\min)(\xi_1,\xi_2)$. Both the
regular $P_{l}^{|m|}(\xi)$ and irregular $Q_{l}^{|m|}(\xi)$ Legendre functions~\cite{Handbook}, 
which are defined in the region
$(1,+\infty)$, are
involved in the expansion as the ``radial" part, while the ``angular" part is
only related to $P_{l}^{|m|}(\eta)$. Note that we chose to work in terms of an un-normalized ``angular" basis rather than the usual spherical harmonics.  The matrix elements of $1/r_{12}$ in a
traditional basis, for example, a \hbox{$B$-spline} or Slater-type basis, can be
computed through the well-known Mehler-Ruedenberg transformation
\cite{Vanne2004,Mehler1969}. 
Due to the discontinuous 
derivative along the line of $\xi_1=\xi_2$ in the Neumann expansion, 
the straight\-forward computation of the matrix element of $1/r_{12}$,
using the value of this interaction potential at the mesh points, 
is very slowly convergent. We seek a more robust representation of 
the $1/r_{12}$ matrix
which retains both the underlying Gauss quadrature and the DVR property of all potentials being
exactly diagonal with respect to the highly localized DVR basis.

In the following, we use the simplified notation
$|ijk\ell m_1m_2\rangle=|f_{i}(\xi_1)f_{j}(\xi_2)g_{k}(\eta_1)g_{\ell}
(\eta_2)\Phi_ { m_1m_2} (\varphi_1 , \varphi_2)\rangle$ to denote the basis.
Essentially, we need the integral 
\begin{align}
&\Big\langle ijk\ell m_1m_2 \Big|
P_{l}^{|m|}(\xi_<)Q_{l}^{|m|}(\xi_>)P_{l}^{|m|}(\eta_1)P_{l}^{|m|}
(\eta_2) \\
&\hspace{1.6cm} \times
e^{im(\varphi_1-\varphi_2)} \Big| i'k'j'\ell' m'_1m'_2 \Big\rangle. \notag
\end{align}
After integrating over $\varphi_1$ and $\varphi_2$, the matrix element of
$1/r_{12}$ can be written as
\begin{align}
&\Big\langle ijk\ell m_1m_2|\frac{1}{r_{12}}| i'j'k'\ell' m'_1m'_2\Big\rangle
 \\ \notag
& =\frac{1}{a}\sum_{l=|m|}^{\infty}
(-1)^{|m|}(2l+1)
\bigg(\frac{(l-|m|)!}{(l+|m|)!} \bigg)^2 {\cal I}_{ijk\ell}^{i'j'k'\ell}(l),
\end{align}
where the selection rule $m=m_1-m'_1=m'_2-m_2$ has been used. Hence 
$m$ is uniquely determined for a given pair of angular partial waves. Above we
introduced the reduced ($\xi,\eta$) integral 
\begin{align}
& {\cal I}_{ijk\ell}^{i'j'k'\ell'}(l) \\ \notag
&=\big\langle ijk\ell \big|
P_{l}^{|m|}(\xi_<)Q_{l}^{|m|}(\xi_>)P_{l}^{|m|}(\eta_1)P_{l}^{|m|}(\eta_2) \big|
 i'j'k'\ell' \big\rangle.
\end{align}
Since $|m|$ is fixed in the above equation, we omitted it in 
${\cal I}_{ijk\ell}^{i'j'k'\ell'}(l)$ and will do so in the related quantities below as well.
It is now worthwhile to define the two electron densities~\cite{McCurdy2004}:
\begin{equation}
\begin{split}
\rho_A(\xi,\eta) &= f_i(\xi)g_{k}(\eta)f_{i'}(\xi)g_{k'}(\eta), \\
\rho_B(\xi,\eta) &= f_j(\xi)g_{\ell}(\eta)f_{j'}(\xi)g_{\ell'}(\eta).
\end{split}
\end{equation}
After truncating the radial integral to the edge of the box, $\xi_{\rm max}$, this yields
\begin{align}
\begin{split}
{\cal I}_{ijk\ell}^{i'j'k'\ell'}(l) &=\iint dV_\xi dV_{\xi'} 
\rho_B(\xi,\eta)P_{l}^{|m|}(\xi_<)Q_{l}^{|m|}(\xi_>) \\ 
&\times P_{l}^{|m|}(\eta) P_{l}^{|m|}(\eta')
\rho_A(\xi',\eta') \\
&=\int_{1}^{\xi_{\rm max}}dV_\xi  P_{l}^{|m|}(\eta) \rho_B(\xi,\eta)
{\cal U}_l(\xi).
\label{intei}
\end{split}
\end{align}
Here a convention for the volume element was made in such a way that, 
for any function $F(\xi,\eta)$, we define $dV_\xi F(\xi,\eta) \equiv 
d\xi a^3\int_{-1}^{+1}(\xi^2-\eta^2)F(\xi,\eta)d\eta$ to simplify the
notation. Most importantly, the function ${\cal U}_l(\xi)$ is defined by
\begin{align}
{\cal U}_l(\xi) = &
Q_{l}^{|m|}(\xi)\int_{1}^{\xi}dV_{\xi'}\rho_{A}(\xi',\eta')P_{l}^{|m|}(\xi')
P_{l}^{ |m|} (\eta') \\ \notag 
+& P_{l}^{|m|}(\xi)\int_{\xi}^{\xi_{{\rm
max}}}dV_{\xi'}\rho_{A}(\xi',\eta')Q_{l}^{|m|}(\xi')P_{l}^{|m|}(\eta').
\label{defyl}
\end{align}

Instead of evaluating the above integrals directly, we solve the differential
equation satisfied by ${\cal U}_l(\xi)$.  
As will become apparent later, this equation can be shown to be the 
``radial" Poisson equation in the prolate spheroidal coordinate system.  The differential equations satisfied by the
Legendre functions $P_{l}^{|m|}(\xi)$ and $Q_{l}^{|m|}(\xi)$ suggests that we introduce the operator
\begin{equation}
\nabla_\xi^2=\frac{d}{d\xi}(\xi^2-1)\frac{d}{d\xi}-l(l+1)-\frac{m^2}{\xi^2-1}
\end{equation}
for given quantum numbers $l$ and $m$. This is the
one-dimensional Laplacian operator in the $\xi$ coordinate.

After some algebra, we obtain
\begin{align}
\frac{d}{d\xi}{\cal U}_{l}(\xi)=&
\frac{dQ_{l}^{|m|}(\xi)}{d\xi}\int_{1}^{\xi}dV_{ t }
\rho_A(t,\tau)P_{l}^{|m|}(t)P_{l}^{|m|}(\tau)  \\  \notag
+&
\frac{dP_{l}^{|m|}(\xi)}{d\xi}\int_{\xi}^{\xi_{\rm max}}dV_{t}
\rho_A(t,\tau)Q_{l}^{|m|}(t)P_{l}^{|m|}(\tau).
\end{align}
and 
\begin{align}
\begin{split}
\frac{d^2}{d\xi^2}{\cal U}_{l}(\xi) = &
\frac{d^2Q_{l}^{|m|}(\xi)}{d\xi^2}\int_{1}^{\xi}dV_{t}
\rho_A(t,\tau)P_{l}^{|m|}(t)P_{l}^{|m|}(\tau) \\ 
+& \frac{d^2P_{l}^{|m|}(\xi)}{d\xi^2}\int_{1}^{\xi}dV_{t}
\rho_A(t,\tau)Q_{l}^{|m|}(t)P_{l}^{|m|}(\tau)+ \\
+W(P_{l}^{|m|} & (\xi),Q_{l}^{|m|}(\xi)) a^3\int_{-1}^{+1}\!\!d\tau
(\xi^2\!-\!\eta^2)\rho_{A}(\xi,\tau)
P_{l}^{|m|}(\tau).
\end{split}
\end{align}
Here the Wronskian of the Legendre functions $P_{l}^{|m|}(\xi)$ and
$Q_{l}^{|m|}(\xi)$
is given by
\begin{equation}
W(P_{l}^{|m|}(\xi),Q_{l}^{|m|}(\xi)) =
\frac{(-1)^{|m|}}{(1-\xi^2)}
\frac{(l+|m|)!}{(l-|m|)!}.
\end{equation}
Consequently, we obtain
\begin{align}
\nabla_\xi^2 {\cal U}_{l}(\xi)= \varrho(\xi).
\label{poissoneq}
\end{align} 
This is the second-order inhomogeneous Poisson equation satisfied by
${\cal U}_{l}(\xi)$ with the ``source" term given by 
\begin{align}
\varrho(\xi) = &(-1)^{|m|+1}\frac{(l+|m|)!}{(l-|m|)!}a^3 \\ \notag 
 \times &\int_{-1}^{+1}d\eta'(\xi^2-\eta'^2)
\rho_{A}(\xi,\eta')P_{l}^{|m|}(\eta').
\end{align}
After carrying out the integral over $\eta'$ via
Gauss quadrature, we recast the source term as
\begin{align}
\begin{split}
\varrho(\xi) = &
\delta_{kk'}(-1)^{|m|+1}\frac{(l+|m|)!}{(l-|m|)!}a^3f_{i}(\xi)f_{i'} (\xi) \\
\times & (\xi^2-\eta_{k}^2) P_{l}^{|m|}(\eta_k).
\end{split}
\label{boundcond}
\end{align}
As one might expect, the inhomogeneous equation reduces to the homogeneous one
if $k\neq k'$. The solution to the Poisson equations (\ref{poissoneq}) and
(\ref{boundcond}) can be uniquely determined by enforcing the 
boundary conditions
\begin{equation}
{\cal U}_l(1) = P_{l}^{|m|}(1)\int_{1}^{\xi_{\rm
max}}dV_{\xi'}\rho_A(\xi',\eta')Q_{l}^{|m|}(\xi')P_{l}^{|m|}(\eta')
\label{bdl}
\end{equation}
at $\xi=1$ and
\begin{equation}
{\cal U}_l(\xi_{\rm max})= Q_{l}^{|m|}(\xi_{\rm max})\int_{1}^{\xi_{\rm
max}}dV_{\xi'}\rho_A(\xi',\eta)P_{l}^{|m|}(\xi')P_l^{|m|}(\eta')
\label{bdr}
\end{equation}
at $\xi=\xi_{\rm max}$, respectively. 

There are two important points to realize, namely: First, on the right-hand boundary, the
function ${\cal U}_l(\xi)$ assumes a non\-zero value, which is given by Eq.~(\ref{bdr}),
for all possible $|m|$ values. Its asymptotic behavior relies on the function $Q_{l}^{|m|}(\xi)$ at 
large $\xi$, which behaves like $1/\xi_{\rm max}^{l+1}$. This indicates that it
is {\em non\-zero} generally, although it could be small at the large
$\xi_{\rm max}$ values used in practical calculations. 
Second, on the left-hand boundary, the situation depends on the value of
$|m|$. ${\cal U}_{l}(\xi)$ takes a {\em non\-zero} value if $|m|=0$, while it
becomes {\em zero} if $|m|\neq 0$. 

Following the philosophy employed to handle the spherical case~\cite{McCurdy2004}, we
first seek a solution, ${\cal U}_l^0(\xi)$, to the Poisson equations
(\ref{poissoneq})-(\ref{bdr}) that satisfies the {\em zero}-value boundary condition at $\xi_{\rm max}$ by
using exactly the same $\xi$ mesh points as those for the wave functions.
In other words, we have $\nabla^2_\xi {\cal U}_l^0(\xi)=\varrho(\xi)$ with
${\cal U}_l^0(1)={\cal U}_l(1)$ and ${\cal U}_l^0(\xi_{\rm max})=0$.
After substituting the DVR expansion
${\cal U}_{l}^{(0)}(\xi)=\sum_{\mu}c_{\mu}f_{\mu}(\xi)$
of the solution into the differential
equation, we obtain a system of linear equations for the unknown
coefficients $\{c_{\mu}\}$:
\begin{align}
\begin{split}
\sum_{\mu'}c_{\mu'}T_{\mu\mu'}^{|m|}=&
(-1)^{|m|}\frac{(l+|m|)!}{(l-|m|)!}\frac{1}{\sqrt{\omega_{\xi}^{i}}}a^3
\delta_{\mu i}\delta_{ii'}\delta_{kk'} \\
\times &(\xi^2_i-\eta_k^2)P_{l}^{|m|}(\eta_k).
\end{split}
\end{align} 
The matrix $T$ is defined by its elements
\begin{align}
T_{\mu\mu'}^{|m|}= - &\int_{1}^{\xi_{\rm max}}d\xi
f_\mu(\xi)\bigg[(\xi^2-1)\frac{d^2}{d\xi^2}
+2\xi\frac{d}{d\xi}  \\ \notag
- & l(l+1)-\frac{m^2}{\xi^2-1} 
\bigg]f_{\mu'}(\xi).
\end{align}
Therefore, the coefficient $c_{\mu}$ can formally be written as
\begin{equation}
c_\mu= \frac{[T^{|m|}]^{-1}_{\mu i}}{\sqrt{\omega_\xi^i}}
(-1)^{|m|}\frac{(l\!+\!|m|)!}{(l\!-\!|m|)!}a^3\delta_{ii'}\delta_{kk'}
(\xi_i^2-\eta_k^2)P_ { l } ^ { |m| } (\eta_k),
\label{cpc}
\end{equation}
where $[T^{|m|}]^{-1}$ denotes the inverse of the matrix~$T^{|m|}$.
In this case ${\cal U}_l^0(\xi)$ fulfills the
left-hand boundary condition ${\cal U}_{l}^0(1)=0$, and so does ${\cal U}_{l}(1)$.
Recall, however, that its right-hand boundary condition differs from those of
${\cal U}_{l}(\xi_{\rm max})$. This suggests that the final answer to the
function ${\cal U}_l(\xi)$ can be constructed as ${\cal U}_{l}(\xi)={\cal
U}_{l}^0(\xi)+ F_l(\xi)$, i.e., we add the difference function $F_l(\xi)$ to
${\cal U}_l^0(\xi)$. The function $F_l(\xi)$
is also a solution to the homogeneous Poisson equation, subject to the 
boundary condition $F_{l}(1)=0$ and $F_{l}(\xi_{\rm max})={\cal
U}_l(\xi_{\rm max})$.  After writing it as a linear combination of
$P_{l}^{|m|}(\xi)$ and $Q_{l}^{|m|}(\xi)$, and imposing the boundary
conditions, $F_l(\xi)$ takes the form 
\begin{align}
\begin{split}
F_l(\xi) = &
\delta_{ii'}\delta_{kk'}a^3(\xi_i^2-\eta_k^2)P_{l}^{|m|}(\xi_i)P_{l
}^{|m|}(\eta_k) \\ 
 \times & \frac{Q_{l}^{|m|}(\xi_{\rm max})}{P_{l}^{|m|}(\xi_{\rm
max})}
P_{l}^{|m|}(\xi).
\end{split}
\end{align}
We finally arrive at
\begin{align}
\begin{split}
{\cal U}_{l}(\xi)  =& 
\frac{(-1)^{|m|}}{\sqrt{\omega_\xi^i}}
\frac{(l+|m|)!}{(l-|m|)!}a^3\delta_{ii'}\delta_{kk'}
(\xi_i^2-\eta_k^2)P_ { l } ^ { |m| } (\eta_k)   \\
\times &\sum_{\mu} [T^{|m|}]^{-1}_{\mu i}
f_{\mu}(\xi)   
+\delta_{ii'}\delta_{kk'}(\xi_i^2-\eta_k^2)P_{l}^{|m|}(\xi_i) \\
\times & P_{l}^{|m|}(\eta_k)
\frac{Q_{l}^{|m|}(\xi_{\rm max})}{P_{l}^{|m|}(\xi_{\rm max})}
P_{l}^{|m|}(\xi).
\label{solutionu}
\end{split}
\end{align}
At this point, the DVR version of the solution ${\cal U}_l(\xi)$ is ready for all
possible values of~$|m|$, either $|m|\neq 0$ or $|m|=0$. Substituting
Eq.~(\ref{solutionu}) into Eq.~(\ref{intei}) allows us to obtain the kernel
integral,
\begin{align}
\begin{split}
{\cal
I}_{ijk\ell}^{i'j'k'\ell'}(l)&=\delta_{ii'}\delta_{jj'}\delta_{kk'}
\delta_{\ell\ell'}a^6(\xi_i^2-\eta_k^2)(\xi_j^2-\eta_\ell^2) P_{l}^{|m|}(\eta_\ell)
\\ 
\times & \Bigg[
\frac{(-1)^{|m|}}
{\sqrt{\omega_\xi^i\omega_{\xi}^{j}}}\frac{(l+|m|)!}{(l-|m|)!}
[T^{|m|}]^{-1}_{ji}P_{l}^{|m|}(\eta_k)+ \\
+ & P_{l}^{|m|}(\xi_i)P_{l}^{|m|}(\xi_j)P_{l}^{|m|}(\eta_k)
\frac{Q_{l}^{|m|}(\xi_{\rm max})}{P_{l}^{|m|}(\xi_{\rm max})}\Bigg].
\end{split}
\end{align}
The matrix element of $1/r_{12}$ can finally be written as
\begin{widetext}
\begin{align}
\begin{split}
\big<ijk\ell m_1m_2\big|\frac{1}{r_{12}}\big|i'j'k'\ell'm'_1m'_2\big>= &
\delta_{ii'}\delta_{jj'}\delta_{kk'}
\delta_{\ell\ell'}a^5(\xi_i^2-\eta_k^2)(\xi_j^2-\eta_\ell^2)
\sum_{l \geqslant |m|}^{l_{\rm max}}(2l+1)
\frac{(l-|m|)!}{(l+|m|)!} P_{l}^{|m|}(\eta_k)P_{l}^{|m|}(\eta_\ell) \\ 
\times & \Bigg [\frac{1}{\sqrt{\omega_\xi^i\omega_\xi^j}}
[T^{|m|}]^{-1}_{ji}+(-1)^{|m|} \frac{(l-|m|)!}{(l+|m|)!} 
P_{l}^{|m|}(\xi_i) P_{l}^{|m|}(\xi_j)
\frac{Q_{l}^{|m|}(\xi_{\rm max})}{P_{l}^{|m|}(\xi_{\rm max})}
\Bigg],
\end{split}
\label{eq:1/r12}
\end{align}
\end{widetext}
where we truncated the $l$~summation in the Neumann expansion to~$l_{\rm max}$.
The above equation can be converted to the normalized $(\xi,\eta)$ bases with
the help of Eq.~(\ref{eq:overall-basis}).
This results in a diagonal representation of the 
matrix elements of the electron-electron Coulomb interaction and thus considerably
simplifies the FE-DVR discretization procedure. 
The above treatment of the $1/r_{12}$ matrix was successfully  applied to the
two-photon double ionization of H$_2$~\cite{Guan2010}.
The implementation of this representation will be illustrated below.

\section{Time Evolution and Extraction of Cross Sections}
\label{sec.5}
The time-dependent laser-driven electronic wave packet in the hydrogen molecule 
is obtained by solving the TDSE on the $(\xi,\eta)$ grid. Launched from
the previously determined ground state, the time
evolution of the system is achieved 
by using our recently developed SIL method
\cite{Guan2008,Guan2009}. The ground
state is determined by relaxing the system in imaginary time from an
initial guess of the wave function on the grid points.
At each time step we only need to generate the
values of the discretized wave function on the selected grid points. If desired, 
the information at arbitrary points within the spatial box can be obtained from the 
interpolation procedure in terms of the DVR bases. 

A few remarks seem appropriate
regarding the efficient implementation of the SIL algorithm.  The highest energy,
$E_{\rm max}$, which essentially depends on the
smallest  separation between the $(\xi,\eta)$ mesh points and also on the maximum values of $|m_{1}|$ and $|m_{2}|$,
determines the largest time step $\Delta t$ for the propagation in real time.
Typically, $E_{\rm max}$ is about $6,000$ atomic units (a.u.) in our
calculations. Although the chances of electrons populating
states with such high energies are practically negligible for short time scales 
of the laser-molecule interaction, 
we generally require $\Delta t\lesssim 2\pi/E_{\rm max}$ in order to resolve the 
most rapid oscillations in the time evolution. This means that at least a few time steps
are needed during one period of $2\pi/E_{\rm max}$. 
We refer readers to Refs.~\cite{Guan2008, Guan2009,Taylor2003} for further
details and discussions behind
the SIL method. 

In order to ensure that the double-ionization wave packet is sufficiently far away from the
nuclei, and also that the two photo\-electrons are well separated, we allow the
system to evolve for a few more cycles in the field-free Hamiltonian,
i.e., after the laser pulse has died off. This is the wave packet we use to extract
the physical information. The ionization probabilities and the corresponding
cross sections are extracted by projecting the time-dependent wave packet onto 
uncorrelated two-electron continuum states satisfying the standard incoming boundary
conditions. The latter states of H$_2$ are constructed from the one-electron 
continuum state of the H$_2^+$ ion described in the following subsection.  

\subsection{Continuum states of H$_2^+$}

The field-free wave function $\Phi(\xi,\eta,\varphi)$ of the
one-electron
molecular ion is completely separable in prolate spheroidal coordinates. For a
given, and conserved,
magnetic quantum number $m$, the wave function takes the form
$\Phi(\xi,\eta,\varphi)=T_m(\xi)\Xi_m(\eta)\Phi_m(\varphi)$, where the
azimuthal dependence,  is given by $\Phi_m(\varphi)\equiv e^{im\varphi}/\sqrt{2\pi}$. The ``radial" part $T_{mq}(\xi)$ and the ``angular" part
$\Xi_{mq}(\eta)$ of the wave function satisfy the
equations
\begin{equation}
\bigg[
\frac{\partial}{\partial \xi}
(\xi^2\!-\!1)\frac{\partial}{\partial \xi}
-\frac{m^2}{(\xi^2\!-\!1)} 
+2R\xi+c^2\xi^2-A_{mq}\bigg]T_{mq}(\xi)=0
\label{eq:xi}
\end{equation}
and
\begin{equation}
\bigg[\frac{\partial}{\partial \eta}
(1-\eta^2)\frac{\partial}{\partial \eta}
-\frac{m^2}{(1-\eta^2)}-c^2\eta^2+A_{mq}\bigg]\Xi_{mq}(\eta)=0,
\label{eq:eta}
\end{equation}
respectively.
Here 
$c=kR/2$
for the continuum state whose momentum vector has the magnitude~$k$.
In addition, we need to introduce another quantum number~$q$, which denotes the number
of nodes of $\Xi_m(\eta)$ in the region $\eta\in [-1,+1]$, to label the
states, and finally the separation constant~$A_{mq}$.  

When the angular function  $\Xi_m(\eta)\Phi_m(\varphi)$ is discretized in
terms of the relevant DVR bases, a few ``spurious" solutions might be
encountered. This is caused by the residual errors associated with the Gauss
quadratures. Consequently, we expand the angular function, or
``spheroidal harmonics" function  
${\cal Y}_{\ell m}(\eta,\varphi)\equiv \Xi_{mq}(\eta)\Phi_{mq}(\varphi)$
with $\ell=|m|+q$
instead in terms of 
spherical harmonics.
These functions are normalized according to 
\begin{equation}
\int_{-1}^{+1}d\eta \int_0^{2\pi} d\varphi
{\cal Y}_{\ell m}^{*}(\eta,\varphi) {\cal Y}_{\ell'm'}(\eta,\varphi)  
= \delta_{mm'}\delta_{\ell\ell'}.
\end{equation}
After obtaining the separation constant $A_{mq}$ by solving Eq.~(\ref{eq:eta}), 
the ``radial" function $T_{mq}(\xi)$ is again expanded in terms of the DVR bases.
The last DVR point at $\xi=\xi_{\rm max}$ needs to be kept for the continuum
state. Asymptotically, the radial function behaves like 
\begin{equation}
T_{mq}(\xi) \rightarrow \frac{1}{\xi R}\sqrt{\frac{8}{\pi}}\sin\bigg[c\xi+
\frac{R}{c}\ln(2c\xi)-\frac{\ell\pi}{2}+\Delta_{mq}(k)\bigg]
\label{eq:asym}
\end{equation}
as $\xi\rightarrow+\infty$. Here 
$\Delta_{mq}(k)$ is the
two-center Coulomb phase shift. The normalization factor on either the energy or
the momentum scale and the Coulomb phase shift can be determined by matching the
numerical solution of $T_{mq}(\xi)$ according to its asymptotic behavior given
in Eq.~(\ref{eq:asym}).

The plane wave in prolate spheroidal
coordinates can be written as~\cite{Flammer}
\begin{equation}
e^{i\bm{k}\cdot\bm{r}} = 4\pi \sum_{\ell m}
i^{\ell}{\cal Y}_{\ell m}(\eta_r,\varphi_r) {\cal
Y}_{\ell m}^{*}(\eta_k,\varphi_k)
R_{\ell m}^{(k)}(\xi),
\end{equation}
where
$R_{\ell m}^{(k)}(\xi)\rightarrow
1/(c\xi)\sin\big[c\xi-\ell\pi/2\big]$
in the asymptotic region. Note that $\eta_k$ and~$\eta_r$ are related to the
directions of $\bm{k}$ and~$\bm{r}$ in spherical coordinates through
$\eta_{k,r}=\cos\theta_{k,r}$. The partial-wave expansion of
the plane wave $e^{i\bm{k}\cdot\bm{r}}$ reminds us that the two-center Coulomb
wave satisfying the incoming boundary condition can be expanded as
\begin{align}
\Phi_{\bm{k}}^{(-)}(\bm{r}) = &\frac{1}{k}\sum_{m=-\infty}^{+\infty}\sum_{\ell 
\geqslant |m|}i^{\ell} e^{-i\Delta_{mq}(k)} \notag \\
&\times {\cal Y}_{\ell
m}^{*}(\bm{k}){\cal Y}_{\ell m}(\eta_r,\varphi_r) T_{mq}^{(k)}(\xi).
\label{eq:con-one}
\end{align}
This function is normalized in momentum space according to
$\langle\Phi_{\bm{k}}^{(-)}|\Phi_{\bm{k}'}^{(-)}
\rangle=\delta(\bm{k}-\bm{k}')$, provided the asymptotic solution in
Eq.~(\ref{eq:asym}) is satisfied.

Uncorrelated two-electron continuum
states with total spin angular momentum~$S$ ($S=0$ in our case)  can generally be constructed as follows:
\begin{align}
\Phi_{\bm{k}_1 \bm{k}_2}^{(-)}(\bm{r}_1,\bm{r}_2)=& \\
\nonumber
&\hspace{-1.5cm}\frac{1}{\sqrt{2}}\Big[
\Phi_{\bm{k}_1}^{(-)}(\bm{r}_1) \Phi_{\bm{k}_2}^{(-)}(\bm{r}_2) + (-1)^S
\Phi_{\bm{k}_2}^{(-)}(\bm{r}_1) \Phi_{\bm{k}_1}^{(-)}(\bm{r}_2)
\Big].
\label{eq:two-continuum}
\end{align}
With the help of Eq.~(\ref{eq:con-one}), its partial-wave representation can be
written as
\begin{align}
&\Phi_{\bm{k}_1\bm{k}_2}^{(-)}(\bm{r}_1,\bm{r}_2) = \Big(\frac{2}{R}\Big)^3 
\frac{1}{\sqrt{2}}\frac{1}{k_1k_2} \sum_{\ell_1 m_1 \ell_2 m_2}i^{\ell_1+\ell_2} \nonumber \\
& \hspace{6mm} \times \sum_{ijk\ell}b_{ijk\ell}^{m_1m_2}(1,2)\sqrt{
(\xi_i^2-\eta_k^2)(\xi_j^2-\eta_\ell^2)}  \\ \nonumber
&\hspace{6mm}\times \Bigg[ 
e^{-i\big(\Delta_{|m_1|\ell_1}(k_1)+\Delta_{|m|_2\ell_2}(
k_2)\big)}
{\cal Y}_{\ell_1m_1}^*(\bm{k}_1)
{\cal Y}_{\ell_2m_2}^*(\bm{k}_2) \\ \nonumber
&\hspace{13mm}C_{ijk\ell}^{\ell_1m_1\ell_2m_2}(k_1,k_2)
 + (-1)^S (\bm{k}_1\leftrightarrow \bm{k}_2) \Bigg].
\label{eq:two-conti}
\end{align}
Here we introduced
\begin{align}
C_{ijk\ell}^{\ell_1 m_1 \ell_2 m_2}(k_1,k_2) = &\\
& \hspace{-2cm}\notag {\tilde T}_{\ell_1 |m_1|}^{(k_1)}(\xi_i)
{\tilde T}_{\ell_2 |m_2|}^{(k_2)}(\xi_j)
{\tilde \Xi}_{\ell_1 |m_1|}^{(k_1)}(\eta_k)
{\tilde \Xi}_{\ell_2 |m_2|}^{(k_2)}(\eta_\ell),
\end{align}
by representing the radial and angular parts on the $(\xi,\eta)$ grid points:
\begin{equation}
T_{\ell m}^{(k)}(\xi) =\sum_{i} f_i(\xi) {\tilde T}_{\ell m}^{(k)}(\xi_i),
\end{equation}
\begin{equation}
\Xi_{\ell m}^{(k)}(\eta) =\sum_{\mu} g_{\mu}(\eta) {\tilde \Xi}_{\ell
m}^{(k)}(\eta_{\mu}).
\end{equation}
Here the exchange symmetry
\begin{equation}
C_{ji\ell k}^{\ell_2 m_2 \ell_1 m_1}(k_2,k_1)
=C_{ijk\ell}^{\ell_1 m_1 \ell_2 m_2}(k_1,k_2)
\end{equation}
is satisfied.

\subsection{Extraction of double-ionization cross sections}

It has been demonstrated \cite{Colgan2001,Guan2008,Guan2010}
that using 
uncorrelated two-electron continuum
states is a good approximation in a time-dependent propagation approach, provided the two
ejected electrons are well separated from each other.  
The probability amplitude of double ionization is then given by
\begin{align}
\langle 
\Phi_{\bm{k}_1\bm{k}_2}^{(-)} |\Psi(t) \rangle
=& \\  \nonumber
&\hspace{-2cm} \frac{1}{k_1k_2}
\sum_{m_1\ell_1m_2\ell_2}(-i)^{\ell_1+\ell_2} 
e^{i\big(\Delta_{|m_1|\ell_1}(k_1)+\Delta_{|m|_2\ell_2}(
k_2)\big)}  \\ \nonumber
&\hspace{-1cm}\times {\cal Y}_{\ell_1m_1}(\bm{k}_1)
{\cal Y}_{\ell_1m_2}(\bm{k}_2)
 \mathfrak{F}_{\ell_1m_1\ell_2m_2}(k_1,k_2),
 \label{eq:project}
\end{align}
where 
 \begin{align}
 \mathfrak{F}_{\ell_1m_1\ell_2m_2}(k_1,k_2)= &\\ \nonumber
 &\hspace{-2cm}\sqrt{2}\sum_{ijk\ell}C_{ijk\ell}^{
\ell_1m_1\ell_2m_2* } (k_1 , k_2)
 {X}_{ijk\ell}^{m_1m_2}(t).
 \end{align}
Here the exchange symmetry
\begin{equation}
 \mathfrak{F}_{\ell_2m_2\ell_1m_1}(k_2,k_1)
 =(-1)^S\mathfrak{F}_{\ell_1m_1\ell_2m_2}(k_1,k_2)
\end{equation}
is satisfied.
We also see that the probability amplitude formulated in prolate
spheroidal coordinates takes a similar form as for the atomic
case in spherical coordinates.  However, a subtle difference from the atomic case
is worth pointing out. Strictly speaking, the 
spheroidal harmonics involved in the probability amplitude generally depend on
the magnitude of the momenta $k_1$ and $k_2$, in addition to their directions. 

The energy sharing of the two photo\-electrons can be specified by introducing the
hyper\-angle $\alpha=\tan^{-1}(k_2/k_1)$.
This describes the double-ionization reaction with kinetic
energies $E_1=E_{\rm exc}\cos^2\hspace{-0.8mm}\alpha$ and $E_2=E_{\rm
exc}\sin^2\hspace{-0.8mm}\alpha$ for the two ionized electrons, respectively.
Here $E_{\rm exc}$ is the available excess energy above the double-ionization 
threshold.  In the present work, specifically, $E_{\rm exc}=23.6$~eV for absorption of a
\hbox{$75$-eV} photon. 

For double ionization by one-photon absorption, the triple
differential cross section with respect to one of the kinetic energies and the
two solid angles $\hat{\bm{k}}_1$ and $\hat{\bm{k}}_2$ can be extracted using
the same formalism as in the corresponding He case~\cite{Colgan2001,Guan2008}, i.e.,   
\begin{align}
\frac{\mbox{d}^{3}\sigma}{\mbox{d}E_1\mbox{d}\hat{\bm{k}}
_1\mbox {d}\hat{\bm{k}}_2} =\frac{1}{k_1k_2\cos^2\hspace{-0.8mm}\alpha}
\frac{\omega}{I_0}\frac{1}{T_{\rm
eff}^{(1)}}
\int_{0}^{k_{\rm max} }\mbox{d} k'_1\int_{0}^{k'_1}\mbox{d} k'_2 &  \notag  \\
&\hspace{-8cm}\times k'_1\delta(k_2'-k_1'\tan\alpha)
\Bigg|
\sum_{m_1\ell_1m_2\ell_2}\hspace{-3mm}
(-i)^{\ell_1+\ell_2}
e^{i\big(\Delta_{|m_1|\ell_1}+\Delta_{|m|_2\ell_2}\big)} \notag \\
&\hspace{-7cm}\times
{\cal Y}_{\ell_1m_1}(k'_1,\hat{\bm{k}}_1)
{\cal Y}_{\ell_2m_2}(k'_2,\hat{\bm{k}}_2)
 \mathfrak{F}_{\ell_1m_1\ell_2m_2}(k'_1,k'_2)
\Bigg|^2.
\label{eq:tdcs}
\end{align}
Here $\omega$ and $I_0$ are the central photon energy and  the peak intensity of
the laser pulse, respectively, while $T_{\rm {eff}}^{(1)}$ denotes the {\em effective}
interaction time between the temporal electric laser field and the electrons in
the one-photon absorption process. 
For a laser pulse of time 
duration~$\tau$ with a sine-squared envelope for the field amplitude, $T_{\rm {eff}}^{(1)}=(3/8)\tau$.
Note that $T_{\rm eff}^{(1)}$ corresponds to the special
case of the generalized $N$-photon effective interaction time 
$T_{\rm eff}^{(N)}$ \cite{Nikolo2006} for a
one-photon reaction. Generalized cross sections
for two-photon double ionization of the hydrogen molecule were extracted 
in the same formalism~\cite{Guan2010}. 

In the TDCC treatment~\cite{Colgan2007},
different strategies were employed to describe the linear one-photon and the non\-linear
two-photon double ionization processes of atoms and molecules. For the one-photon case, 
the cross sections were obtained through the time derivative of the
double-ionization probability, $\partial P_{\rm ion}^{2+}(t) /\partial t$. 
The laser field
does not need to be turned off in this case. On the other
hand, a true laser pulse was used for the two-photon case and an effective
time, defined as the time
integral under a flat-top pulse with a smooth turn-on and turn-off, was
introduced. In the present work, we employed a unified formulation through
an effective interaction time for both 
one-photon and multi-photon ionization in laser pulses.

For the one-photon double ionization initialized from the lowest 
$X\,^1\Sigma_g$ state,  the two ejected electrons can only populate the final
$^1\Sigma_u$ and $^1\Pi_u$ continuum states, with the specifics depending on the relative
orientation of the molecular axis and the laser polarization vector. Consequently
only partial waves with 
{\it ungerade} parity [i.e., $(-1)^{\ell_1+\ell_2}=-1$] need to be included 
in Eq.~(\ref{eq:tdcs}). 

\section{RESULTS}
\label{sec.6}

\subsection{Preparation of the initial electronic $X\,^1\Sigma_g$ state}
For the non\-sequential double-ionization process induced by one- or two-photon
absorption, 
electronic correlation plays a dominant role, as the two photo\-electrons must
share the available excess energy  $E_{\rm exc}$. 
Double ionization by a single photon would not 
occur at all if the two-electron atom or molecule were approximated by an independent-electron
model. Therefore, the quality of the description of electron-electron correlation in a
laser-driven system is crucially important for accurate results to be obtained. The
Coulomb interaction between the two electrons has to be described in a consistent
manner for both the initial bound state and the time-evolved wave packet.
Before we go any further, it is worth discussing how we prepare the
initial $X\,^1\Sigma_g$ state at the equilibrium distance of $R=1.4$ bohr. 
 
As seen from Eq.~(\ref{eq:1/r12}), the magnetic quantum number in the Neumann
expansion of the matrix element of $1/r_{12}$ is uniquely determined by the
angular bases. However, this is not the case for the index $l$, if we choose to
discretize the coordinate $\eta$, rather then expanding that part of the wave function into spherical
harmonics. In practice, the summation over~$l$ must be truncated at a finite value 
of~$l_{\rm max}$.
In principle, the higher-order expansion terms
always guarantee well-converged results.  However, as mentioned earlier,
we approximate the relevant $\eta$-integrals by using
Gauss-Legendre quadrature. This is the price we have to pay for making
the dielectronic Coulomb potential diagonal in the DVR bases. As a
consequence, we need to determine how the approximation
introduced in the $\eta$-integrals for the two-electron integrals affects the
results for the cross sections of interest.

\begin{figure}[thb]
\centering
\includegraphics[width=8.6cm,clip=]{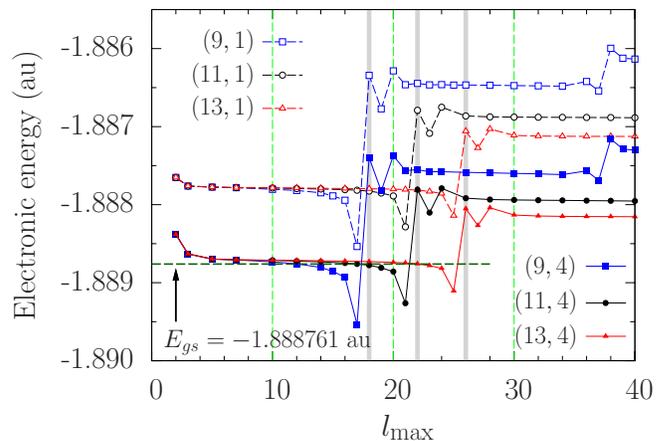}
\caption{(Color online) Energy of the lowest electronic $X\,^1\Sigma_g$ state at
$R=1.4$ bohr as a function of the $l_{\rm max}$ value used in the Neumann expansion of $1/r_{12}$.
The number of $\eta$ points, $n_\eta$, and the largest magnetic quantum number, 
$|m|_{\rm max}$, are labeled as $(n_\eta,|m|_{\rm max})$. The open symbols
correspond to $|m|_{\rm max}=1$, while the filled symbols are for $|m|_{\rm max}=4$. The
benchmark energy ($E_{gs}$) from Ref.~\cite{Sims2006} is shown as well.} 
\label{fig:energy}
\end{figure}

To answer this question, we first investigate the dependence of
the energy obtained for the initial $X\,^1\Sigma_g $ state on the value of  
$l_{\rm max}$ used in the Neumann expansion. Figure \ref{fig:energy} shows the
variation of the initial-state electronic energy of the hydrogen molecule with
respect to $l_{\rm max}$, obtained with a $\xi$ setup of ten elements in the
region of  $1 < \xi\leqslant 15.82$, five in the region $1 < \xi \leqslant 5$ and another
five in $5 \leqslant \xi \leqslant 15.82$. 
Each element, in turn, contains five~DVR points to further
discretize the
configuration space. Furthermore, we employ \hbox{$9$th-order} DVR points for~$\eta$. 
For a given number of $\eta$ mesh
points ($n_\eta$) and $|m|_{\rm max}=|m_1|_{\rm max}=|m_2|_{\rm max}$, 
we observe that the
resulting energy 
typically exhibits a plateau-like behavior with increasing $l_{\rm max}$. 
For given $n_\eta$, when $l_{\rm max}$ is relatively small, the $\eta$-integral
can be computed very accurately by using Gauss quadrature. However, when $l_{\rm max}$
is too large, the numerical errors introduced from the Gauss quadrature
cause the energy value to fluctuate. This occurs when $l_{\rm max}$ approaches $2n_\eta$ and
is shown by the grey stripes in Fig.~\ref{fig:energy}. In this region of~$l_{\rm max}$ 
an unphysically {\em low} energy can be produced. Beyond 
that point, the calculated energy increases to the next plateau. 

Ultimately, this is not too
surprising, since any Gauss quadrature is only reasonably accurate 
up to a limited polynomial order of the integrand. Consequently, if we want to keep more
terms in the Neumann expansion, we have to increase $n_\eta$
correspondingly. This finding is further substantiated by the dependence of the energy 
found for $n_\eta=11$ and $13$. The plateaus are indeed extended to the correspondingly larger 
values of $2n_\eta$. Most importantly,  the amplitude of the energy fluctuation is
systematically reduced with increasing $n_\eta$. The error in the energy is
lowered from $2.13\times 10^{-3}$ to $1.45\times 10^{-3}$ and finally 
$1.05\times 10^{-3}$~a.u., when $n_\eta$ increases from~$9$ to~$11$ and 
then~$13$ for $|m|_{\rm max}=4$. We obtained the energy at $R=1.4$ bohr as
$-1.8887324$~a.u.~for $l_{\rm max}=10$, $n_\eta=9$, and $|m|_{\rm max}=4$,
resulting in a double-ionization potential of $51.394$~eV.
Keeping the other parameters unchanged, we obtained an energy of $-1.8887128$~a.u.~for 
$n_\eta=11$. The benchmark energy in the literature is $-1.888761428$~a.u.~at 
the same $R$~\cite{Sims2006}, after we take out the nucleus-nucleus
interaction of $1/1.4$~a.u. 

To summarize: 
Unlike for other expansion parameters, it is important to be consistent in the size of the angular quadrature and the largest $l_{\rm max}$ employed in the Neumann expansion in practical calculations, if we discretize the
coordinate $\eta$. However, this provides a way to examine a potential sensitivity 
of the physical observables of interest (here the differential cross sections) to the
ground-state wave functions generated by varying $l_{\rm max}$ and other
parameters. This will be further discussed below.

\subsection{Convergence of the TDCS}
\begin{figure}[thb]
\centering
\includegraphics[width=8.0cm,clip=]{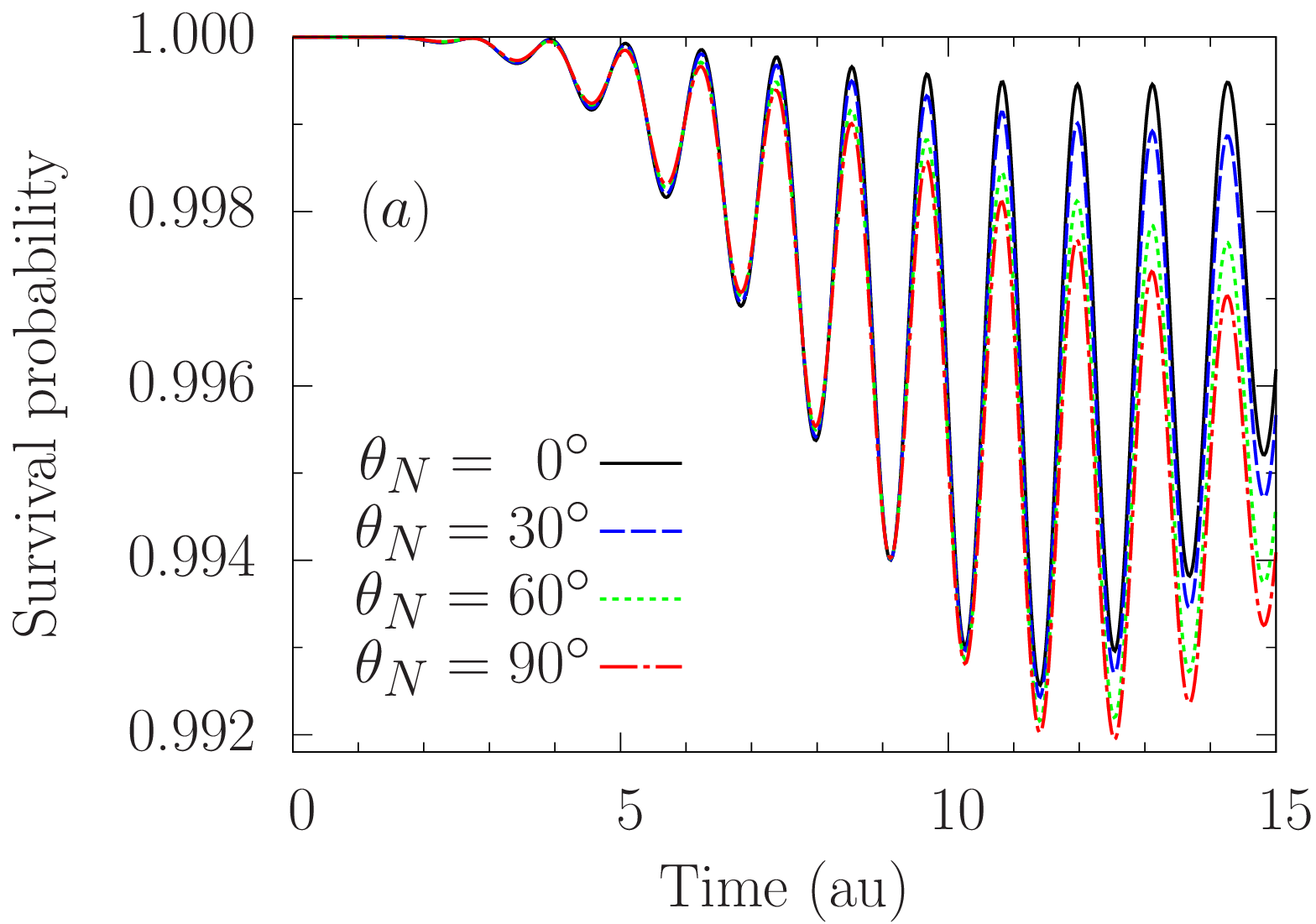}
\includegraphics[width=8.0cm,clip=]{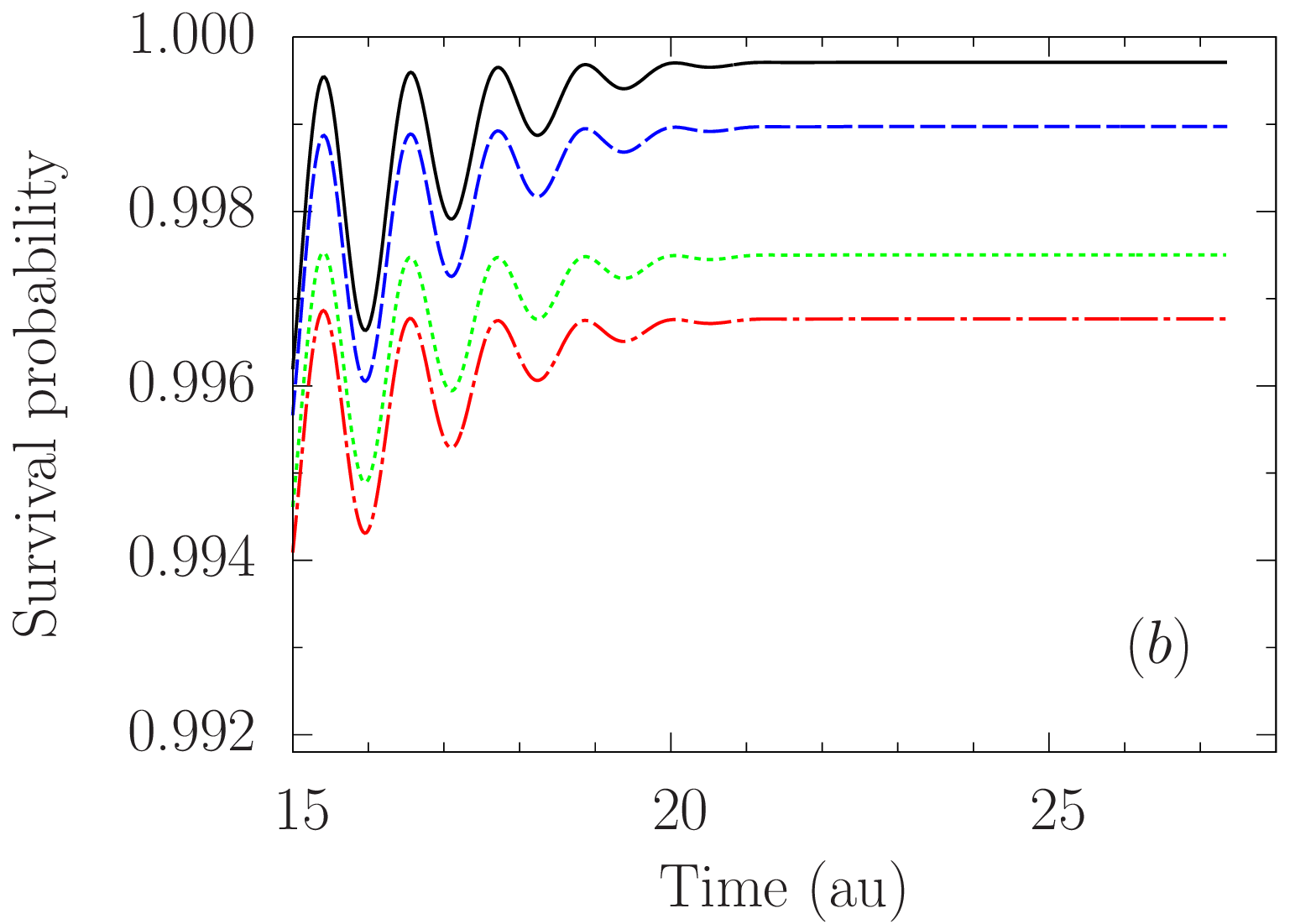}
\caption{(Color online) Survival probability of the hydrogen molecule subjected to a
sine-squared laser pulse with a peak intensity of $10^{15}$ W/cm$^2$. 
The laser pulse lasts for $10$ optical cycles and the system is followed for a period of another
$2$~cycles of field-free propagation. The central photon energy of the laser pulse is
$75$~eV.} 
\label{fig:surv}
\end{figure}

Before we present our results for the cross sections, let us take a closer look at the
survival probability 
\begin{equation}
P_{\rm surv} = |\langle \Psi_{\rm gs}|\Psi(t) \rangle|^2
\end{equation}
of the aligned H$_2$
molecule in its ground state~$\Psi_{\rm gs}$.  This is  
shown in Fig.~\ref{fig:surv}. 
For homo\-nuclear molecules, the independent alignment angle~$\theta_N$ between
the molecular axis and the polarization vector can be confined to the region from
$0^\circ$ to $90^\circ$.  In the xuv regime, we observe that the 
hydrogen molecule shows a larger probability of being ionized or excited 
(i.e., a lower probability of staying in the initial state)
at the end of the pulse in an aligned geometry. This indicates that the perpendicular 
component of the temporal
electric field exerts more influence on the ionization process due to the larger
dipole momentum. Interestingly, at the earlier stages of the time
evolution (e.g., $t\lesssim 9$~a.u.), when the ionized wave packet is driven
back by the change in direction of the electric field, the tilted molecule has a
larger probability of staying in its ground state. This happens near the various 
minima in $P_{\rm surv}$. However, once the electric field has become
sufficiently strong ($t\gtrsim 9$~a.u.), the wave packet is driven out and spread
into a larger space. This leads to lower minima in $P_{\rm surv}$ for the tilted
molecule.

When the wave packet is driven back to the nuclear region and
therefore has a
chance to recombine with the H$_2^+$ ion, a maximum in $P_{\rm surv}$ appears.
Not surprisingly, the parallel geometry always has the
largest probability for this to happen.  
Although the wave
packet can also be scattered for the untilted molecule in the plane perpendicular to the
molecular axis, the probability is undoubtedly larger if the laser electric
field is perpendicular to the molecular axis.  
A similar behavior of H$_2^+$ in xuv pulses was observed in 
Ref.~\cite{Hu2009}.

For most calculations performed in this study, we expose the hydrogen molecule to a laser pulse with a peak
intensity of $10^{15}$ W/cm$^2$.
Looking at Fig.~\ref{fig:surv} we see that the depletion of the initial ground state can be safely
neglected for our typical interaction times. Even for
$\theta_N=90^\circ$, $P_{\rm surv}=0.99677$ remains very close to unity. 
The negligible depletion of the ground state suggests that the concept of cross sections
is valid and applicable. On the other hand, it also presents a numerical challenge to
predict the cross sections accurately from a time-dependent treatment, due to the 
generally small ionization probability.

At first glance, a peak intensity of $10^{15}$ W/cm$^2$ might seem very intense for
most atomic and molecular targets. Here, however, we consider an 
xuv rather than an IR pulse. For an xuv pulse with central photon
energy of $75$~eV, such laser fields 
definitely fall into the ``weak-field" regime. 
The ponderomotive energy in the xuv regime is much smaller than the photon
energy of
interest.
 
\begin{table}
\caption{\label{tab:parameters} The discretization and expansion parameters of
the H$_2$ wave function in prolate spheroidal coordinates. Here $\xi_b$ stands
for the border between the inner and outer regions in the $\xi$ coordinate, while
$\xi_{\rm max}$ is the size of the $\xi$~box. In addition,
$n_\xi$ denotes the
number of $\xi$ mesh points in each element. 
The numbers of $\xi$ elements in the inner and
outer region are $n_{\rm inn}$ and $n_{\rm out}$, respectively. These $\xi$
parameters produce the total number of $\xi$ mesh points $N_\xi$. The $\xi$ grid~I 
and $\xi$ grid~II are used to
examine the convergence of our results.}
\begin{ruledtabular}
\begin{tabular}{lcccccc}
 & $\xi_b$& $n_{\rm inn}$& $n_{\rm out}$& $\xi_{\rm max}$& $n_\xi$&
$N_\xi$ \\ \hline
$\xi$ grid~I &     $5$&     $5$&  $67$&   $150$&  $5$&   $288$\\
$\xi$ grid~II&     $9$&     $1$&  $11$&   $97$&  $14$&  $156$\\
\end{tabular}
\end{ruledtabular}
\end{table}

In this work, we are mainly interested in the triple-differential cross section, 
since it reveals the fine details of possible energy sharings and preferred
directions of the ejected electrons in the double-ionization process.
Given the discrepancies between results from various calculations found in the literature, 
we carried out comprehensive convergence tests for our predictions of the TDCSs. 
These tests are essentially divided into two groups. 
The first group concerns the laser parameters, while the second one 
deals with the discretization and expansion parameters. An example of 
two different parameter sets for the $\xi$ grid is given in Table \ref{tab:parameters}
and will be further discussed below. 

In order to obtain a good handle on the sensitivity of the results to the various parameters and
the resulting level of ``convergence", we try to only vary a single
parameter while keeping all others fixed if possible. For the dependence
on the laser parameters, we use the $\xi$ grid~I combined with 
$(n_\eta,|m|_{\rm max},l_{\rm max})=(9,4,10)$. For the tests regarding the 
discretizations and expansions, the peak intensity of laser was fixed at
$10^{15}$ W/cm$^2$ and a time scale of ``$10+2$'' optical cycles (o.c.) was used.
Here ``$10+2$'' refers to a \hbox{$10$-cycle} laser pulse with a sine-squared
envelope for the field amplitude, followed by a \hbox{$2$-cycle} field-free
propagation.

\begin{figure}[thb]
\centering
\includegraphics[width=4.27cm,clip=]{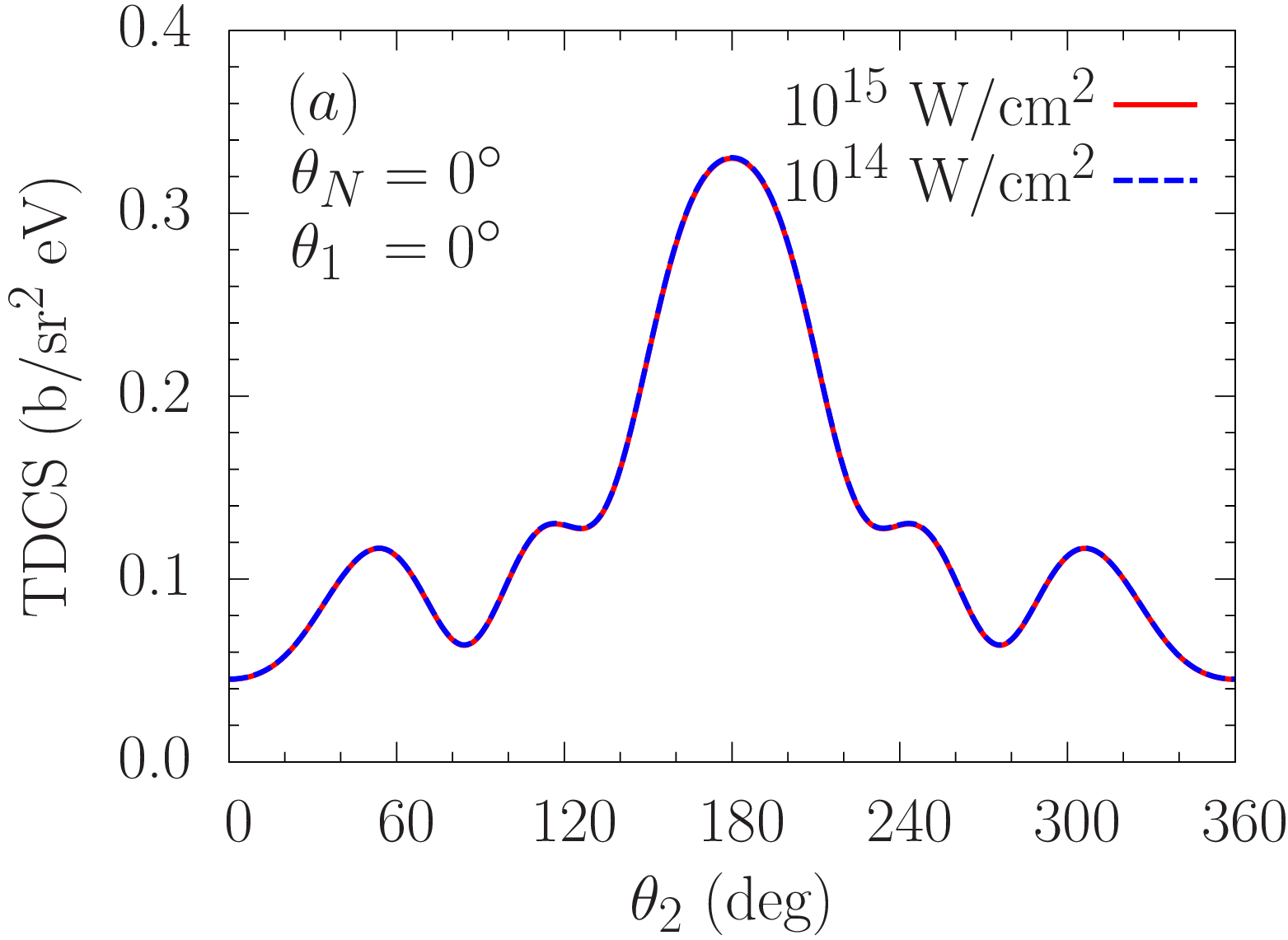}
\includegraphics[width=4.27cm,clip=]{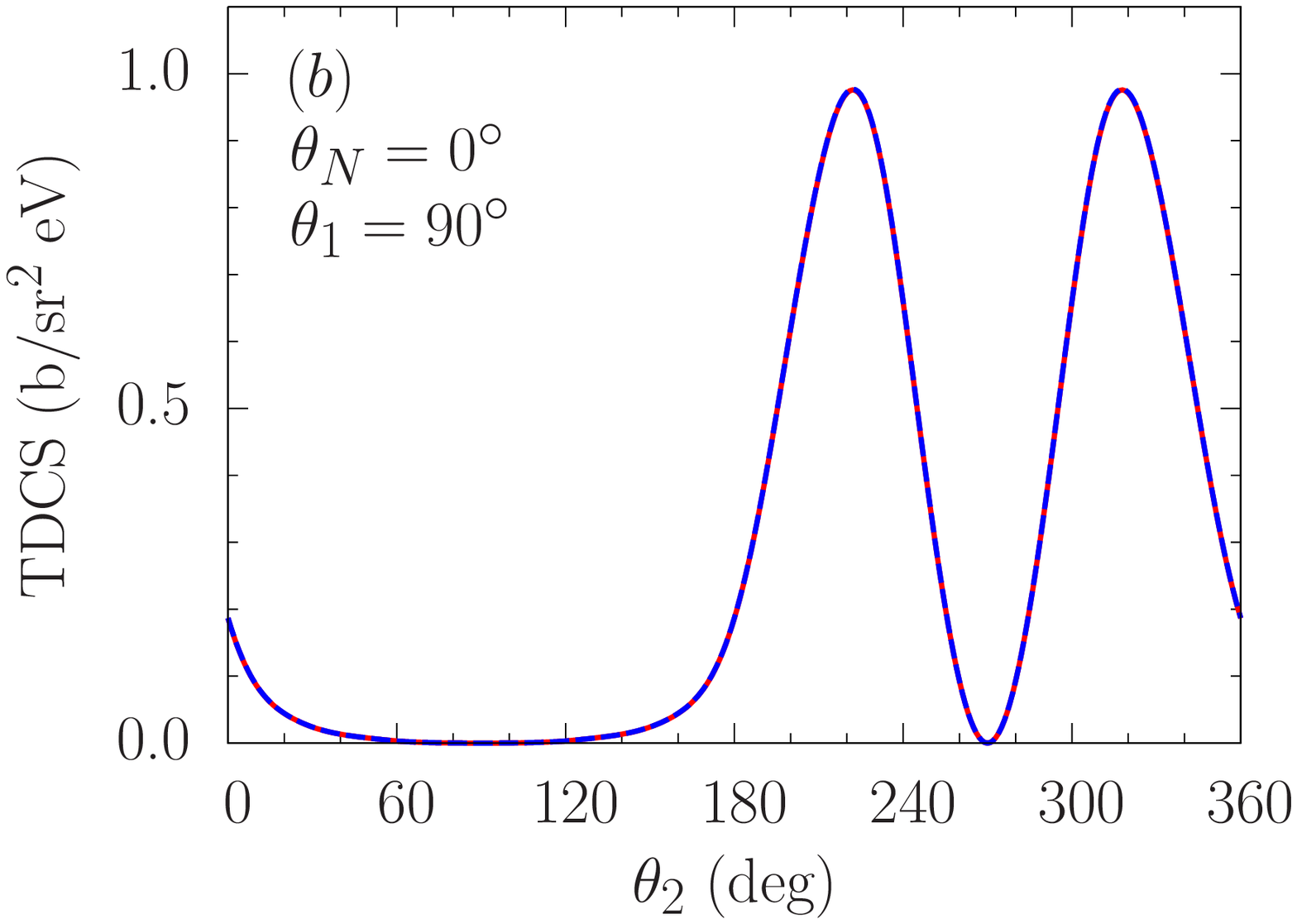} \\
\includegraphics[width=4.27cm,clip=]{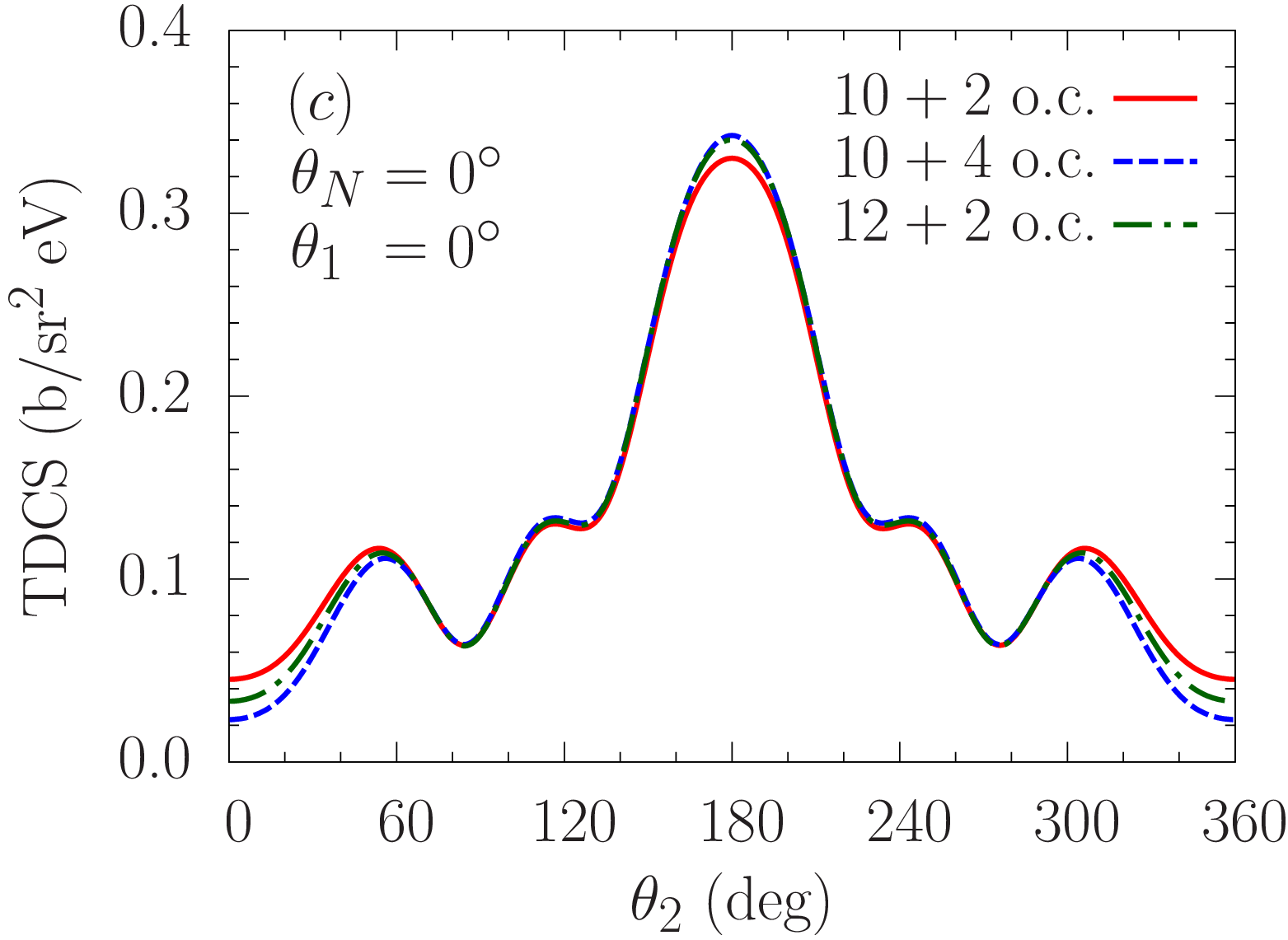}
\includegraphics[width=4.27cm,clip=]{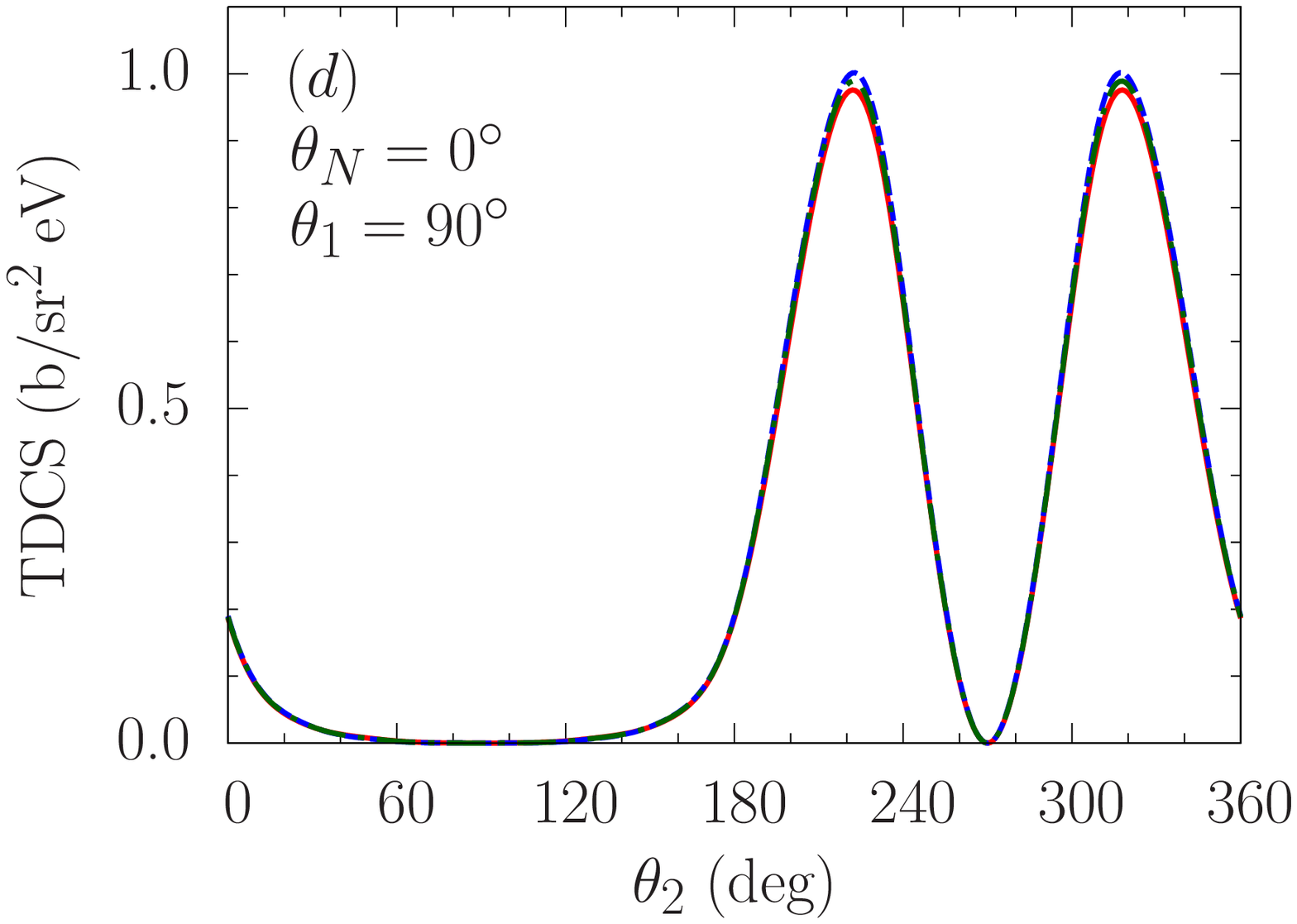}
\caption{(Color online) Convergence of the coplanar TDCS results for the hydrogen molecule
for asymmetric energy sharing with respect to the laser peak intensity and the time
scale. The central photon energy is $75$~eV. The slow
reference electron, observed at the fixed angle~$\theta_1$, 
takes away $20\%$ of the available excess energy $(E_1=4.7$~eV), while the other electron takes $80\%$ of 
$E_{\rm exc}$ $(E_2=18.9$~eV).  The peak laser intensity in
panels $(b)$-$(d)$ is $10^{15}$~W/cm$^2$. The two columns show the corresponding convergence of the TDCS for
$\theta_1 = 0^\circ$ (left) and $\theta_1 = 90^\circ$ (right), respectively. $1$~barn~(b) $=10^{-24}$ cm$^2$. 
} 
\label{fig:tdcs-laser}
\end{figure}

\begin{figure}[thb]
\centering
\includegraphics[width=4.27cm,clip=]{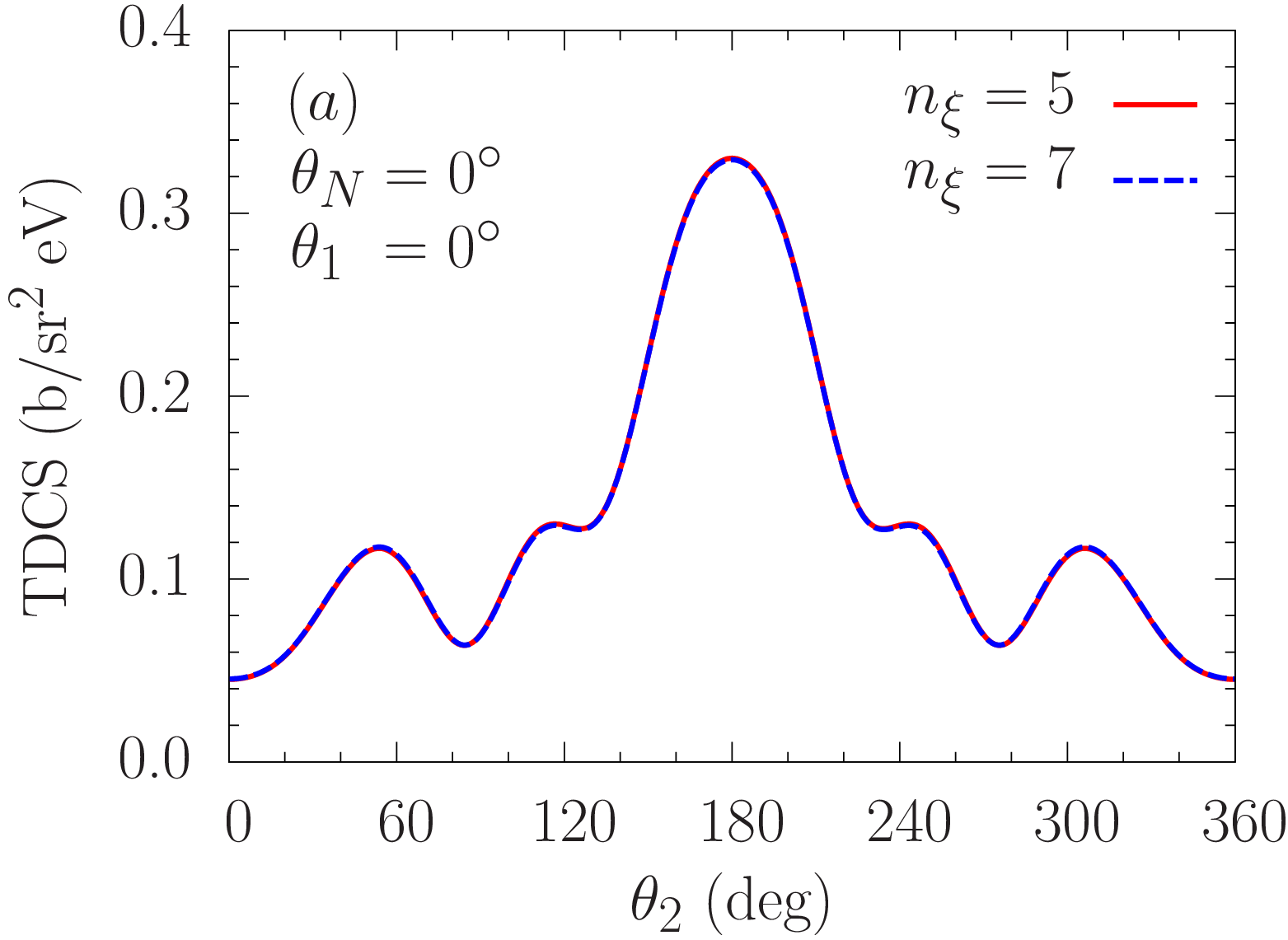}
\includegraphics[width=4.27cm,clip=]{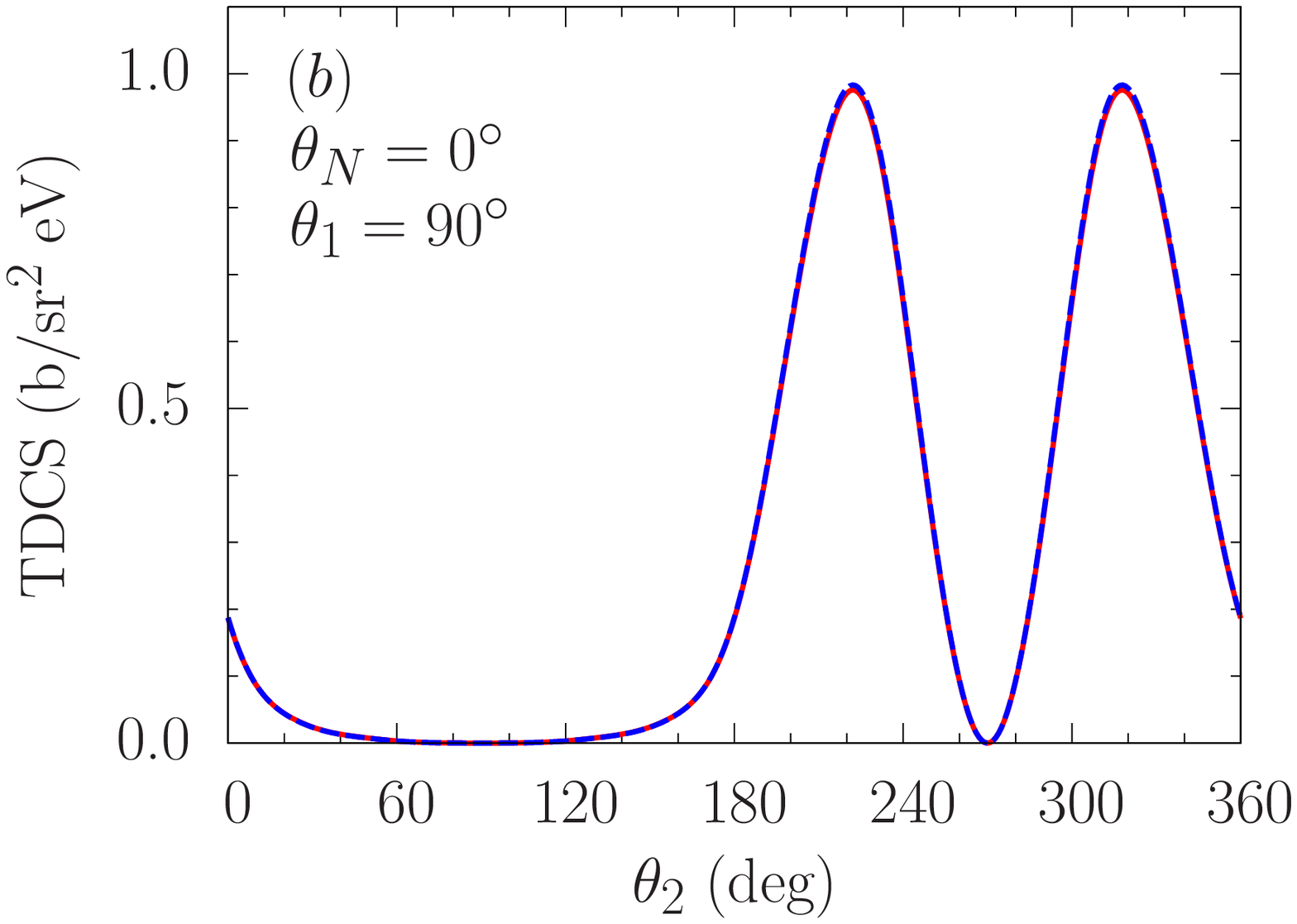} \\
\includegraphics[width=4.27cm,clip=]{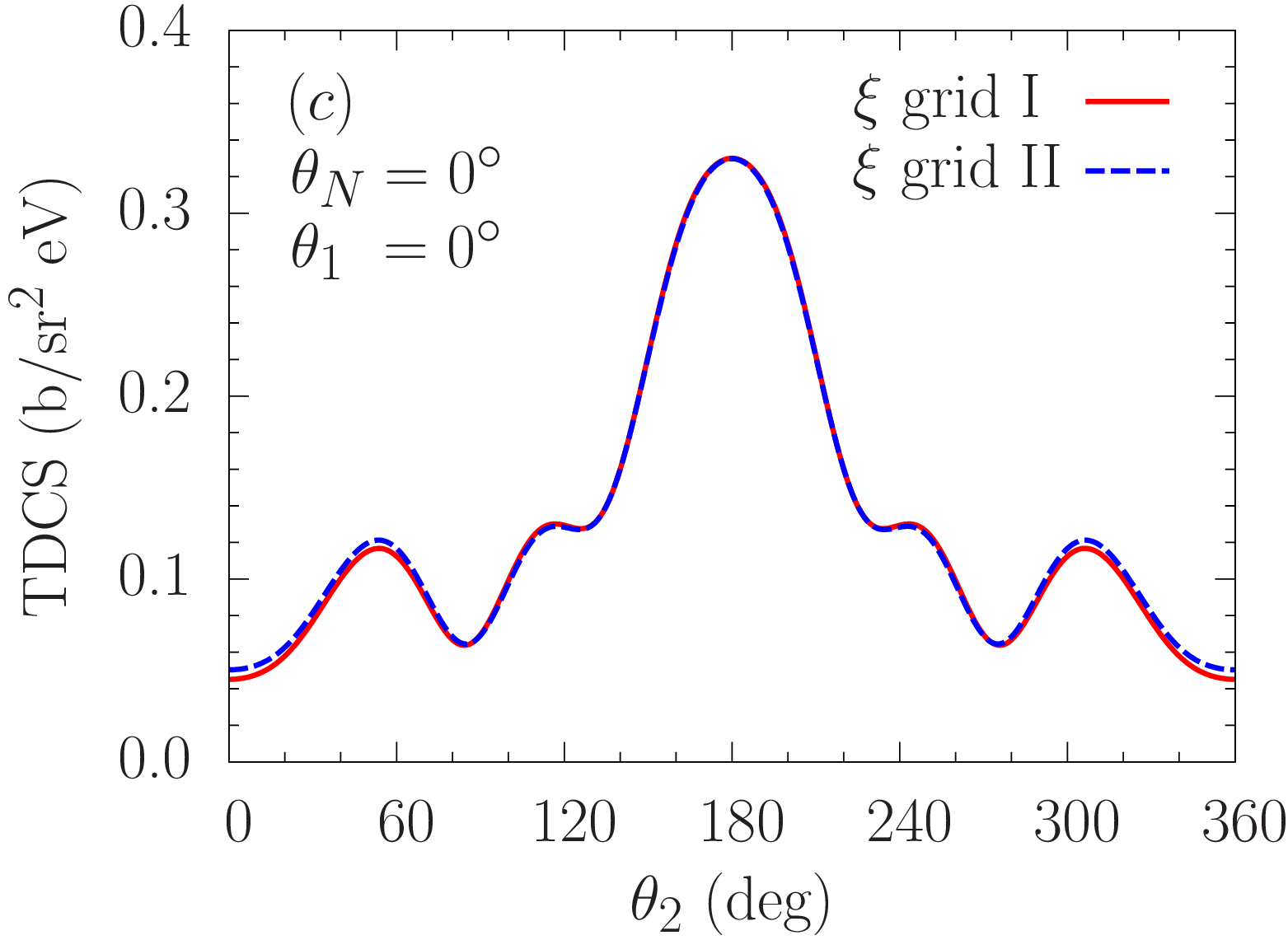}
\includegraphics[width=4.27cm,clip=]{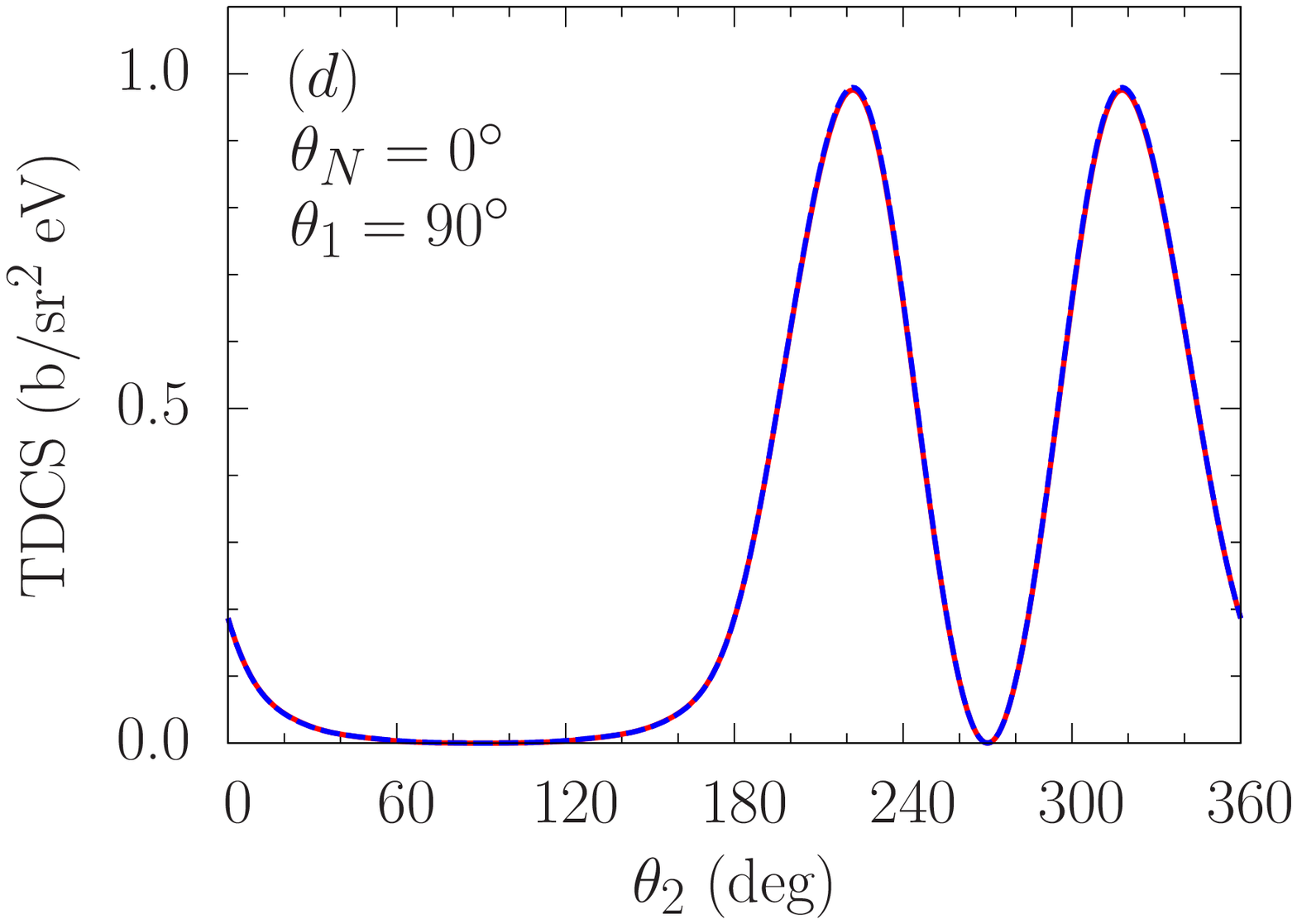} \\
\includegraphics[width=4.27cm,clip=]{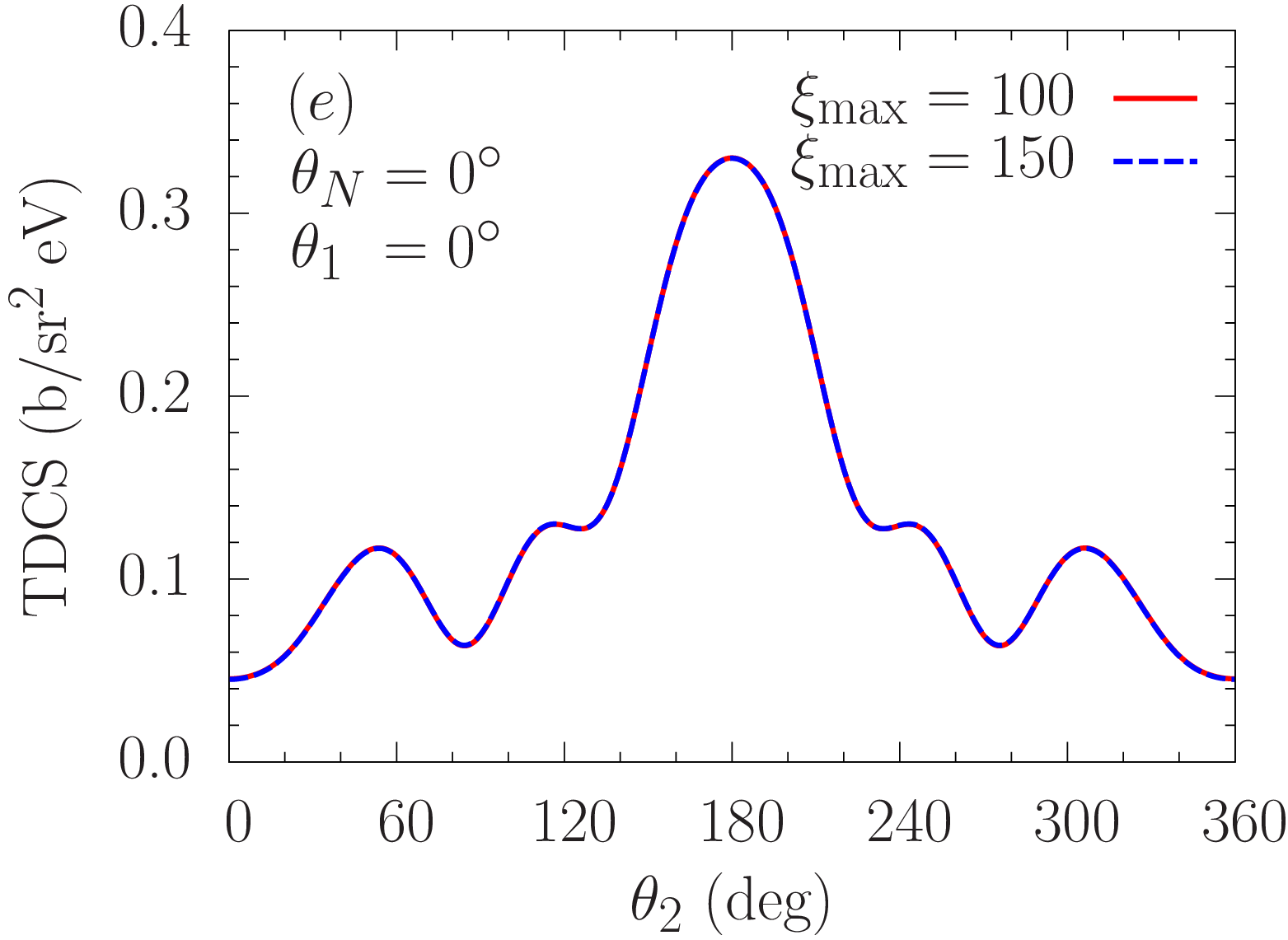}
\includegraphics[width=4.27cm,clip=]{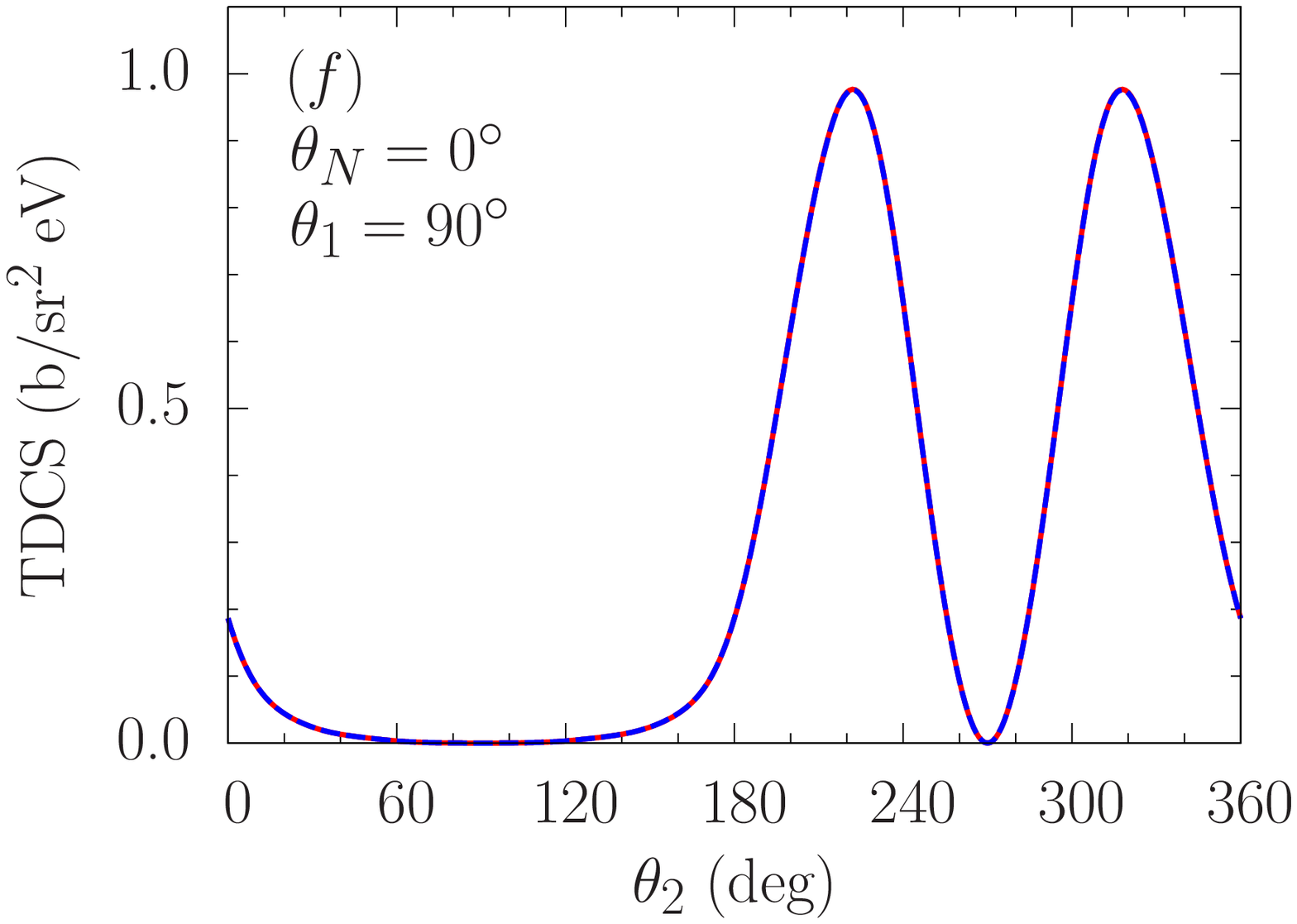} \\
\includegraphics[width=4.27cm,clip=]{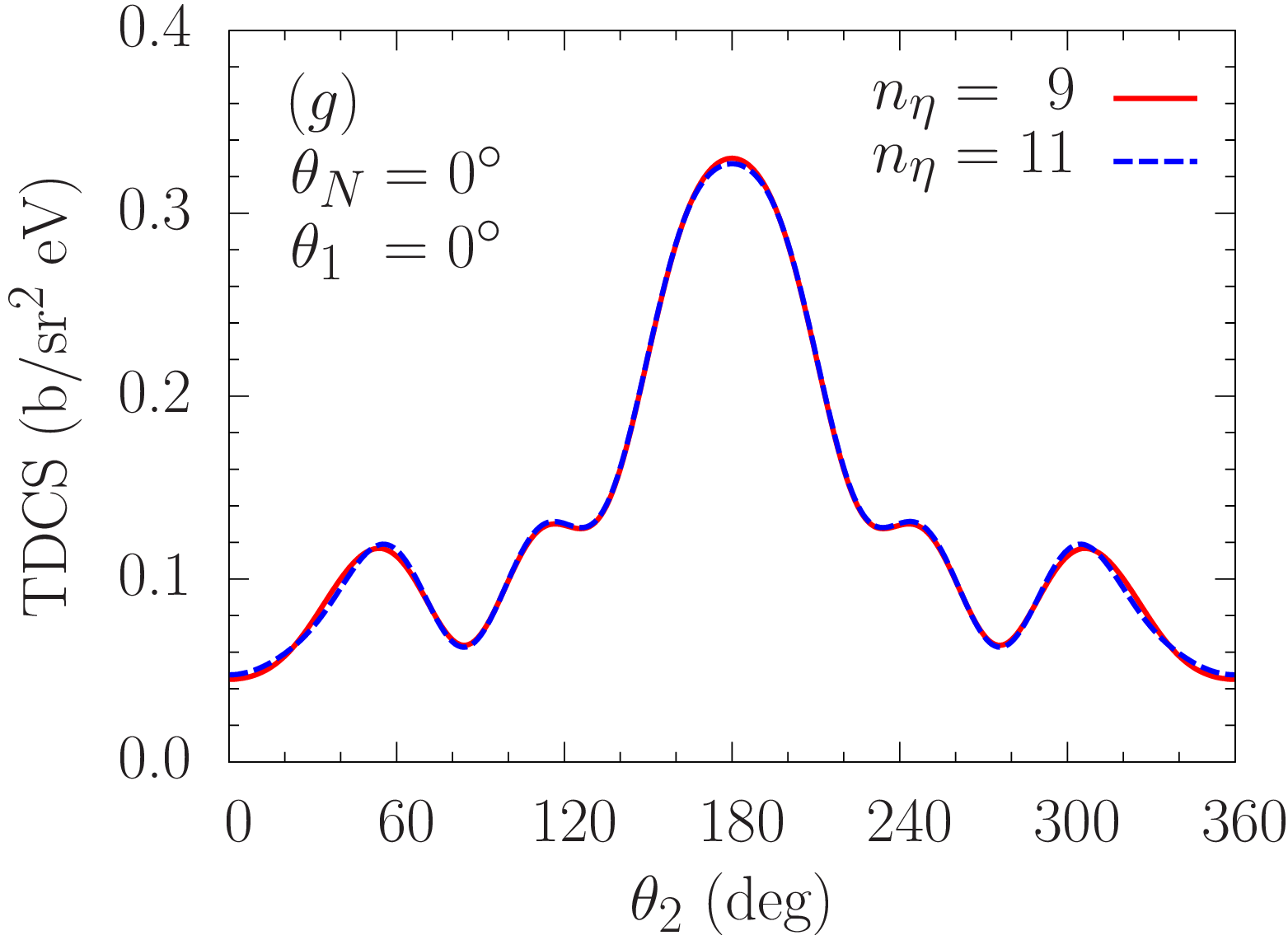}
\includegraphics[width=4.27cm,clip=]{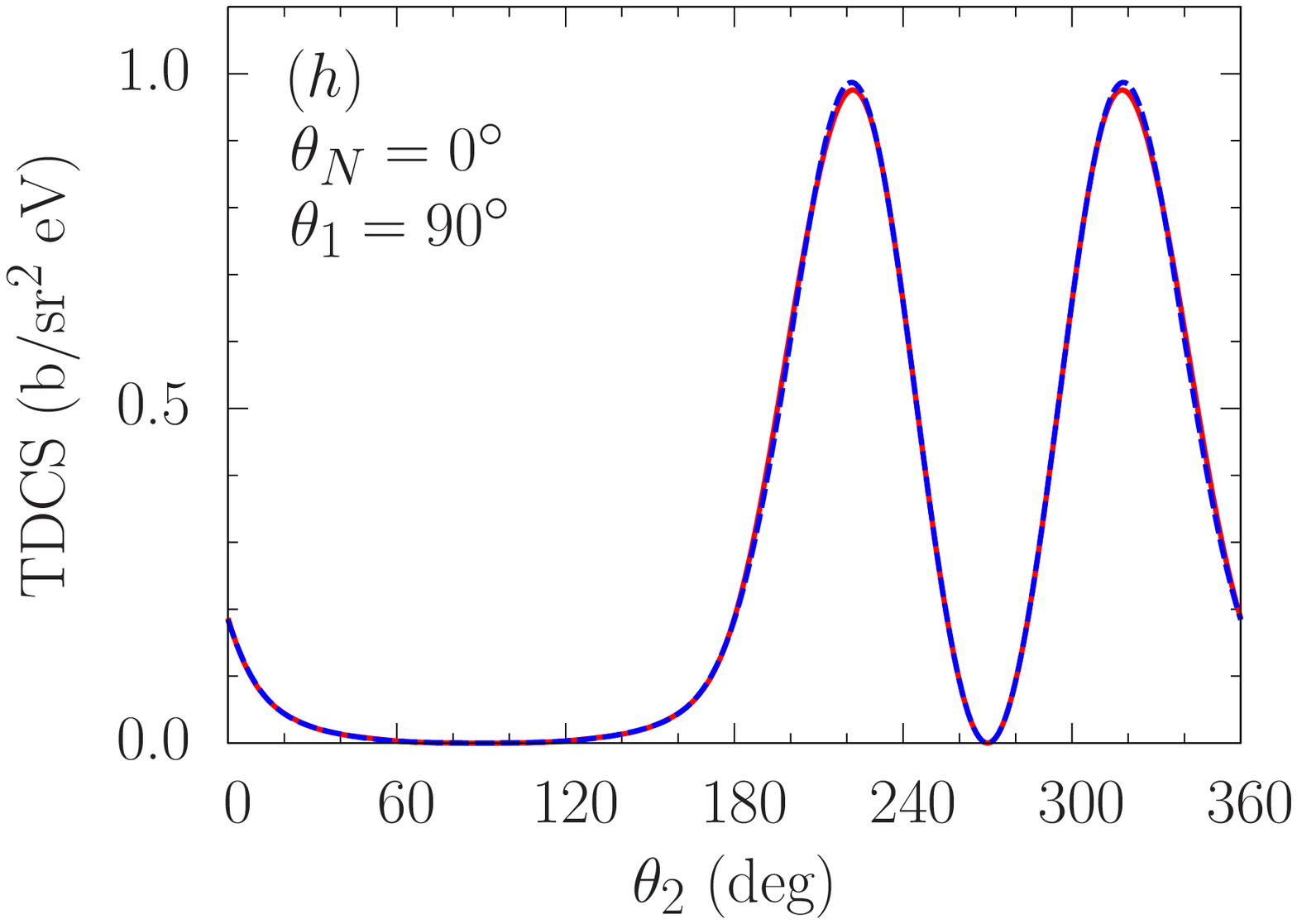} \\
\includegraphics[width=4.27cm,clip=]{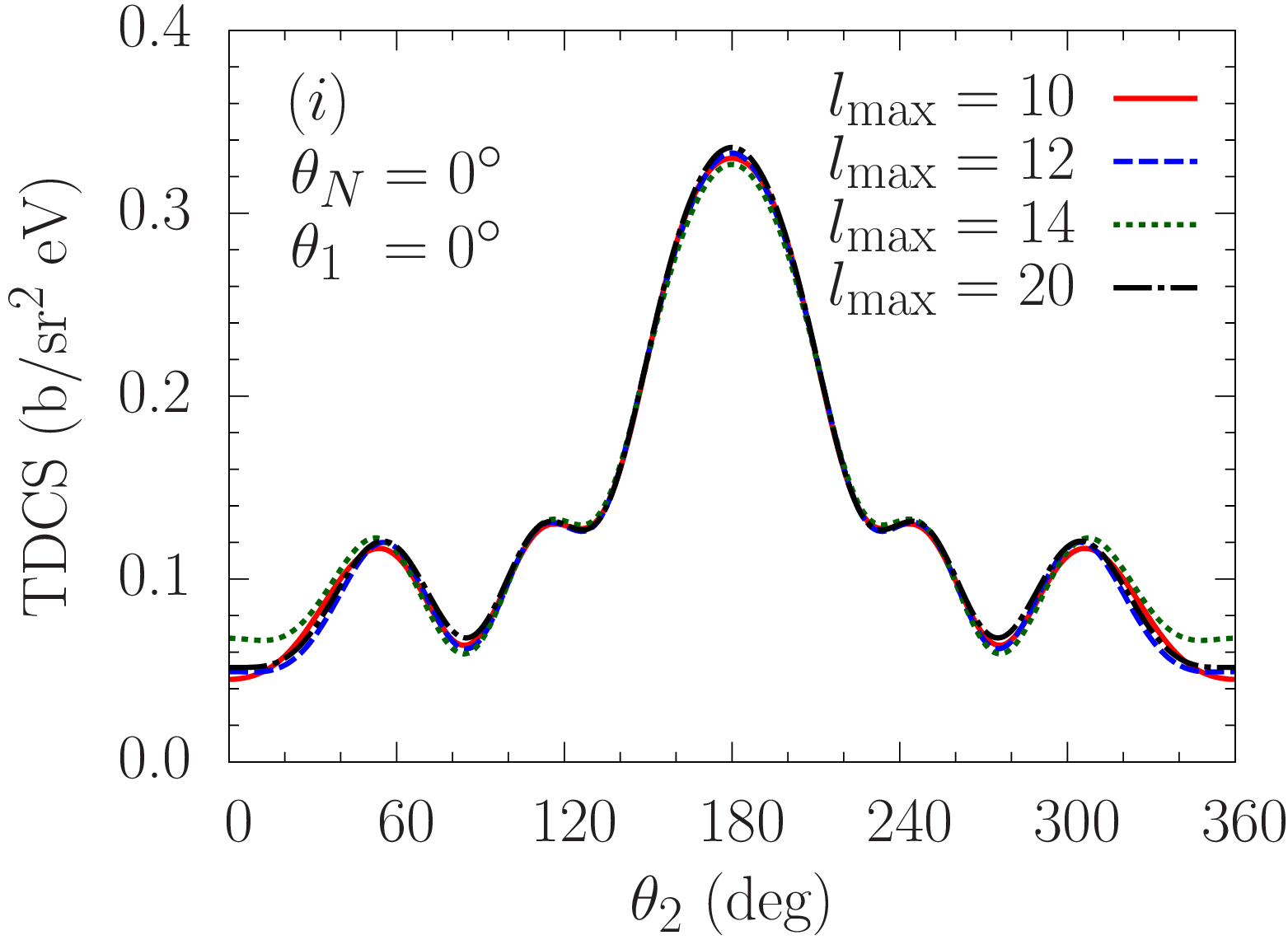}
\includegraphics[width=4.27cm,clip=]{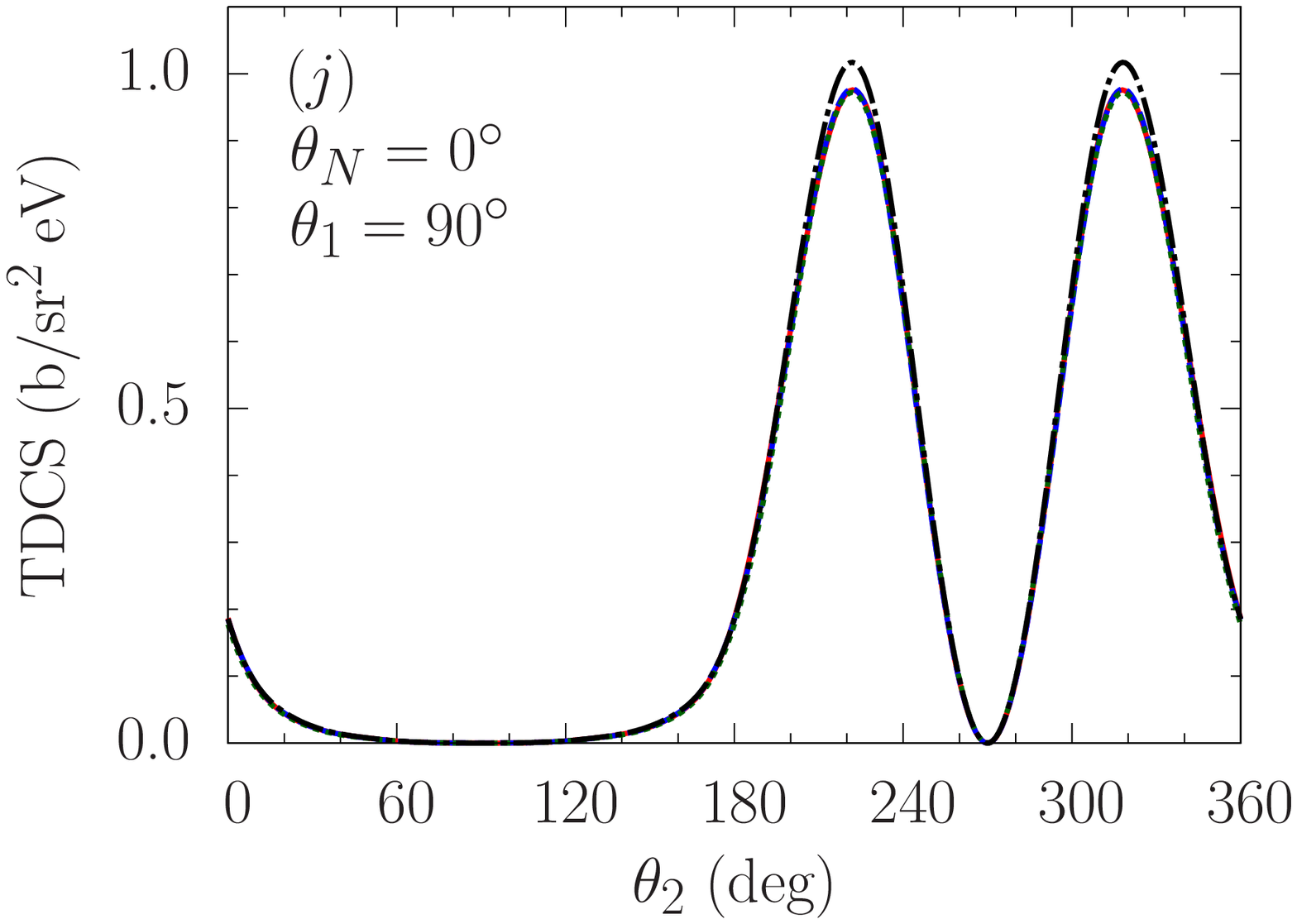} \\
\includegraphics[width=4.27cm,clip=]{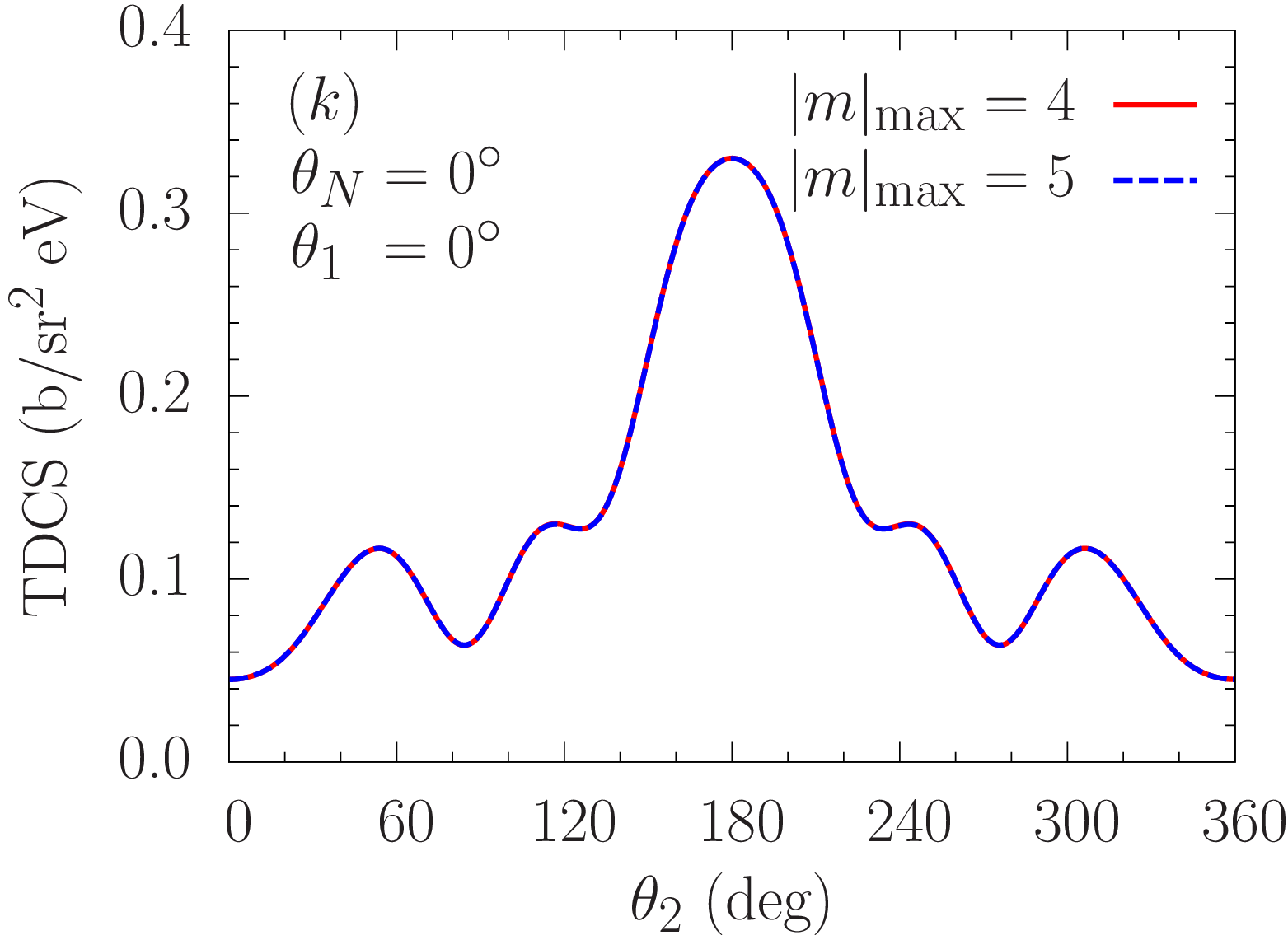}
\includegraphics[width=4.27cm,clip=]{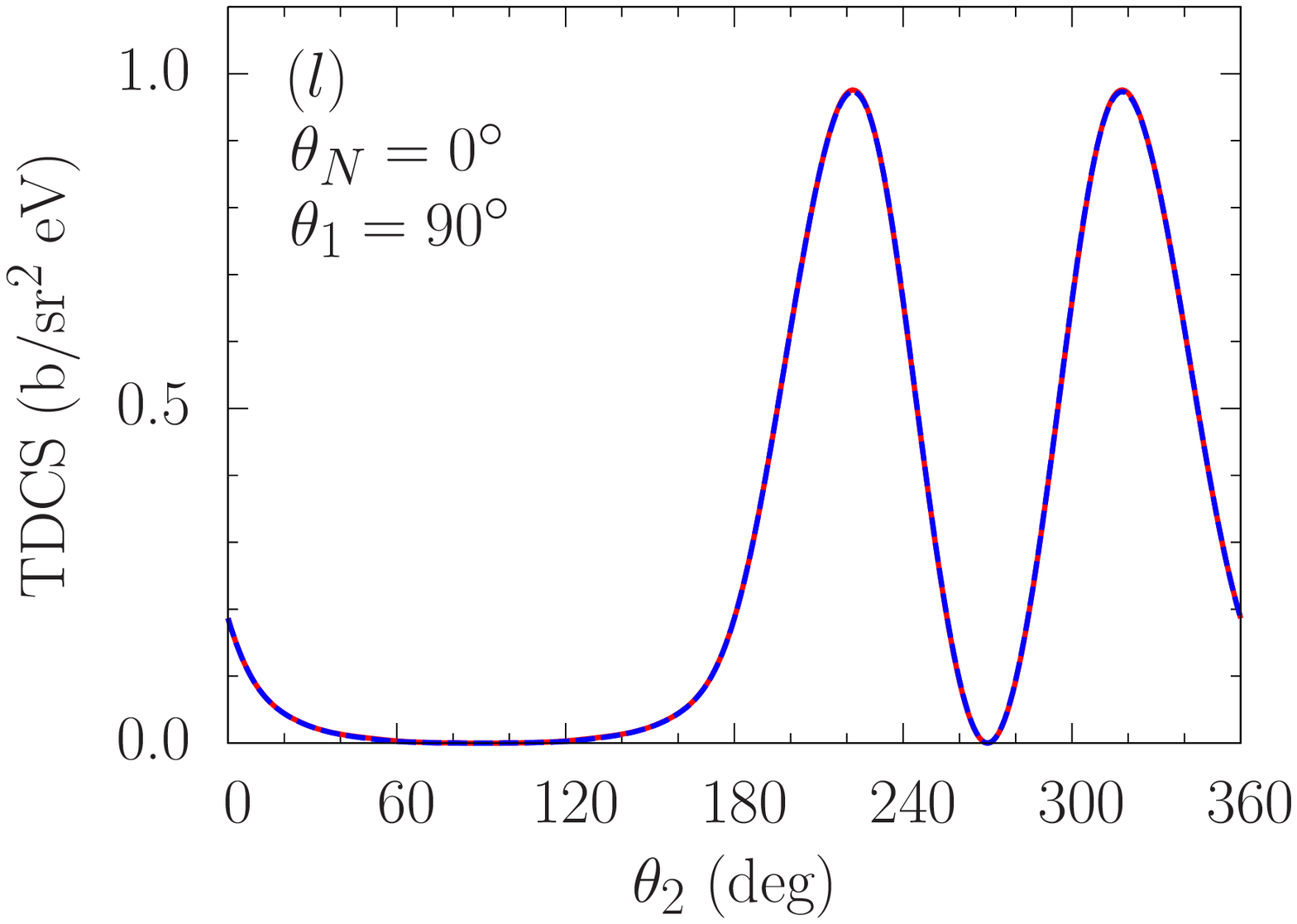}
\caption{(Color online) Same as Fig.~\ref{fig:tdcs-laser}, but for the 
convergence of the TDCS results with respect to the discretization and expansion parameters. See
text for the details.} 
\label{fig:tdcs-discret}
\end{figure}

Figures~\ref{fig:tdcs-laser} and \ref{fig:tdcs-discret} show the convergence
pattern of our TDCS results for asymmetric energy sharing in the parallel geometry
($\theta_N=0^\circ$). The energy sharing between 
electron~$1$ (observed at the fixed angle~$\theta_1$) and electron~$2$ (observed
at the variable angle~$\theta_2$) is $20\%:80\%$. 
Only the electron that takes away $20\%$ of the excess energy
is recorded at fixed
positions either parallel or perpendicular to the polarization axis.

Since the laser pulse is explicitly involved in our time-dependent
treatment, we first have to be sure that the extracted cross sections are
essentially independent of the laser intensity and the time scales.  Only then
are the calculations of cross sections meaningful. This also allows us to compare
the physical information extracted from our time-dependent scenario to that 
obtained through conventional time-independent treatments, which are
effectively equivalent to the weak-field approximation and ``infinitely" long
interaction times.
Rather than computing the
cross sections, it would be more appropriate to consider ionization rates if the
cross sections were found to be sensitive to the laser parameters. 

In Fig.~\ref{fig:tdcs-laser}, we display the dependences of our TDCS results upon
the laser parameters. Note that the TDCSs extracted from
$I_0=10^{15}$ W/cm$^2$ and $10^{14}$ W/cm$^2$ at fixed time evolution of ``$10+2$" cycles
are nearly identical and agree with each other to better than the thickness of the line. 
When we turn to the dependence of time scales at
a fixed intensity of $10^{15}$ W/cm$^2$, we use the same pulse, but
allow the system to freely evolve for a few additional cycles to extract the TDCS.
This corresponds to the time scale of ``$10+4$"~o.c. Also, we may increase the
laser-molecule interaction time, but extract the TDCSs at the same cycles of 
field-free time evolution after the pulse died off. This gives the scenario of 
``$12+2$"~o.c. Since the total time durations are the same ($14$~o.c.), they allow us to
examine the extracted TDCSs from different perspectives. The increased interaction
time yields a reduced bandwidth of the photon energy, while the longer field-free
propagation ensures that the double-ionization wave packet is
further away from the nuclear region~\cite{Madsen2007}. The calculated TDCSs
indeed show a slight, though in our opinion acceptable, sensitivity to the time scales. 
Not surprisingly, the
sensitivity is most visible for the smaller cross sections, when the two
ejected electrons travel nearly parallel along the same direction (c.f.~Fig.~\ref{fig:tdcs-laser}$(c)$).

\begin{figure*}[thb]
\centering
\includegraphics[width=8.cm,clip=]{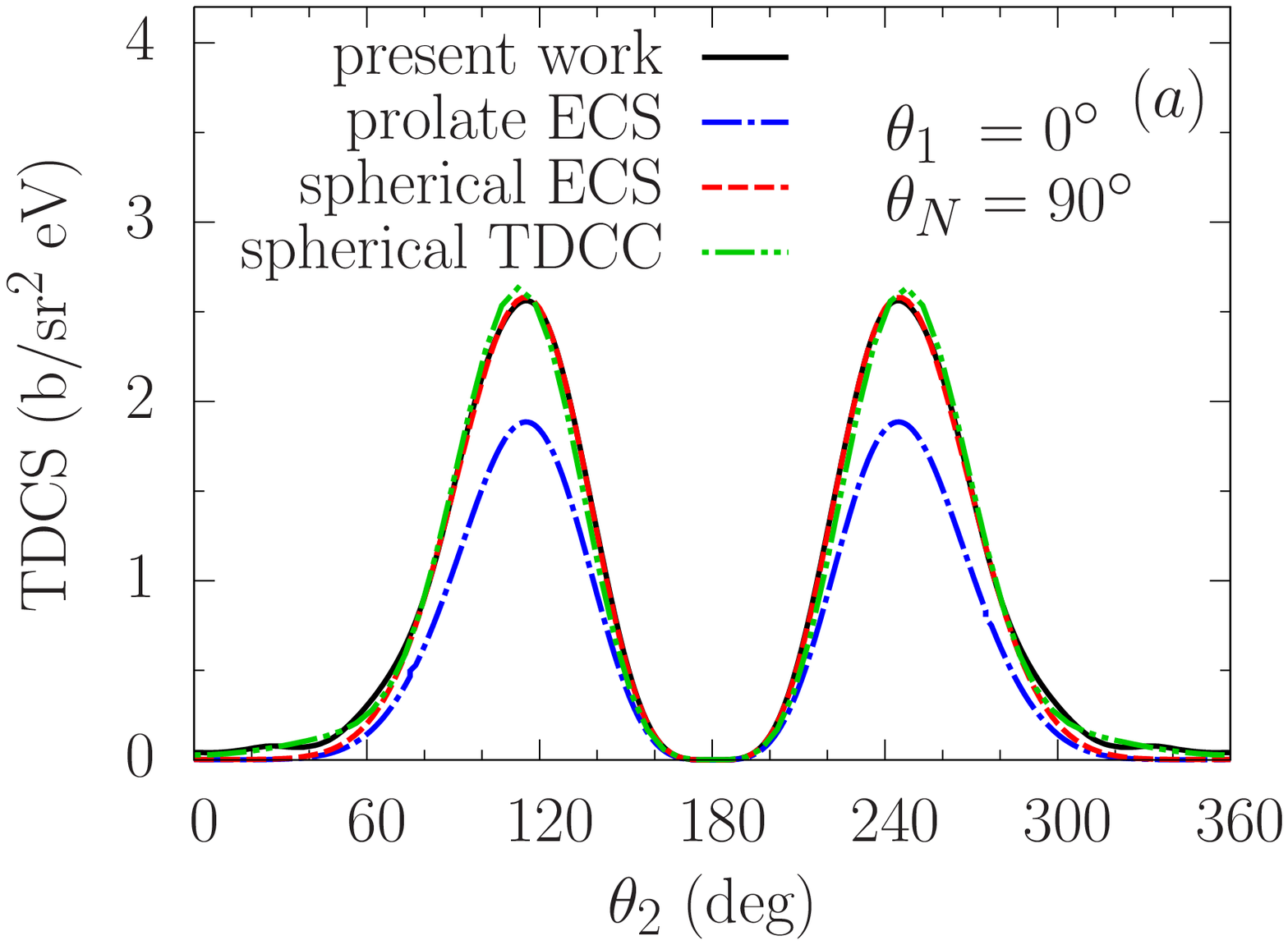}
\includegraphics[width=8.cm,clip=]{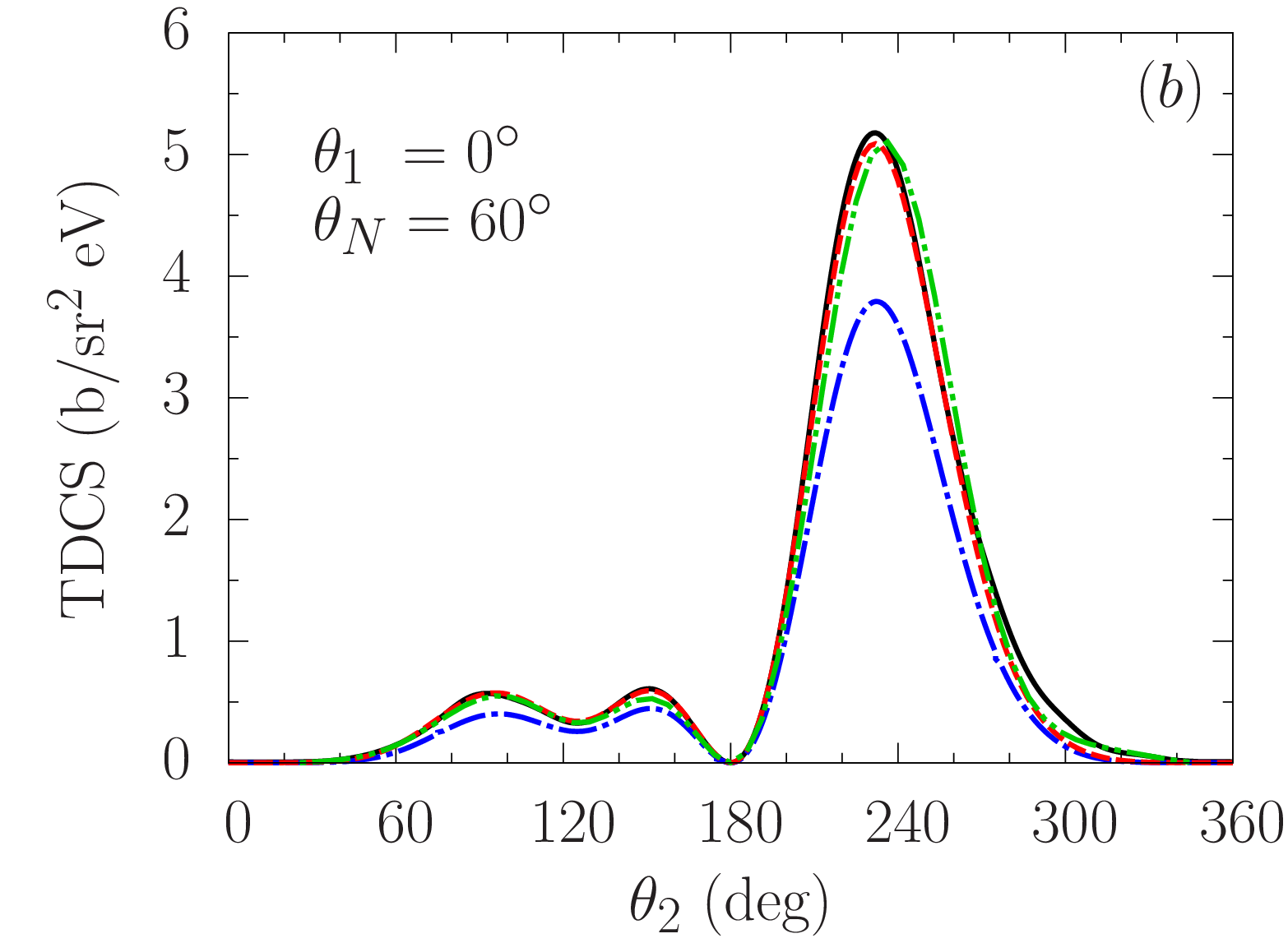} \\
\includegraphics[width=8.cm,clip=]{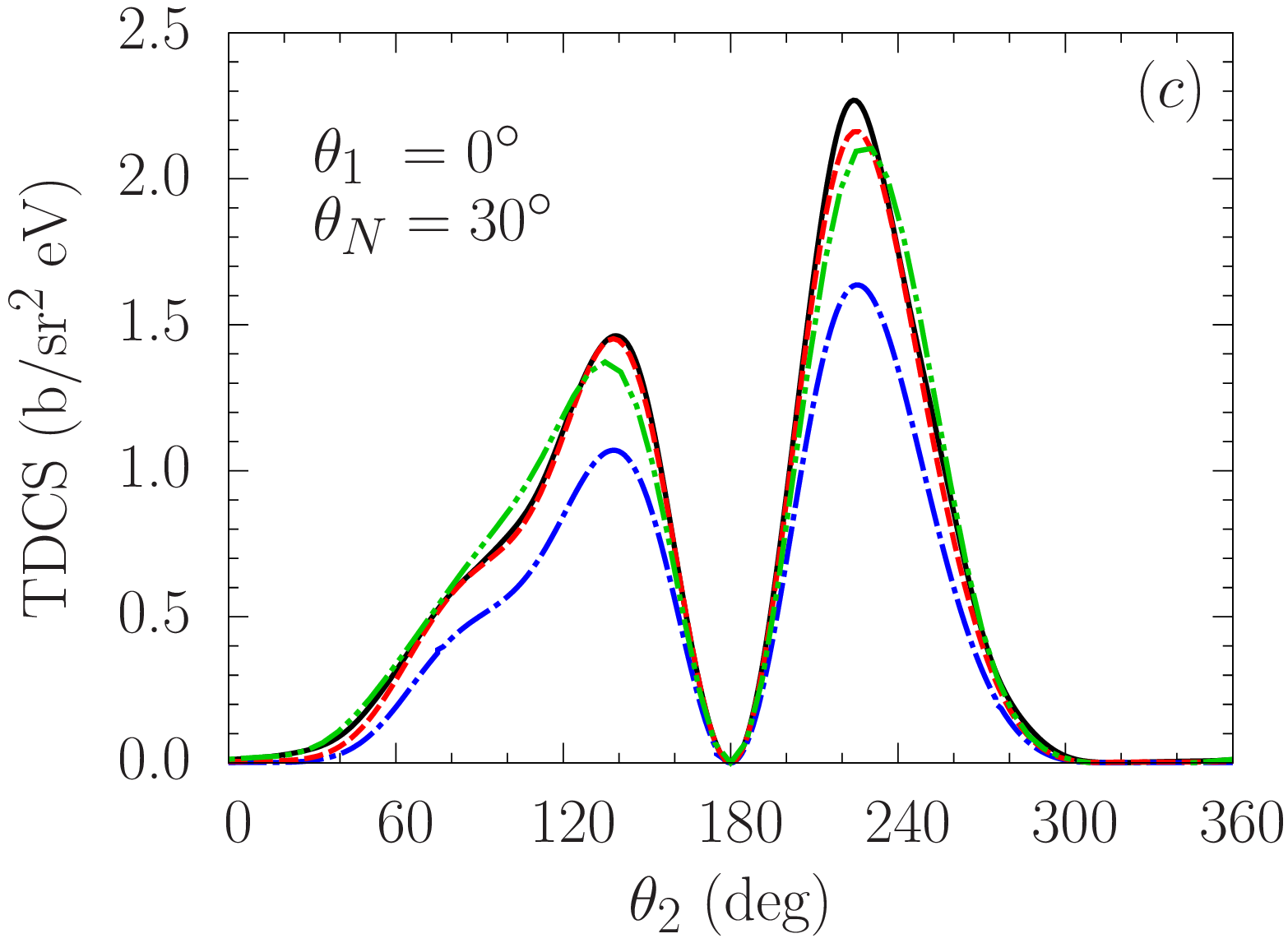}
\includegraphics[width=8.cm,clip=]{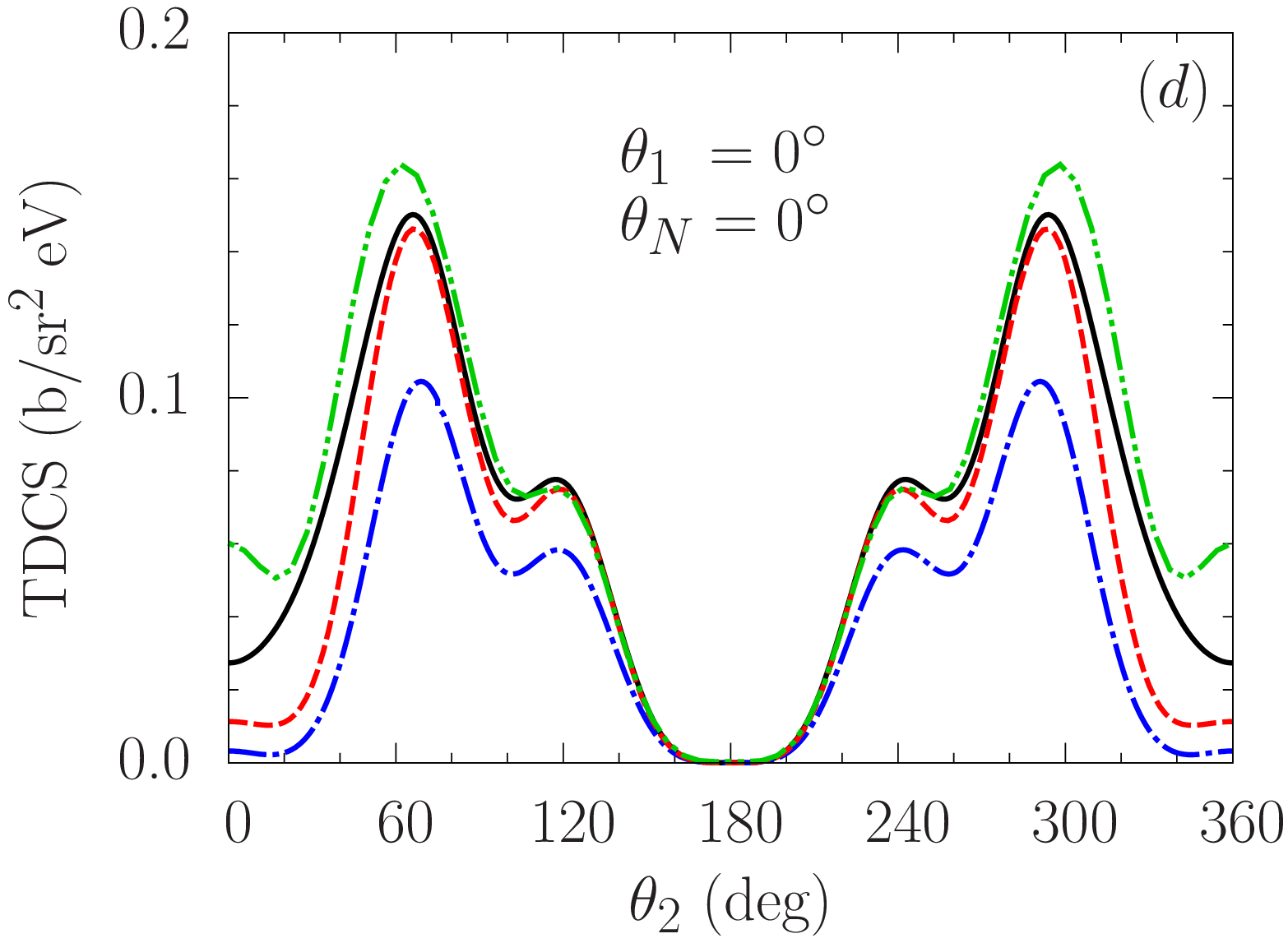}
\caption{(Color online) Coplanar TDCS of the aligned hydrogen molecule for equal
energy sharing ($E_1=E_2=11.8$~eV). The central photon
energy is $75$~eV. One electron is detected
at the fixed direction of $\theta_1=0^\circ$ with respect to the laser polarization axis. 
Also shown are the one-center spherical ECS results~\cite{Van2006-1}, the
two-center prolate spheroidal results~\cite{Tao2010}, and one-center spherical
TDCC results~\cite{Colgan2010}. 
} 
\label{fig:tdcs-equal-sharing-ele1-00}
\end{figure*}

\begin{figure*}[thb]
\centering
\includegraphics[width=8.cm,clip=]{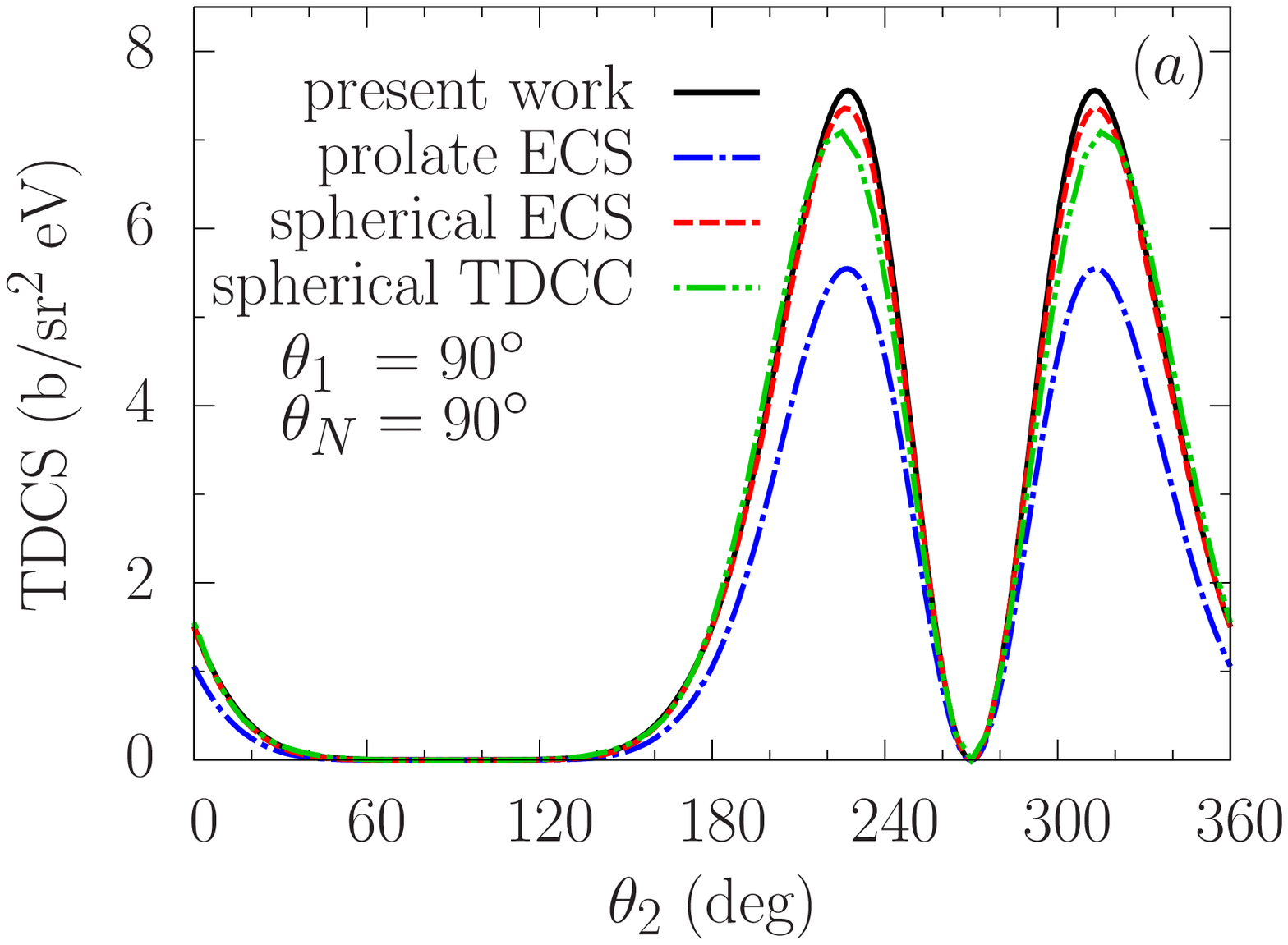}
\includegraphics[width=8.cm,clip=]{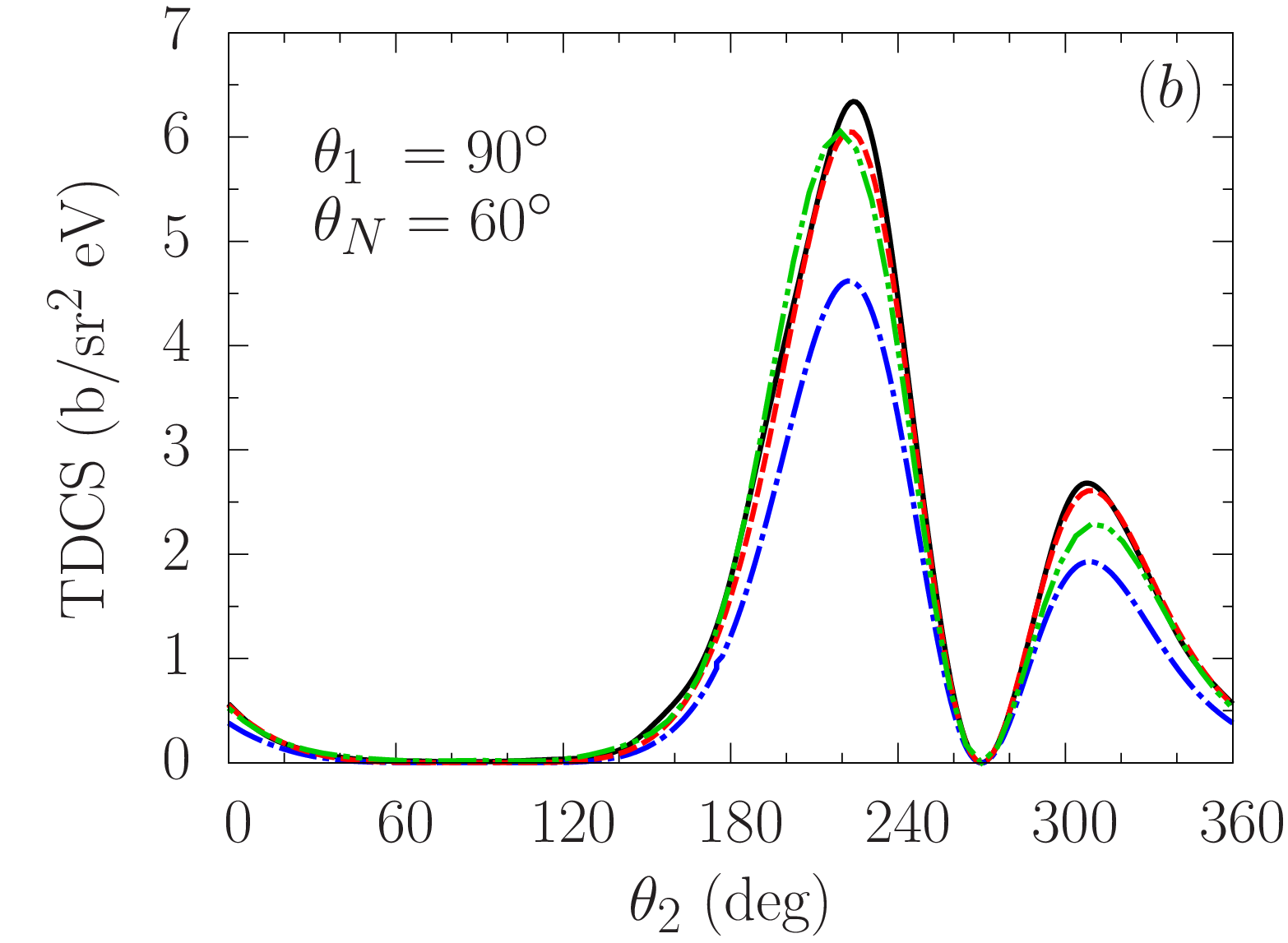} \\
\includegraphics[width=8.cm,clip=]{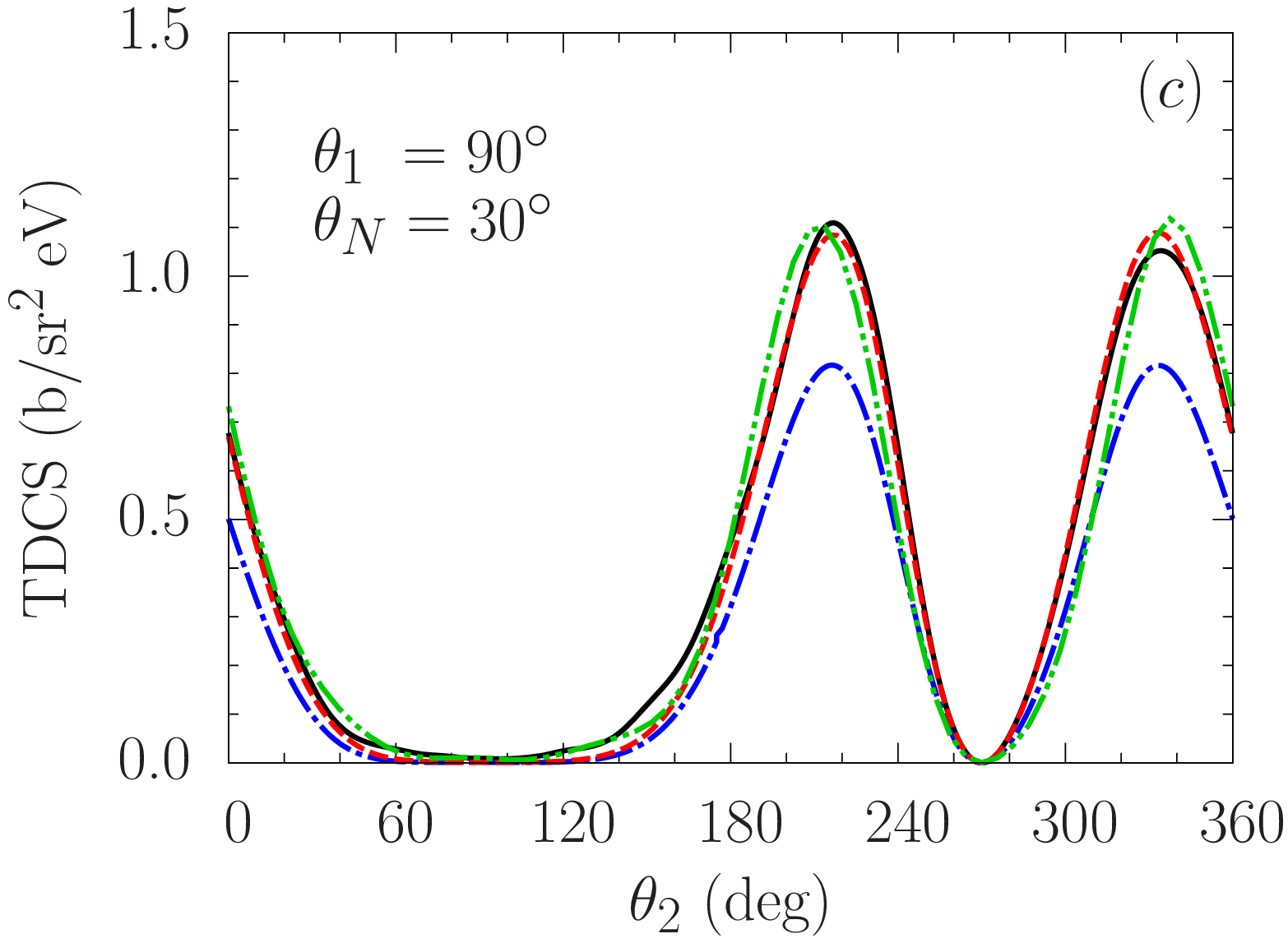}
\includegraphics[width=8.cm,clip=]{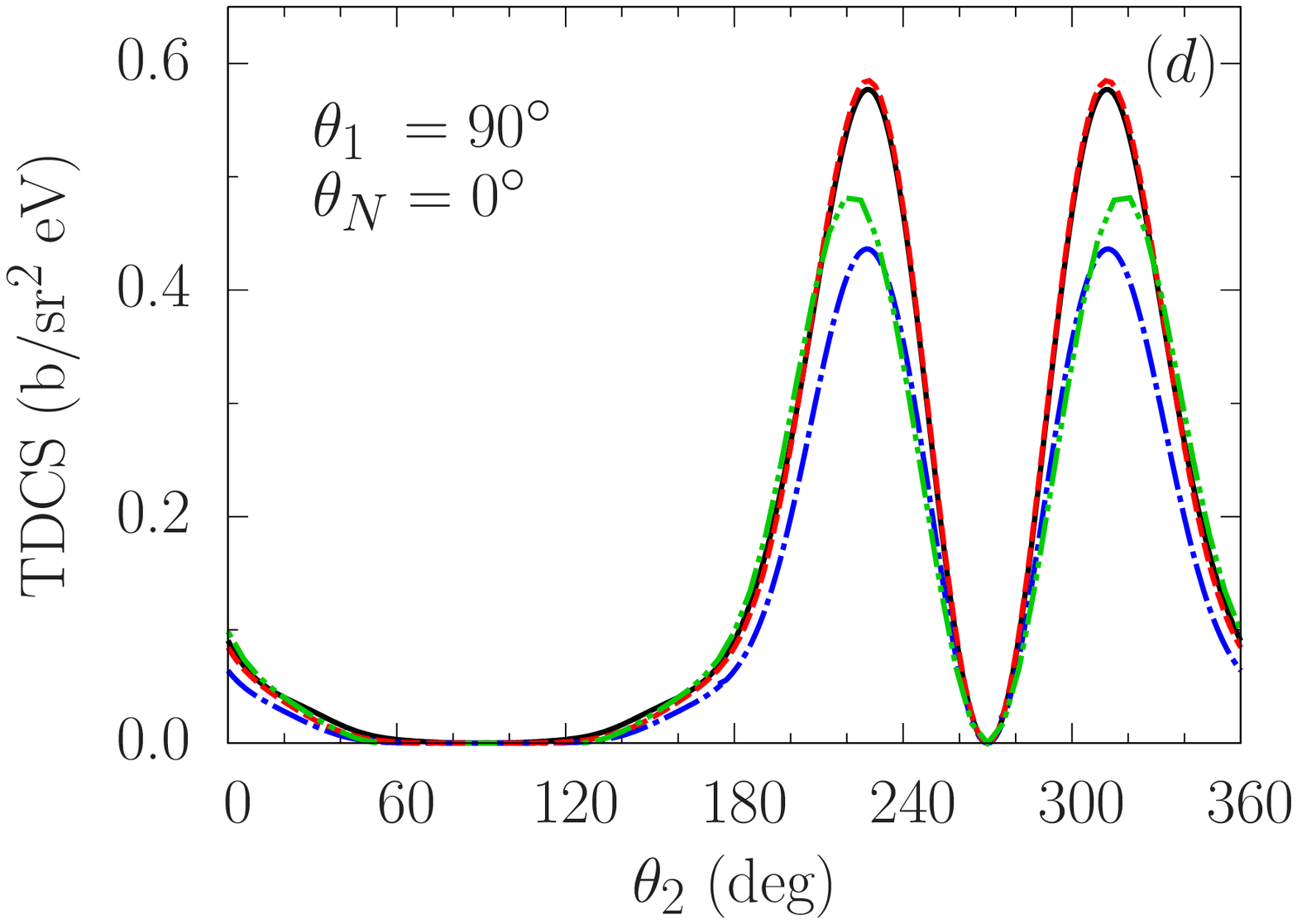}
\caption{(Color online) Same as Fig.~\ref{fig:tdcs-equal-sharing-ele1-00}, except that
the fixed electron is detected at the angle $\theta_1=90^\circ$ with
respect to the laser polarization axis. Since there was a plotting error in Fig.~3 of Tao~{\it et al.}~\cite{Tao2010},
we are comparing here with the proper numbers~\cite{Rescigno2010} from that calculation.} 
\label{fig:tdcs-equal-sharing-ele1-90}
\end{figure*}

\begin{figure*}[thb]
\centering
\includegraphics[width=8.cm,clip=]{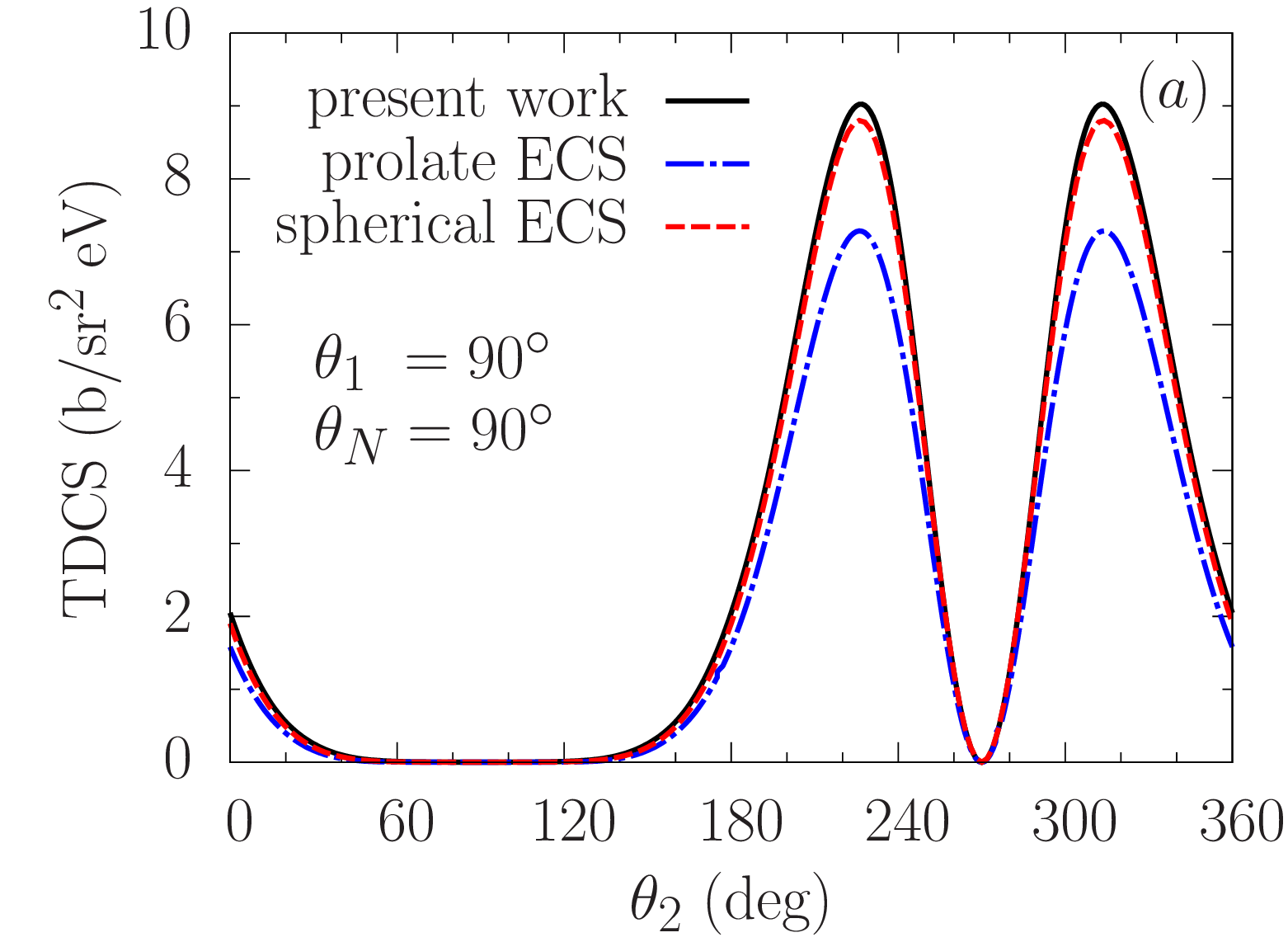}
\includegraphics[width=8.cm,clip=]{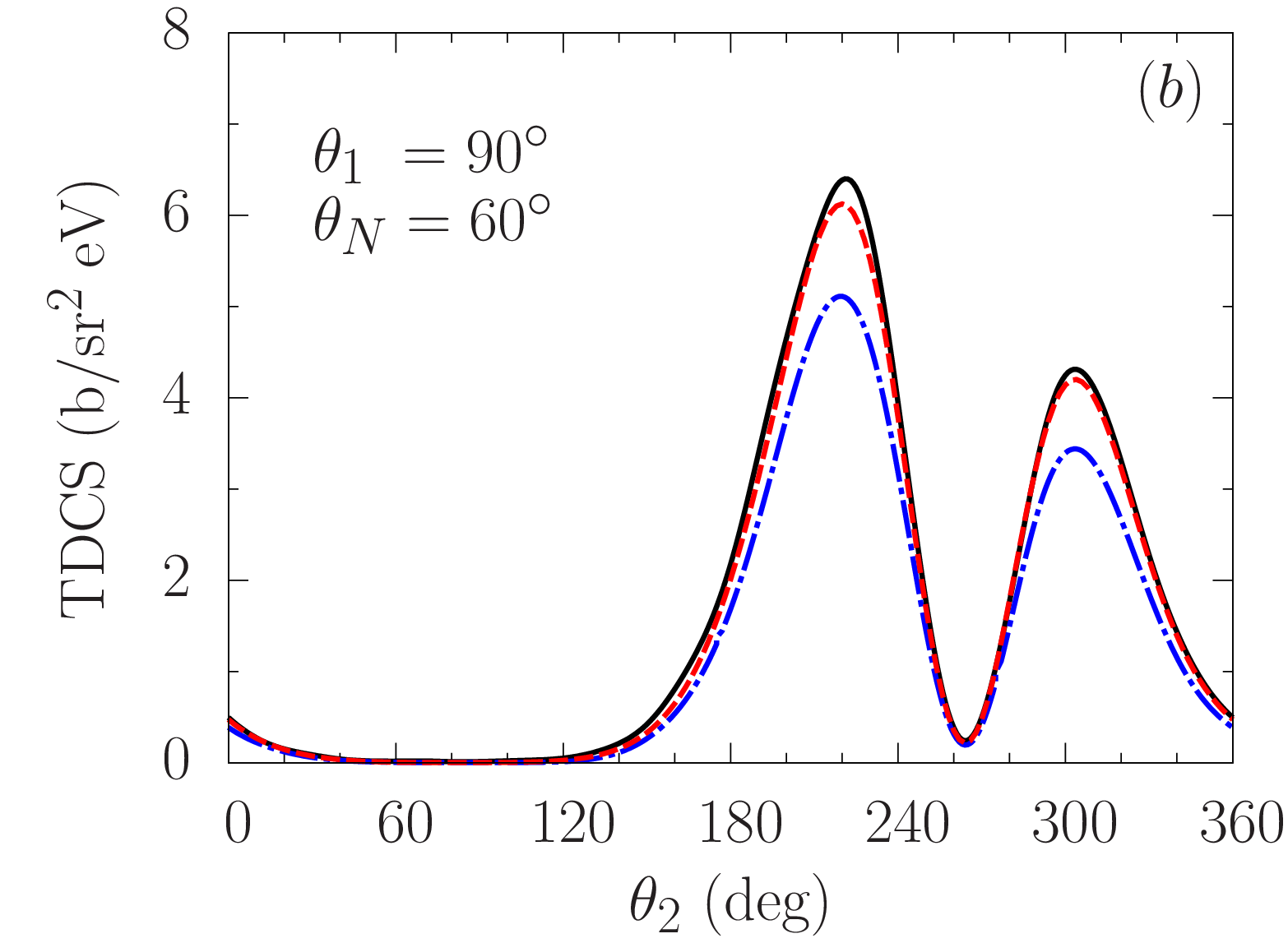} \\
\includegraphics[width=8.cm,clip=]{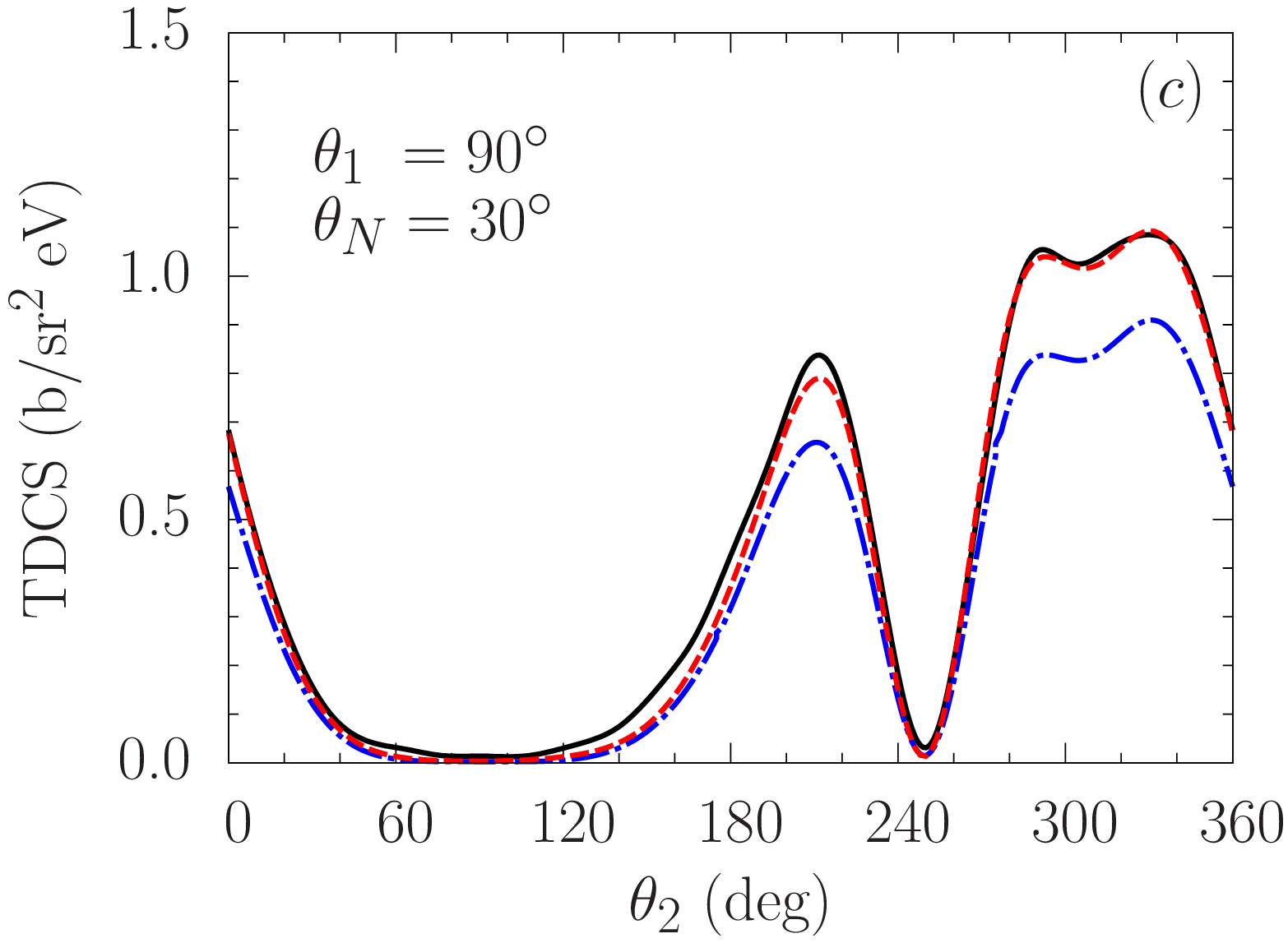}
\includegraphics[width=8.cm,clip=]{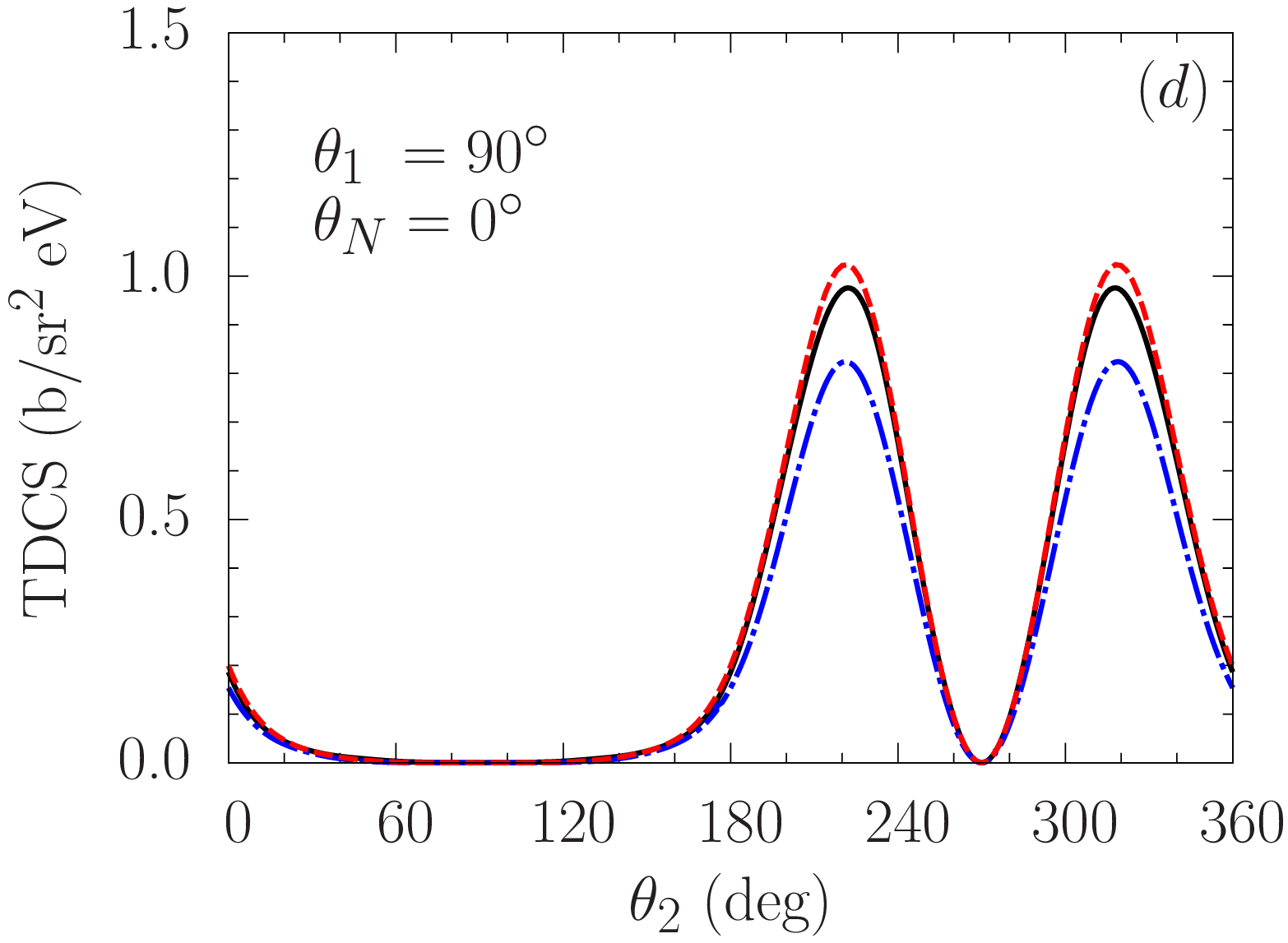}
\caption{(Color online) Coplanar TDCS of the aligned hydrogen molecule for
asymmetric energy sharing. The electron detected at the fixed angle~$\theta_1 = 90^\circ$ takes away $20\%$
of the available excess
energy, while the second electron takes away $80\%$ of $E_{\rm exc}$. The present
time-dependent FE-DVR results are compared with those from
time-independent one-center spherical ECS~\cite{Van2006-1} and two-center
prolate spheroidal ECS~\cite{Tao2010} calculations. } 
\label{fig:tdcs-nonequal-sharing-ele1-90}
\end{figure*}

\begin{figure*}[thb]
\centering
\includegraphics[width=4.cm,clip=]{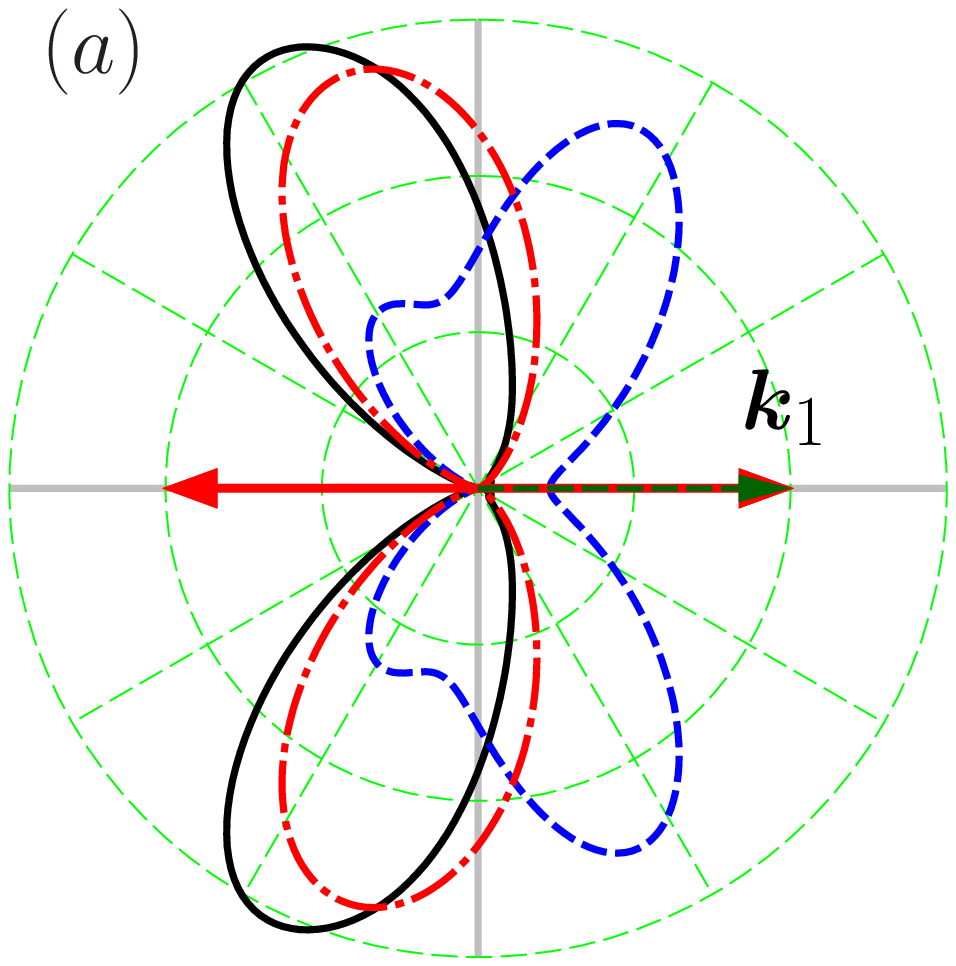}
\includegraphics[width=4.cm,clip=]{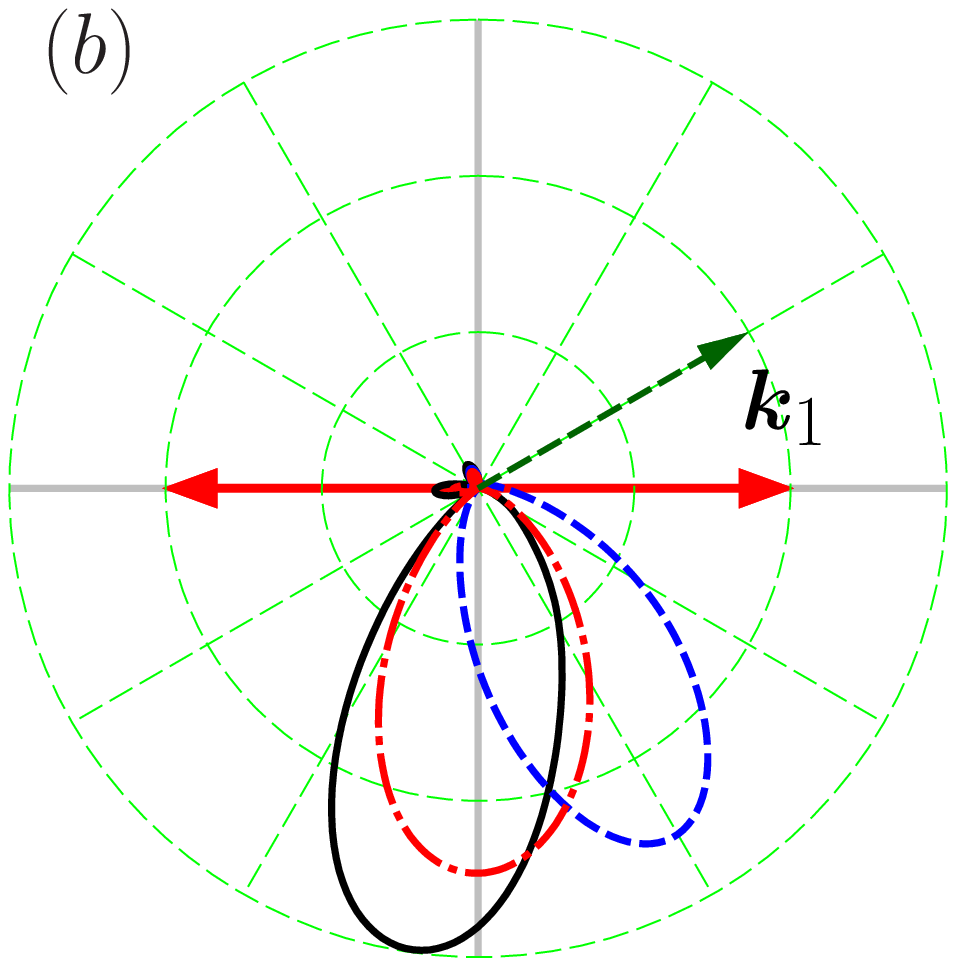}
\includegraphics[width=4.cm,clip=]{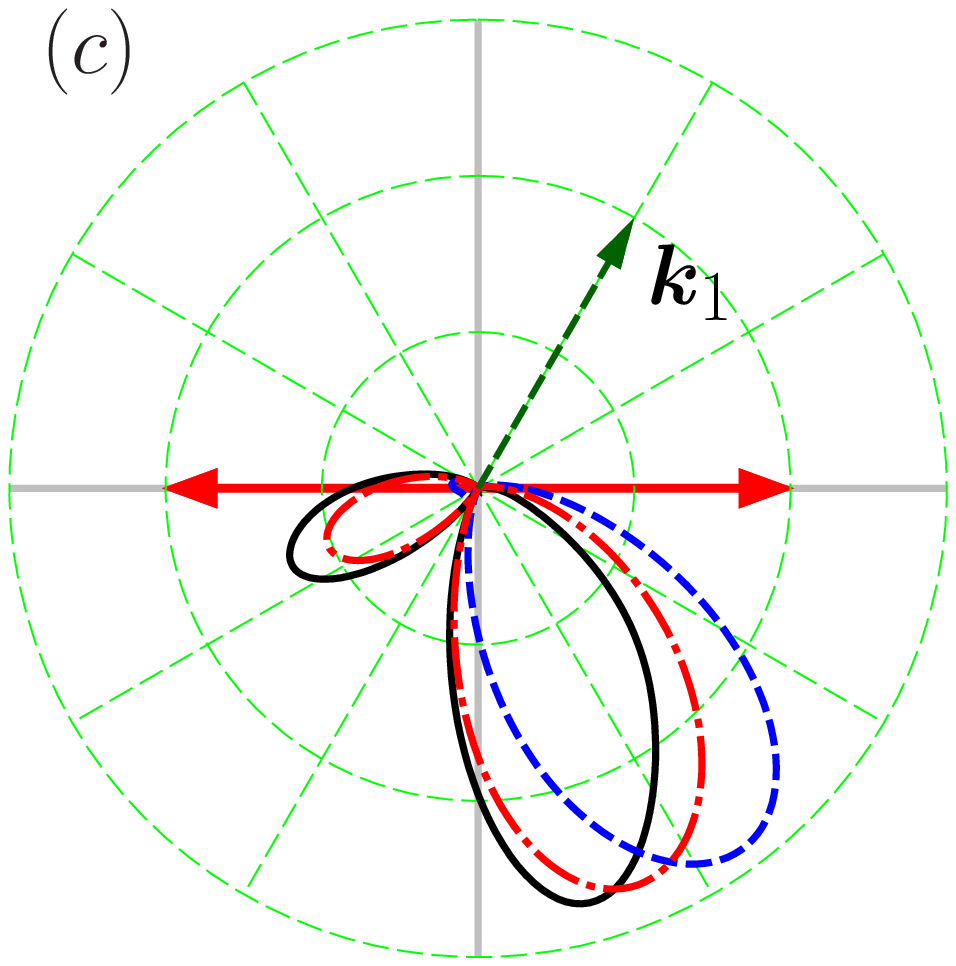}
\includegraphics[width=4.cm,clip=]{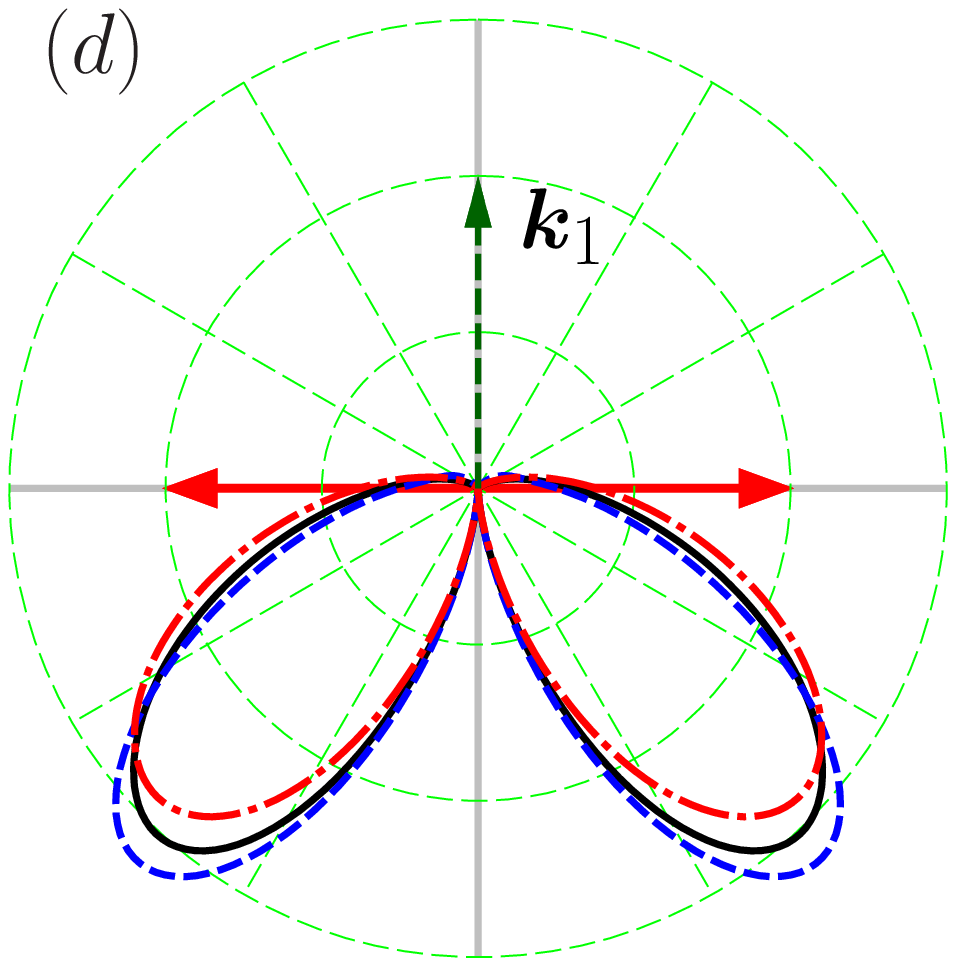}
\caption{(Color online) 
Comparison of predicted relative coplanar TDCSs between H$_2$ 
in the perpendicular (solid lines) and parallel (dashed lines) geometries, and 
He (chain lines)~\cite{Guan2008-2} for equal-energy sharing in polar
coordinates.
The polarization axis is taken along the horizontal direction. 
The photon energies for H$_2$ and He are $75$~eV and $99$~eV, respectively.
The fixed observation angles for one of the
electrons are $0^\circ$~(a), $30^\circ$~(b), $60^\circ$~(c), and $90^\circ$~(d) with
respect to the laser polarization vector.
Scaling factors were used to emphasize the shape comparison.} 
\label{fig:tdcs-h2-helium}
\end{figure*}

\begin{figure*}[thb]
\centering
\includegraphics[width=7.cm,clip=]{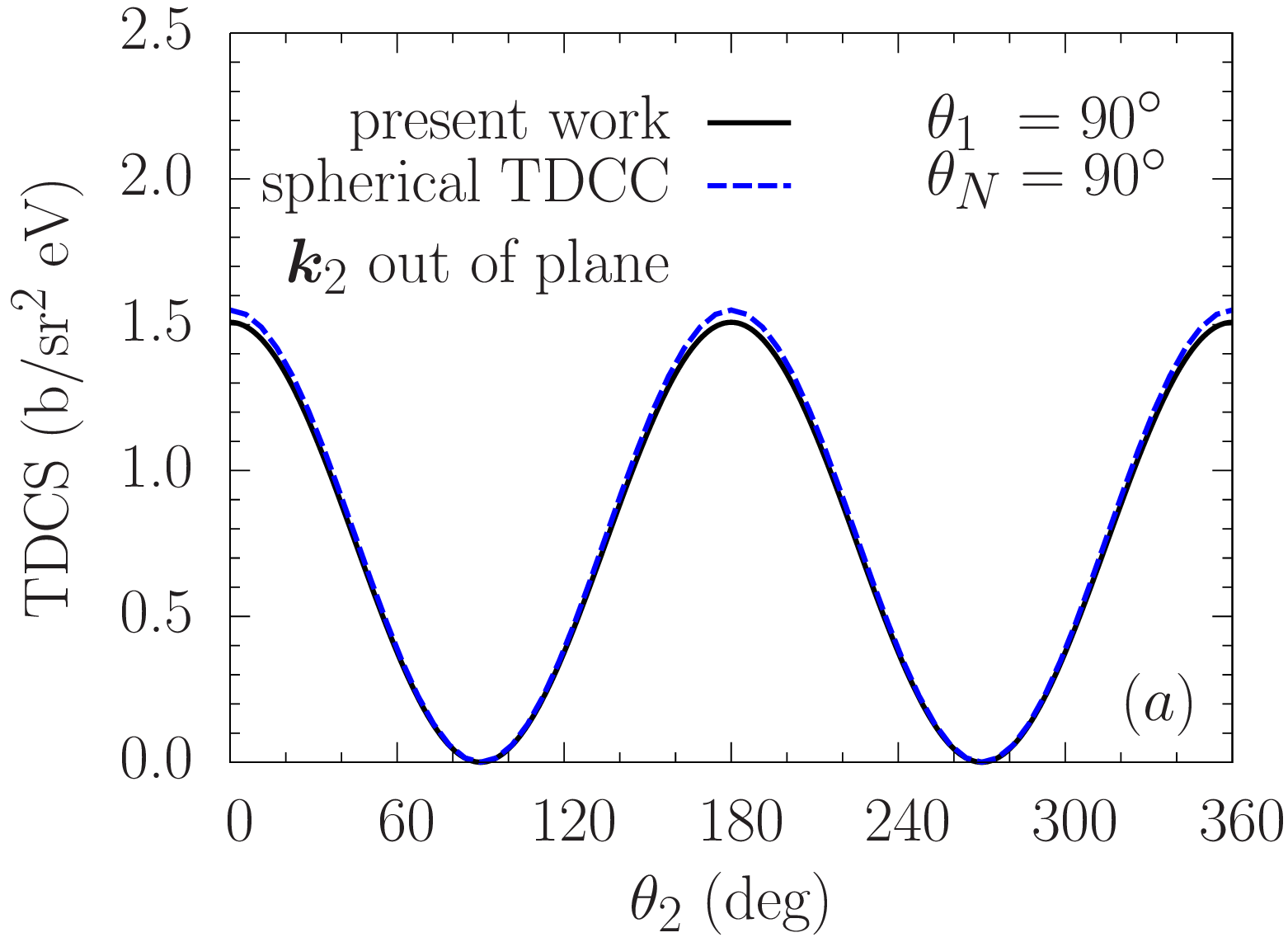}
\includegraphics[width=7.cm,clip=]{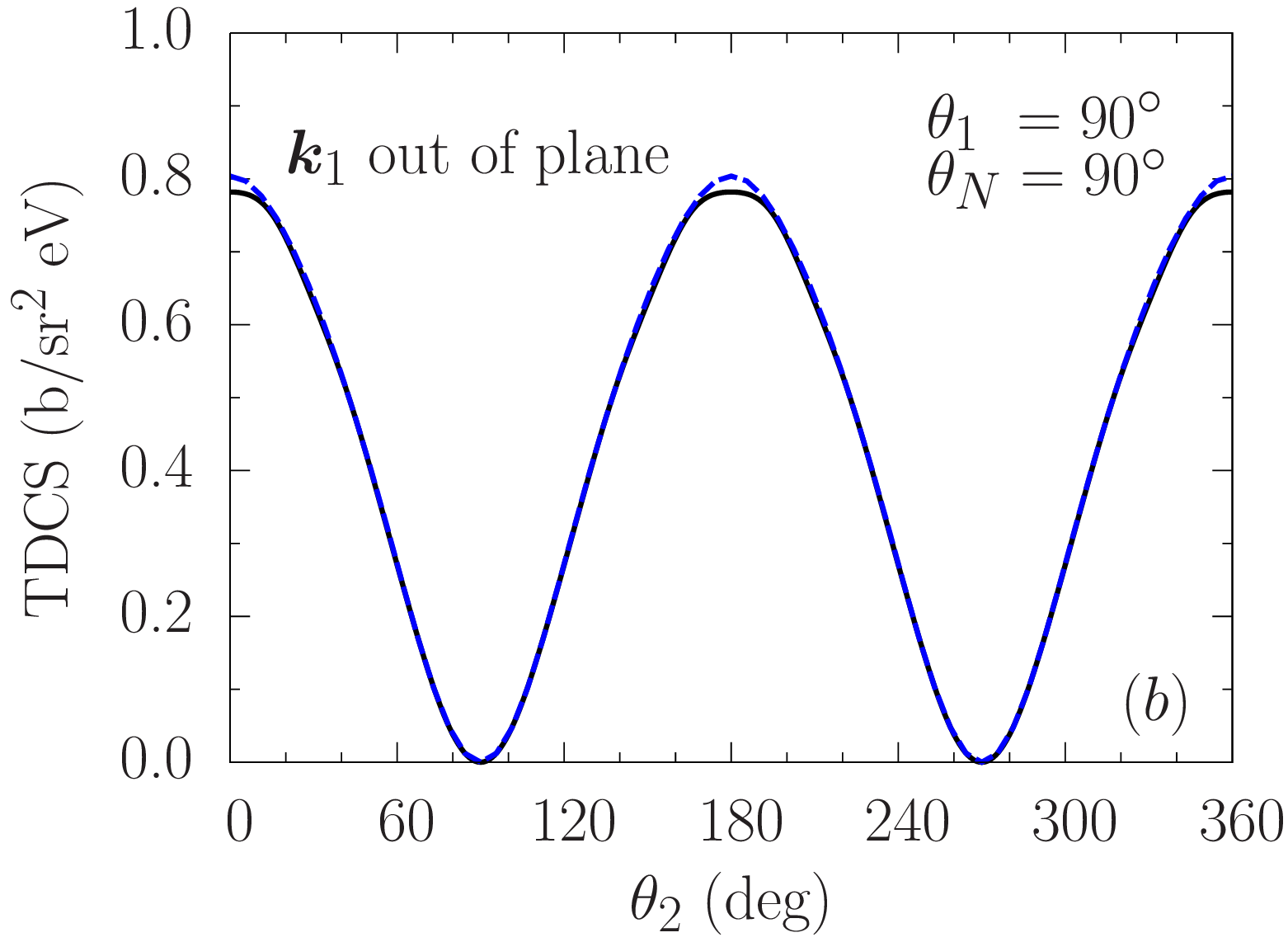}
\includegraphics[width=7.cm,clip=]{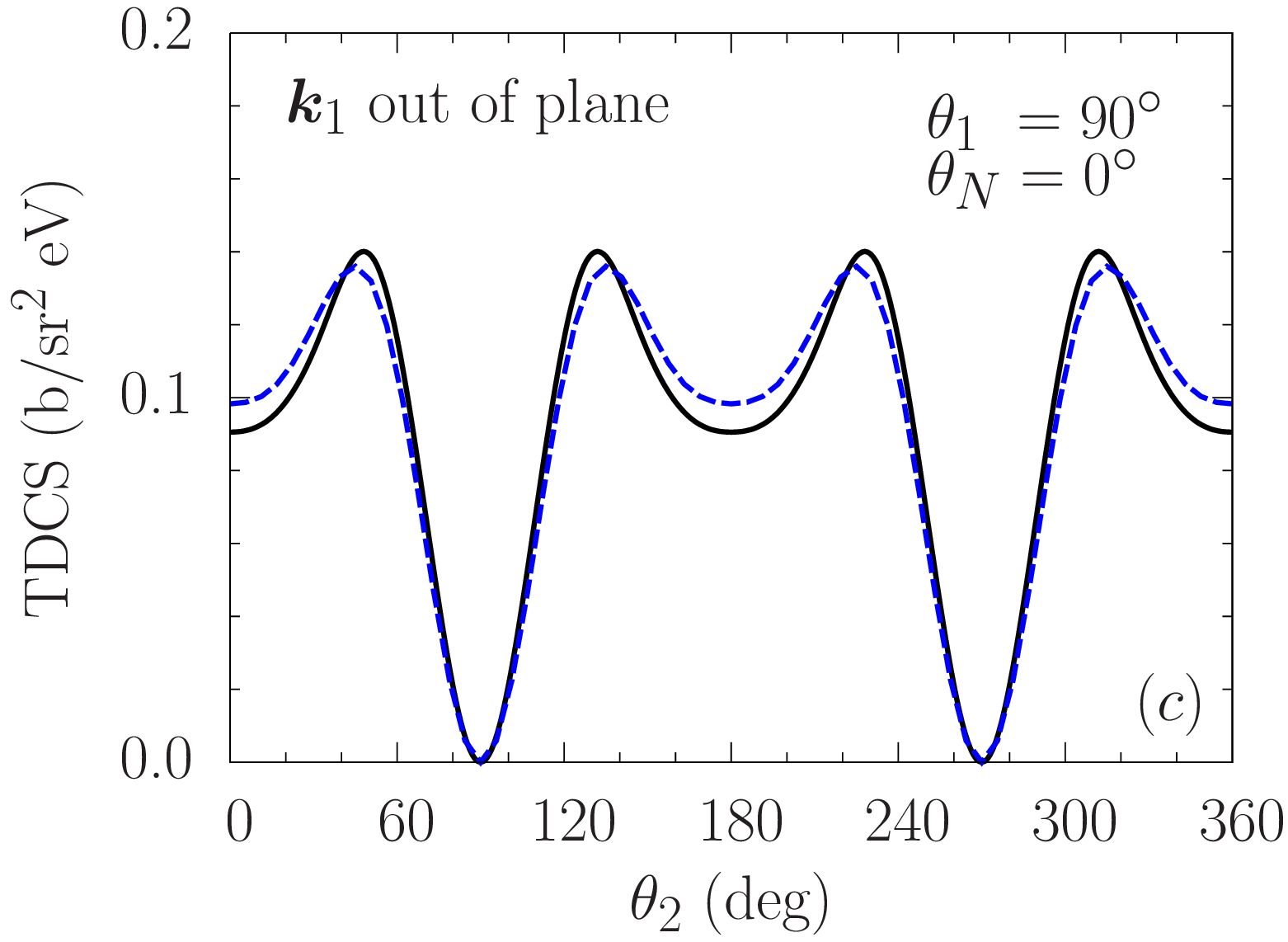}
\caption{(Color online) Non\-coplanar TDCS of the aligned hydrogen molecule for
equal energy sharing. The present time-dependent FE-DVR results are compared
with TDCC predictions~\cite{Colgan2010}.} 
\label{fig:tdcs-out-plane}
\end{figure*}
 
Having confidence in using the current sets of laser parameters, we now turn
our attention to the scheme of spatial discretization ($n_\xi$, $\xi_{\rm max}$, $n_\eta$) 
and the convergence of the expansion ($l_{\rm max}$, $|m|_{\rm max}$). 
The results are displayed in Fig.~\ref{fig:tdcs-discret}.

For the discretization parameters, we obtain well-converged TDCSs by
increasing $n_{\xi}$ from $5$ to $7$, $n_{\eta}$ from $9$ to $11$, and 
extending the spatial box of $\xi_{\rm max}$ from $100$ to $150$.  Most
importantly, however, we consider two sets of $\xi$ mesh points: $\xi$ grid~I 
and $\xi$ grid~II (see Table \ref{tab:parameters}).  The principal 
motivation was to see whether or not 
we can reproduce the much lower TDCS values (by about $20\%$ compared to the
one-center spherical results) that were recently obtained in an ECS 
calculation in two-center elliptical
coordinates by Tao {\it et al.}~\cite{Tao2010}.

We emphasize that these two grids in the
``radial" $\xi$ coordinate are completely different regarding both the distribution of
the elements and the number of grid points per element. In the $\xi$ grid~I, we
divide the $\xi$ space into two parts, an inner and an outer region with a  
border at $\xi_b=5$. We place a narrow span of elements in the inner
region, and then wider elements in the outer region. In contrast to that, 
$\xi$ grid~II does not distinguish between inner and outer regions, i.e., the
elements uniformly span the region from $1$ to $\xi_{\rm max}$. The
mesh setup in $\xi$ grid~II is the same as that used in
Ref.~\cite{Tao2010}, except for the complex rotation.
The $\xi$ grid~I has a much denser distribution of mesh points than $\xi$ grid~II.
Nevertheless, the extracted TDCSs from both sets of $\xi$ grids are in
excellent agreement with each other, even for the smallest cross sections. This strongly
suggests that the results are well converged at least with regard to the $\xi$ grid. Both $\xi$
sets are good enough to capture the physics of interest. 
Differences at the $20\%$ level are unlikely to be caused by
using different sets of $\xi$ meshes. 

Finally, we discuss the convergence of our results with respect to the
expansion parameters, 
$|m|_{\rm max}$ and $l_{\rm max}$. As expected for a one-photon process,
$|m|_{\rm max}=4$ produced well-converged results. 

Recall the discussion above regarding the ground state, 
especially how the truncated Neumann expansion of $1/r_{12}$ in our present FE-DVR 
implementation affects the initial-state energy and therefore the quality of the wave function.
For consistency,  we use the same $l_{\rm max}$
in the real-time propagation and in the
ground-state wave function.
As seen from Figs.~\ref{fig:tdcs-discret}$(i)$ and \ref{fig:tdcs-discret}$(j)$,
our truncated Neumann expansion has little effect on the calculated TDCS values.
Well-converged results can be obtained even with an inappropriately large value of
$l_{\rm max}=20$, which
yields a slightly higher energy of the ground state (c.f.~Fig.~\ref{fig:energy}).
 
Overall, our detailed convergence tests only reveal a very 
weak sensitivity of the TDCS results to both the time scales and the values
of~$l_{\rm max}$. 
Well-converged TDCS results can be obtained by using either $\xi$ grid~I or
$\xi$ grid~II combined with 
$(n_\eta,|m|_{\rm max},l_{\rm max})=(9,4,10)$. 
In the production calculations for the TDCSs shown in the next subsection, we used the
$\xi$ grid~I to discretize the two-electron wave packet and a ``$10+2$" sine-squared
laser pulse with a peak intensity of $10^{15}$ W/cm$^2$.

\subsection{TDCSs for the aligned H$_2$ molecule}
Figures
\ref{fig:tdcs-equal-sharing-ele1-00}, \ref{fig:tdcs-equal-sharing-ele1-90}, and
\ref{fig:tdcs-nonequal-sharing-ele1-90} display the coplanar TDCSs of the
aligned hydrogen molecule at equal and asymmetric ($E_1:E_2=20\%:80\%$) energy sharing.
The two electrons are detected in the same (coplanar) plane defined by the
$\bm{\zeta}$ and $\bm{\epsilon}$ axes. The angles $\theta_1$, $\theta_2$, and $\theta_N$
are all measured with respect to the laser linear polarization axis.  
We compare our TDCS predictions with those obtained in the
time-independent one-center spherical ECS calculation~\cite{Van2006-1}, the time-independent
two-center spheroidal
ECS model~\cite{Tao2010}, and the time-dependent one-center spherical TDCC approach~\cite{Colgan2010}.
The TDCC numbers were recently recalculated with a bigger box size and differ, in some cases substantially, 
from those published originally~\cite{Colgan2007}.
Except for the recent two-center prolate spheroidal ECS results of Tao {\it et al.}~\cite{Tao2010,Rescigno2010}, 
the agreement between the other three sets of results is very satisfactory.  Once again, the
largest relative differences occur when the cross sections are small 
(see Figs.~\ref{fig:tdcs-equal-sharing-ele1-00}$(d)$ and \ref{fig:tdcs-equal-sharing-ele1-90}$(d)$).

Using spheroidal coordinates as well, as an illustrative
example of their two-center ECS approach, Serov and Joulakian~\cite{Serov2009} recently presented the
TDCS at the same photon energy, but only for a single geometry of 
\hbox{$\theta_N=20^\circ$} and \hbox{$\theta_1=40^\circ$} for asymmetric energy sharing of 
$E_1:E_2=80\%:20\%$. 
Although not shown here, there is again good agreement between their results,
Vanroose {\it et al.}'s one-center spherical ECS numbers~\cite{Van2006-1}, and our
time-dependent FE-DVR predictions.
 
It is also interesting to investigate the dominant escape modes for the various scenarios.
These modes are strongly dependent on how the electrons share the excess
energy.  In an arbitrary geometry ($0^\circ \leqslant \theta_N\leqslant
90^\circ$), for example, the back-to-back escape mode $(\theta_{12}=180^\circ)$ is forbidden for 
equal energy sharing. On the other hand, it becomes the dominant mode for significantly asymmetric 
energy sharing, including the 20\%:80\% scenario discussed in the present paper (see Fig.~\ref{fig:tdcs-laser}).

These results can be understood from a symmetry
analysis~\cite{Maulbetsch1995}. 
Equal-energy sharing and back-to-back emission is equivalent to
$\bm{k}_1=-\bm{k}_2$. When we consider the
exchange and parity operations simultaneously in Eq.~(\ref{eq:two-continuum}), we have
$\Phi_{-\bm{k}_2,-\bm{k}_1} = P (-1)^S\Phi_{\bm{k}_1,\bm{k}_2}$. Here  $P=\pm 1$ is the
parity for the gerade and ungerade states, respectively. For the singlet double-continuum 
state with {\em ungerade} parity, we therefore must 
have \hbox{$\Phi_{-\bm{k},\bm{k}}(\bm{r}_1,\bm{r}_2)=0$} at any configuration of 
$\bm{r}_1$ and $\bm{r}_2$. Although the
magnitudes of the momenta $k'_1$ and $k'_2$ are not exactly conserved in the time-dependent
picture, the
ionization events we collect must satisfy the condition $k'_1=k'_2$
(because of the $\delta$ function in Eq.~(\ref{eq:tdcs})) for the equal-energy
sharing.  This is the reason behind the forbidden back-to-back ($\theta_{12}=180^\circ$)
escape mode for the equal energy sharing, as we observed in Figs.~\ref{fig:tdcs-equal-sharing-ele1-00}
and \ref{fig:tdcs-equal-sharing-ele1-90}.
Since the argument does not involve the relative alignment angles, it is valid
for all possible values of~$\theta_N$. 
On the other hand, this is not the case when the excess energy is not evenly distributed
among the two electrons.  Indeed, Figs.~\ref{fig:tdcs-laser}$(a)$ 
and~\ref{fig:tdcs-laser}$(c)$
show maxima in the back-to-back emission, thereby illustrating the dramatic change in the 
dominant escape mode.

For equal-energy sharing in the parallel geometry ($\theta_N=0^\circ$), the electron-electron Coulomb
repulsion suggests that the TDCS should be dynamically small if the two electrons
travel along the same direction. This is in agreement with the numerically
small cross sections (not exactly zero, however) at $\theta_2=0^\circ$
or $360^\circ$ seen in Fig.~\ref{fig:tdcs-equal-sharing-ele1-00}$(d)$.

Recall that the one-photon double-photoionization process in helium~\cite{Colgan2001, Guan2008-2}
shares the same property.
The back-to-back mode is forbidden for equal-energy sharing, and this can be
explained by the above argument. It is one of the similarities between the
molecular hydrogen and the atomic helium targets for double
photo\-ionization. However, Figs.~\ref{fig:tdcs-equal-sharing-ele1-00}, 
\ref{fig:tdcs-equal-sharing-ele1-90}, and
\ref{fig:tdcs-nonequal-sharing-ele1-90} also reveal
significant molecular effects
in the TDCS results.  These are missing for the helium atom, not only in the shape of the angular
distributions, but also in the magnitudes of the cross sections.  Depending on the
relative orientation $(0^\circ<\theta_N<90^\circ)$, there is interference between the
$\Sigma_u$ and $\Pi_u$ symmetries in H$_2$.  
A nice example of this effect was presented by Reddish {\it et al.}~\cite{Reddish2008}. 
Even without interference (i.e., for $\theta_N=0^\circ$ or $90^\circ$), 
the perpendicular geometry shows much larger magnitudes of the TDCS than the parallel
geometry. Figure \ref{fig:tdcs-h2-helium} shows the three cases of angular
distributions:
H$_2$ $\bm{\zeta}\perp \bm{\epsilon}$, H$_2$ $\bm{\zeta}\parallel
\bm{\epsilon}$, and He at equal energy sharing.
Interestingly, in most cases the angular distributions of the perpendicular
geometry resemble those of helium. The molecular effect 
can definitely not be ignored in the parallel geometry for $\theta_1=0^\circ$ 
(c.f.~Fig.~\ref{fig:tdcs-h2-helium}$(a)$). The
{\em forward} escape mode of the second electron is dominant for the H$_2$
parallel geometry. In contrast, the  {\em backward} mode is dominant for the 
H$_2$ perpendicular geometry and also for helium.

In Fig.~\ref{fig:tdcs-out-plane}, we show the TDCS for non\-coplanar geometries. 
Again, all angles are defined with respect to the polarization vector. 
For the perpendicular geometry, Fig.~\ref{fig:tdcs-out-plane}$(a)$ depicts the
escape modes for the configuration of $\bm{k}_1\parallel \bm{\zeta}$ (the
fixed electron) and at the same time $\bm{k}_2$ in the plane
perpendicular to the plane formed by $\bm{\epsilon}$ and~$\bm{\zeta}$. 
Figure~\ref{fig:tdcs-out-plane}$(b)$ shows the TDCS after exchanging the directions
of $\bm{k}_1$ and $\bm{k}_2$ in Fig.~\ref{fig:tdcs-out-plane}$(a)$.  
With the same directions of $\bm{k}_1$ and $\bm{k}_2$ as in
Fig.~\ref{fig:tdcs-out-plane}$(b)$,
Fig.~\ref{fig:tdcs-out-plane}$(c)$ is for the case of the
molecular axis orientated along the polarization vector.
In the parallel case ($\theta_N=0^\circ$), we observe that any escape modes of 
both electrons ejected in the direction perpendicular to~$\bm{\epsilon}$ 
are forbidden. This can be understood by analyzing the
spheroidal harmonics in Eq.~(\ref{eq:tdcs}). In this case, only the $^1\Sigma_u$
states
can be populated.  Hence, only partial waves with $(m_1,m_2)=(-m,m)$ and
$(-1)^{\ell_1+\ell_2}$ can contribute to the cross sections, since
${\cal Y}_{\ell_1 m_1}(k',\hat{\bm{k}}_1){\cal Y}_{\ell_2
m_2}(k',\hat{\bm{k}}_2)$ at
the angles
of $\theta_1=\theta_2=90^\circ$ vanish in spherical coordinates.
Once again, the agreement between our FE-DVR non\-coplanar TDCSs and the
refined TDCC results~\cite{Colgan2010} is excellent.

\section{Summary}
\label{sec.7}

We have presented calculations for one-photon double ionization of the
hydrogen molecule at a photon energy of $75$~eV by solving the time-dependent
\Schro equation in prolate spheroidal coordinates. The triple-differential
cross sections were extracted through the projection of the
time-dependent wave packet onto uncorrelated two-electron continuum states,
a few cycles of field-free time evolution after the laser pulse died off. 

Exhaustive convergence studies of the TDCS results were performed with respect to a 
number of discretization and expansion parameters, as well as the details of the laser field.
These tests provide a strong indication that 
the results for the triple-differential cross sections presented here are well converged
and numerically accurate.  Excellent agreement was obtained 
between the current time-dependent results in prolate spheroidal coordinates, 
those obtained with the ECS approach in spherical coordinates~\cite{Van2006-1} and, finally,
larger TDCC calculations~\cite{Colgan2010} than those published earlier~\cite{Colgan2007}. 

The present calculations do not confirm the 
significant reduction by about $20\%$ in the TDCS results predicted in recent ECS calculations 
in the two-center prolate spheroidal coordinates~\cite{Tao2010}.
Furthermore, our results did not show the level of 
sensitivity to the description of the ground state that was also 
reported by Tao {\it et al.}~\cite{Tao2010}. 

The detailed analysis reported in this study provides a high level of confidence in the present results. 
We hope that they will be used as benchmarks for comparison in future investigations.  
Tables of these results are available in electronic format from the authors upon request.

\vspace{0.1cm}
\section*{Acknowledgments}
We thank Drs.~T.~N.~Rescigno and J.~Colgan for sending their
results in numerical form and for helpful discussions.  This work was supported by
the NSF under grant 
PHY-0757755 (XG and KB) and generous supercomputer resources through the NSF TeraGrid allocation award
TG-PHY$090031$ (Kraken at NICS, Oak Ridge
National Laboratory) and by the Department of Energy allocation award MPH$006$ (Jaguar at NCCS, Oak Ridge
National Laboratory).  Without these computational resources it would have been impossible to study the 
convergence properties of our cross section results on various physical and computational parameters.

\end{document}